\documentclass[10pt]{article}
\usepackage{amsmath,amsthm,amsfonts,amscd,eucal,latexsym,amssymb,bm,amsbsy,mathrsfs} 
\usepackage{epsfig}  %
\def\sp{\hskip -5pt} 
\def\spa{\hskip -3pt} 
\def\cB{{\ca B}}

\def\cE{{\ca E}}

\def\cH{{\ca H}}

\def\cN{{\ca N}}
\def\cO{{\ca O}}
\def\cP{{\ca P}}

\def\cS{{\ca S}}

\def\cW{{\ca W}}

\def\sH{{\mathsf H}}

\def\sS{{\mathsf S}}
\def\sK{{\mathsf K}}

\def\bC{{\mathbb C}}           

\def\bI{{\mathbb I}}

\def\bN{{\mathbb N}}

\def\bR{{\mathbb R}}

\def\bS{{\mathbb S}}

\newsymbol\rest 1316         

\def\gh{{\mathfrak h}} 
 
\def\gH{{\mathfrak H}}

\def\mA{\mathscr{A}} 
\def\mD{\mathscr{D}} 
 
\def\mS{\mathscr{S}}  
\def\mF{\mathscr{F}}  
  
\def\mM{\mathscr{M}} 
\def\mW{\mathscr{W}} 
\def\mB{\mathscr{B}} 
\def\mK{\mathscr{K}} 
\def\mN{\mathscr{N}}

\def\beq{\begin{eqnarray}}
\def\eeq{\end{eqnarray}}
\def\pa{\partial}
\def\at{\left(}               
\def\ag{\left\{}              

\def\ct{\right)}              
\def\cg{\right\}}             
\newcommand{\ca}[1]{{\cal #1}}         

\def\ga{\gamma}

\def\la{\lambda}

\def\si{\sigma}
\def\om{\omega}
\def\vphi{\varphi}

\def\Ga{\Gamma}

\def\La{\Lambda}
\def\Si{\Sigma}
\def\Om{\Omega}

\def\scri{{\Im^{+}}}
\def\scrim{{\Im^{-}}}

\def\supp{\mbox{supp}\:}

\newcommand{\nref}[1]{(\ref{#1})}

\newcounter{proposition}[section]
\newcounter{theorem}[section]
\newcounter{lemma}[section]
\newcounter{definition}[section]
\newcounter{corollary}[section]
\newcounter{remark}[section]
\def\theproposition{\thesection.\arabic{proposition}}
\def\thetheorem{\thesection.\arabic{theorem}}
\def\thelemma{\thesection.\arabic{lemma}}
\def\thedefinition{\thesection.\arabic{definition}}
\def\thecorollary{\thesection.\arabic{corollary}}
\def\theremark{\thesection.\arabic{remark}}

\newcommand{\se}[1]{\section{#1}}

\def\vsp{\vspace{0.2cm}}
\def\vspp{\vspace{0.1cm}}

\newcommand{\ssb}[1]{\subsection{#1}}

\def\ssa #1 {\ifhmode{\par}\fi\refstepcounter{subsection}
  \noindent {\bf\thesubsection.} {\bf #1.}\quad
  \addcontentsline{toc}{subsection}{\protect\numberline{\thesubsection} #1}%
  }

\def\proposizione #1 {\vsp\ifhmode{\par}\fi\refstepcounter{proposition}
  \vsp\ifhmode{\par}\fi\noindent {\bf Proposition \theproposition}. \quad {\em #1}}
\def\teorema #1 {\vsp\ifhmode{\par}\fi\refstepcounter{theorem}
  \vsp\ifhmode{\par}\fi\noindent {\bf Theorem \thetheorem}. \quad {\em #1}}
\def\lemma #1 {\vsp\ifhmode{\par}\fi\refstepcounter{lemma}
  \vsp\ifhmode{\par}\fi\noindent {\bf Lemma \thelemma}. \quad {\em #1}}
\def\definizione #1 {\ifhmode{\par}\fi\refstepcounter{definition}
  \vsp\ifhmode{\par}\fi\noindent {\bf Definition \thedefinition}. \quad {\em #1}}
\def\corollario #1 {\vsp\ifhmode{\par}\fi\refstepcounter{corollary}
  \vsp\ifhmode{\par}\fi\noindent {\bf Corollary \thecorollary}. \quad {\em #1}}
  \def\remark {\vsp\ifhmode{\par}\fi\refstepcounter{remark}
  \vsp\ifhmode{\par}\fi\noindent {\bf Remark \theremark}. }

\def\proof #1 {\vspp\ifhmode{\par}\fi\noindent {\it Proof.} {#1} $\Box$\vsp\par}

\begin{document} 
 
\hfill{\sl Desy 09-105 - UTM 730 - ZMP-HH/09-13,  July 2009} 
\par 
\bigskip 
\par 
\rm 
 
 
\par 
\bigskip 
\LARGE 
\noindent 
{\bf Rigorous construction and Hadamard property of the Unruh state in Schwarzschild spacetime} 
\bigskip 
\bigskip 
\bigskip 
\par 
\rm 
\normalsize 
 

\large
\noindent {\bf Claudio Dappiaggi$^{1,a}$},
{\bf Valter Moretti$^{2,b}$}, {\bf Nicola Pinamonti$^{3,c}$} \\
\par
\small
\noindent $^1$ 
Dipartimento di Fisica Nucleare e Teorica, Universit\`a di Pavia \& Istituto Nazionale di Fisica Nucleare,  
Sezione di Pavia, via A. Bassi 6 I-27100 Pavia, Italy.\smallskip
\smallskip

\noindent$^2$ Dipartimento di Matematica, Universit\`a di Trento
\&  Istituto Nazionale di Fisica Nucleare -- Gruppo Collegato di Trento, via Sommarive 14  
I-38050 Povo (TN), Italy. 

\smallskip
\noindent$^3$ Dipartimento di Matematica, Universit\`a di Genova, Via Dodecaneso 35, I-16146 Genova, 
Italy.

\bigskip

\noindent E-mail: $^a$claudio.dappiaggi@unipv.it,
 $^b$moretti@science.unitn.it,  $^c$pinamont@dima.unige.it\\ 
 \normalsize


\par 
 
\rm\normalsize 

\rm\normalsize 
 

\bigskip
\noindent 
\small 
{\bf Abstract}. The discovery of the radiation properties of black holes prompted the
search for a natural candidate quantum ground state for a massless
scalar field theory on Schwarzschild spacetime, here considered in the  Eddington-Finkelstein representation. Among the several
available proposals in the literature, an important physical role is
played by the so-called Unruh state which is supposed to be appropriate
to capture the physics of a black hole
formed by spherically symmetric collapsing matter. Within this respect, we shall consider a massless Klein-Gordon 
field and we shall rigorously and globally construct such state, that is
on the algebra of Weyl observables localised in the union of the static external region, the future event 
horizon and the non-static black hole region. 
Eventually,  out of a careful use of microlocal techniques, we prove that the built state fulfils, where
defined, the so-called Hadamard condition; hence, it is perturbatively stable, in other words realizing the 
natural candidate with which one could study purely quantum phenomena such as the role of the back reaction 
of Hawking's radiation.\\
>From a geometrical point of view, we shall make a profitable use of a bulk-to-boundary reconstruction 
technique which carefully exploits the Killing horizon structure as well as the conformal asymptotic 
behaviour of the underlying background. From an analytical point of view, our tools will range from 
H\"ormander's theorem on propagation of singularities, results on the role of passive states, and a detailed
use of the recently discovered peeling behaviour of the solutions of the wave equation in Schwarzschild 
spacetime. 
\normalsize
\bigskip 


\tableofcontents

\se{Introduction}
In the wake of Hawking's discovery of the radiating properties of black holes \cite{Hawking}, several 
investigations on the assumptions leading to such result were prompted. In between them, that of Unruh 
\cite{Unruh} caught the attention of the scientific community, since he first emphasised the need to identify
a physically sensible candidate quantum state which could be called the vacuum for a quantum massless scalar
field theory on the Schwarzschild spacetime. This is especially true when such spacetime is viewed as that of
a real black hole obtained out of the collapse of spherically symmetric matter.

If we adopt the standard notation ({\it e.g.}, see \cite{Wald2}), this spacetime can be identified with the 
union of the regions I and III in the Kruskal manifold including the future horizon, though we must omit the 
remaining two regions together with their boundaries \cite{Wald,Wald2}.
To the date, in the literature, three candidate background states are available, going under the name of 
{\em Boulware} (for the external region),
{\em Hartle-Hawking} (for the complete Kruskal manifold) and {\em Unruh} state (for the union of both the 
external and black hole region, including the future event horizon). 
The goal of this paper is to focus   on the latter, mostly due to its
remarkable physical properties. As a matter of fact, earlier works (see for example \cite{Candelas, Balbinot,
Balbinot2}) showed that such a state could be employed to compute the expectation value
of the regularised  stress-energy tensor for a massless scalar field in the physical region of Schwarzschild
spacetime, above pointed out. The outcome is a regular expression on the
future event horizon while, at future null infinity, it appears an outgoing flux of radiation compatible
with that of a blackbody at the black hole temperature. As pre-announced, this result, together with Birkhoff's theorem,  lead to the conjecture that the very
same Unruh state, say $\om_U$, as well as its smooth perturbations, is the natural candidate to be used in the description of the gravitational 
collapse of a spherically symmetric star. However, to this avail, one is also lead to assume that $\om_U$ 
fulfils the so-called {\em Hadamard property} \cite{KW,Wald2}, a prerequisite for states on curved background
to be indicated as physically reasonable. As a matter of fact, in between the many properties, it is noteworthy to
emphasise that such condition assures the existence of a well-behaved averaged stress energy tensor 
\cite{Wald2}. Therefore, from a heuristic point of view, this condition is tantamount to require 
that the ultraviolet behaviour mimics that of the Minkowski vacuum, leading to a physically clear 
prescription on how to remove the singularities of the averaged stress-energy tensor; this comes at hand
whenever one needs to compute the back-reaction of the quantum matter on the gravitational background 
through Einstein's equations. 

The relevance of the Hadamard condition is further borne out by the analysis in
\cite{Fredenhagen}, where the description of the gravitational collapse of a spherically symmetric 
star is discussed and, under the assumption of the existence of suitable algebraic states of Hadamard form, 
it is shown that the appearance of the Hawking radiation, brought, at large times, by any of the said states, is precisely 
related to the scaling-limit behaviour of the underlying two-point function of the state computed on the $2$-sphere 
determined by the locus where the star radius crosses the Schwarzschild one. 

It is therefore manifest the utmost importance to verify whether $\om_U$ satisfies or not the Hadamard 
property, a condition which appears reasonable to assume at least in the static region of 
Schwarzschild spacetime also in view both of the former analysis in \cite{Candelas} and of
the general results achieved in \cite{SV00} applied to those in \cite{DimockKay}.
Indeed such a check is one of the main purposes to write this paper.
Our goals are, however, broader, as we shall make a novel use of the Killing and  conformal structure of
Schwarzschild spacetime in order to construct rigorously and unambiguously the Unruh state, contemporary in 
the static region, inside the internal region and on the future event horizon. 
To this avail, we shall exploit some techniques which in the 
recent past have been successfully applied to manifolds with Killing  horizons, asymptotically flat 
spacetimes (see also the recent \cite{Schroer}) as well as cosmological backgrounds 
\cite{MP05, DMP, Moretti06, Da08a, Da08, Moretti08, DMP2, DMP3}.
That mathematical technology  also relies upon some ideas essentially due to Ashtekar \cite{As87} and that have also received 
attention for applications to electrodynamics \cite{Her97,Her08} in asymptotically flat spacetimes.
Finally, a mathematically similar procedure  was employed in  \cite{Ho00} to prove the Hadamard property of some relevant states in a different physical context.

{From} the perspective of this manuscript, the above cited papers by the authors of the present work are most notable for their underlying common
``philosophy''. To wit, as a first step, one always identifies a preferred codimension $1$ null submanifolds 
of  the background, one is interested in. Afterwards, the classical solutions of the bulk dynamical system,
one wishes to consider, are projected on a suitable function set living on the chosen submanifold. The most
notable property of this set is that one can associate to it a Weyl algebra of observable, which carries  
a corresponding distinguished quantum algebraic state which can be pulled-back to bulk via the above
projection map. On the one hand this procedure induces a state for the bulk algebra of observables and, on
the other hand, such new state enjoys several important physical properties, related both with the symmetries 
of the spacetime and with suitable notions of uniqueness and energy positivity.
 
Particularly, although at a very first glance, one would be tempted to conclude that the Hadamard property 
is automatically satisfied as a consequence of the construction itself and of the known results for the 
microlocal composition of the wave front sets, actually we face an harsher reality. To wit, this feature has 
to be verified via a not so tantalising case by case analysis since it is strictly intertwined to the 
geometrical details of the background. Unfortunately the case, we analyse in this paper, is no exception and,
thus, we shall be forced to use an novel different procedure along the lines below outlines.\\
As a starting point, we shall remark that, in the Schwarzschild background, the role of the distinguished 
null codimension one hypersurface, on which to encode the bulk data, will be played by the union of 
the complete Killing past horizon and of null past infinity. Afterwards, as far as the state is concerned,
it will be then defined on the selected hypersurfaces just following the original recipe due to Unruh: 
a vacuum defined with respect to the affine parameter of the null geodesics forming the horizon and a vacuum 
with respect to the Schwarzschild  Killing vector $\partial_t$ at past null infinity. At a level of two-point 
function, the end point of our construction takes a rather distinguished shape whenever restricted to the 
subalgebra smeared by compactly supported functions, which coincides with the one already noticed in \cite{
Sewell,DimockKay,KW}. Nonetheless, from our perspective, the most difficult technical step will consist of the
extension of the methods employed in our previous papers, 
the reason being that the full algebra both on the horizon and on null infinity is subject to severe constraints
whose origin can be traced back to some notable recent achievements by Dafermos and Rodnianski \cite{DR08}. 
To make things
worse, a similar problem will appear for the state constructed for the algebra at the null infinity.
Nonetheless we shall display a way to overcome both potential obstructions and the full procedure will
ultimately lead to the implementation of a fully mathematically coherent Unruh state, $\om_U$ for the 
spacetime under analysis.\\
Despite these hard problems, the bright side of the approach, we advocate, lies in the possibility to 
develop a global definition for $\om_U$ for the spacetime which encompasses the future horizon, the external 
as well as the internal region. Furthermore our approach will be advantageous since it allows to avoid most 
of the technical cumbersomeness, encountered in the earlier approaches, the most remarkable in 
\cite{DimockKay} (see also \cite{KayD}), where the Unruh state was
defined via an S-matrix out of the solutions of the corresponding field equation of motion in asymptotic 
Minkowski spacetimes. Alas, the definition was established only for the static region and the Hadamard 
condition was not checked, hence leaving open several important physical questions.  \\
Differently, our boundary-to-bulk construction, as pre-announced, will allow us to make a 
full use of the powerful techniques of microlocal analysis, thus leading to a verification of the Hadamard 
condition using the global microlocal characterisation discovered by Radzikowski \cite{Rada, Radb} and 
fruitfully exploited in all 
the subsequent literature (see also \cite{BFK}). Differently from the proofs of the Hadamard property presented in \cite{Moretti08}
and \cite{DMP3}
here we shall adopt a more indirect procedure, which has the further net advantage to avoid potentially 
complicated issues related to the null geodesics reaching $i^-$
from the interior of the Schwarzschild region. The Hadamard property will be first established in the 
static region 
making use of an extension of the formalism and the results presented in \cite{SV00} valid for 
{\em passive states}. The black hole region together 
with the future horizon will be finally encompassed by a profitable use of the H\"ormander's 
propagation of singularity theorem joined with
a direct computation of the relevant remaining part of wavefront set of the involved distributions, all in 
view of well-established results of microlocal analysis. \\
>From a mathematical point of view, it is certainly worth acknowledging  that the results we present in this paper are obtainable
thanks to several remarkable achievements presented in   a recent series of papers due to 
 Dafermos and Rodnianski 
\cite{DR03, DR07, DR08, DR05}, who discussed in great details the behaviour of a solution $\varphi$ of the 
Klein-Gordon equation in Schwarzschild spacetime improving
a classical result of Kay and Wald \cite{KW2}. Particularly we shall benefit from the obtained peeling 
estimates for $\varphi$ 
both on the horizons and at null infinity, thus proving the long-standing conjecture known as Price law 
\cite{DR03}.\\

\noindent In detail, the paper will be divided as follows. \\
In section 2.1, we recall the geometric properties of
Schwarzschild spherically symmetric solution of Einstein's equations. Particularly, we shall introduce,
characterise and discuss all the different regions of the background which will play a distinguished role in
the paper.\\
 Subsequently, in section 2.2 and 2.3, we shall define  the relevant  Weyl C$^*$-algebras of observables 
respectively in the bulk and in the codimension $1$ submanifolds, we are interested in, namely the past
horizon and null infinity.\\ Eventually, in section 2.4, we shall relate bulk and boundary data by means of an
certain isometric $*$-homomorphism whose existence will be asserted and, then, discussed in detail. \\
Section 3 will be instead devoted to a detailed analysis on the relation between
bulk and boundary states. Particularly we shall focus on the state  defined by Kay and Wald 
for a (smaller) algebra associated with the past horizon $\cH$ \cite{KW}, showing that that state can be extended to the (larger) algebra 
relevant for our purposes. \\
The
core of our results will be in section 4 where we shall first define the Unruh state and, then, we will prove
that it fulfils the Hadamard property. Eventually we draw
some conclusions.\\
Appendix A contains further geometric details on the 
conformal structure of Schwarzschild spacetime, while Appendix C encompasses the proofs of most propositions.
At the same time Appendix B is noteworthy because it summarises several different definitions of the KMS
condition and their mutual relation is briefly sketched.\\

\ssb{Notation, mathematical conventions}\label{secgauge} 

\noindent Throughout, $A\subset B$  (or $A \supset B$) includes the case $A=B$, moreover
$\bR_+\doteq [0,+\infty)$, $\bR^*_+ \doteq (0,+\infty)$,
$\bR_-\doteq (-\infty,0]$, $\bR^*_- \doteq (-\infty,0)$
and $\bN\doteq \{1,2,\ldots\}$. For  smooth manifolds $\mM,\mN$, 
$C^\infty(\mM;\mN)$
 is the space of smooth functions $f: \mM\to \mN$.
$C^\infty_0(\mM;\mN)\subset C^\infty(\mM;\mN)$ is the subspace of compactly-supported functions.  
If $\chi : \mM\to \mN$ is a diffeomorphism, $\chi^*$ is the natural extension to tensor bundles 
(counter-, co-variant and mixed) from $\mM$ to $\mN$ (Appendix C in \cite{Wald}).
A {\bf spacetime} $(\mM,g)$ is a  Hausdorff, second-countable, smooth, four-dimensional  
connected manifold $\mM$, whose smooth metric has signature $-+++$. We shall also assume that 
a spacetime is {\em oriented} and {\em time oriented}. The symbol $\Box_g$ denotes the standard 
{\bf D'Alembert operator}
associated with the unique metric, torsion free, affine connection $\nabla_{(g)}$ constructed out of  the metric $g$.
$\Box_g$ is locally individuated by
$g_{ab}\nabla_{(g)}^a \nabla_{(g)}^b$.
We adopt definitions and results about causal structures as in \cite{Wald,ON}, but we take recent results \cite{BS03,BS06} into account, too. 
  If $(\mM,g)$ and $(\mM',g')$ are spacetimes 
 and $S\subset \mM\cap \mM'$, 
 then
  $J^\pm(S;\mM)$ ($I^\pm(S;\mM)$) and $J^\pm(S;\mM')$
 ($I^\pm(S;\mM')$) indicate the {\bf causal} (resp. {\bf chronological}) 
 {\bf sets} generated by $S$ in the spacetime 
 $\mM$ or $\mM'$, respectively. 
An (anti)symmetric bilinear map over a real vector space $\sigma : V \times V \to \bR$
is {\bf nondegenerate} when $\sigma(u,v)=0$ for all $v\in V$ entails $u=0$.\\

\se{Quantum Field theories - bulk to boundary relations}

\ssb{Schwarzschild-Kruskal spacetime}

\noindent In this paper we will be interested in the analysis of a Klein-Gordon scalar massless field theory 
on Schwarzschild spacetime and, therefore, we shall first 
recall the main geometric properties of the background we shall work with. Within this respect, we shall 
follow section 6.4 of \cite{Wald} and we 
will focus on the {\em physical region} $\mM$ of the full Kruskal manifold  $\mK$ (represented in figure 2 
in the appendix), 
associated with a black hole of mass $m>0$. \\$\mM$ is made of the union of three pairwisely disjoint parts, 
$\mW,\mB$ and $\cH_{ev}$ which we shall proceed to describe. According to figure 1 (and figure 2 in the 
appendix), we individuate 
${\mW}$ as the (open) {\bf Schwarzschild wedge},  the (open) {\bf black hole} region is denoted by ${\mB}$ 
while their common boundary, the {\bf event horizon}, is indicated by $\cH_{ev}$.\\
 The underlying metric is easily described if we  make use of the standard {\bf Schwarzschild coordinates} 
 $t,r,\theta,\phi$,
where $t\in \bR$, $r\in (r_S,+\infty)$, $(\theta,\phi) \in \bS^2$ in   ${\mW}$,
whereas  $t\in \bR$, $r\in (0,r_S)$, $(\theta,\phi) \in \bS^2$ in ${\mB}$. 
Within this respect the metric in both ${\mW}$ and ${\mB}$ assumes the standard Schwarzschild form: 
\beq\label{Schw}
-\left(1-\frac{2m}{r}\right) dt \otimes dt + \left(1-\frac{2m}{r}\right)^{-1} dr \otimes dr
+ r^2 h_{\bS^2}(\theta,\phi)\:,
\eeq 
where $h_{\bS^2}$ is the standard metric on the unit $2$-sphere. Here, per direct inspection, one can 
recognise that the locus $r=0$ corresponds to
proper  metrical singularity of this spacetime, whereas  $r=r_S=2m$ individuates the apparent singularity on 
the event horizon.\\
It is also convenient to work with the {\bf Schwarzschild light} or {\bf Eddington-Finkelstein coordinates} 
\cite{KW,Wald2} $u,v,\theta,\phi$ 
which cover  ${\mW}$ and ${\mB}$ separately, such that $(u,v)\in \bR^2$, $(\theta,\phi)\in \bS^2$ and
\begin{align*}
&u \doteq  t-r^* \mbox{ in ${\mW}$,}  \quad u\doteq-t-r^* \mbox{ in $\mB$,}
  \\
&v \doteq t+r^* \mbox{ in ${\mW}$,} \quad v\doteq t-r^* \mbox{ in $\mB$,}
\\
&r^* \doteq r + 2m \ln \left|\frac{r}{2m}-1 \right| \in \bR\:. 
\end{align*}
A third convenient set of {\bf global null coordinates} $U,V,\theta,\phi$ can be introduced on the whole 
Kruskal spacetime \cite{Wald}:
\beq
&U  =  -e^{-u/(4m)}\:, \quad
V  =  {e^{ v/(4m)}} \quad \mbox{in ${\mW}$,} \\
&U  =  e^{  u/(4m)}\:, \quad
V  =  {e^{ v/(4m)}} \quad \mbox{in $\mB$}\:.
\eeq
In this frame,
\begin{gather*}
 \mW\equiv  \{ (U,V, \theta,\phi) \in \bR^2 \times \bS^2\:| \: U<0,V>0\}\:,\\
\mB \equiv  \{ (U,V, \theta,\phi) \in \bR^2 \times \bS^2\:| \:UV<1 \:,  U, V > 0\}\:,\\
 {\mM} \doteq  \mW \cup \mB \cup \cH_{ev} \equiv  \{ (U,V, \theta,\phi) \in \bR^2 \times \bS^2\:| \: UV<1\:,  V > 0\}\:. 
\end{gather*}
Each of the three mentioned regions, seen as independent spacetimes, is globally hyperbolic.
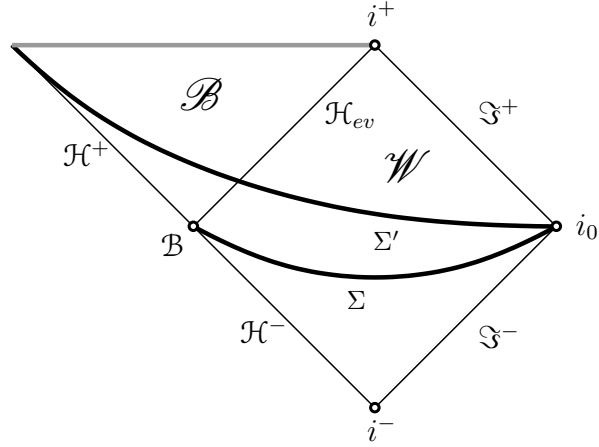
\begin{figure}
\centering
     \begin{picture}(150,130)(0,0) 
       \put(0,10){\includegraphics[height=5cm]{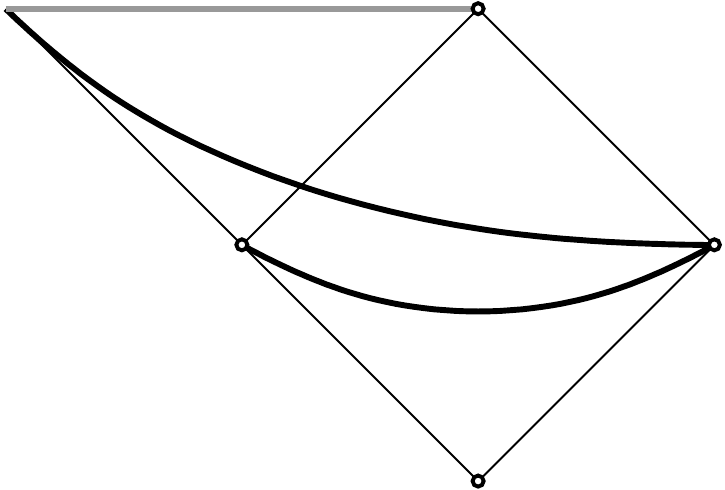}}
       \put(142,98){\LARGE $\mW$}
        \put(65,125){\LARGE $\mB$}
       \put(178,35){\large $\scrim$}
       \put(178,120){\large $\scri$}
       \put(215,78){\large $i_0$}
       \put(136,0){\large $i^-$}    
      \put(136,157){\large $i^+$}
      \put(21,104){\large $\cH^+$}
       \put(58,70){\large $\cB$}
       \put(88,37){\large $\cH^-$}
       \put(120,120){\large$\cH_{ev}$}
       \put(135,72){ $\Si'$}
       \put(125,50){ $\Si$}            \end{picture}
     \caption{The overall picture represents $\mM$.  The regions $\mW$ and $\mB$ respectively correspond to 
     regions $I$ and $III$ in fig 2. The thick horizontal line denotes the metric singularity at $r=0$, 
     $\Si$ is a spacelike Cauchy surface for $\mM$  while $\Si'$ is a spacelike Cauchy surface for $\mW$.}
\end{figure}
The  event horizon of $\mW$,  $\cH_{ev}$ is one of the two horizons we shall consider. The other is the 
complete {\bf past horizon} of $\mM$, $\cH$
which is part of the boundary of $\mM$ in the Kruskal manifold. These horizons are respectively individuated by:
\begin{equation*}
  \cH_{ev} \equiv  \{ (U,V, \theta,\phi) \in \bR^2 \times \bS^2\:| \: U = 0, V>0\}
 \:, \quad 
\cH \equiv  \{ (U,V, \theta,\phi) \in \bR^2 \times \bS^2\:| \: V = 0, U\in \bR\}
\:.
\end{equation*}
For future convenience, we  decompose $\cH$ into the {\em disjoint} union  $\cH = \cH^- \cup \cB \cup \cH^+$ 
where $\cH^\pm$ are defined 
according to $U>0$ or $U <0$ while $\cB$ is the {\bf bifurcation surface} at $U=0$, {\em i.e.}, the spacelike 
$2$-sphere with radius $r_S$ where $\cH$ meets the closure of $\cH_{ev}$.\\
The metric on $\mM$ (and in the whole Kruskal manifold)
takes the form:
\beq g = -\frac{16 m^3}{r} e^{-\frac{r}{2m}} (dU\otimes 
 dV + dV \otimes dU) + r^2 h_{\bS^2}(\theta,\phi)\:,
\label{g}
 \eeq
where the apparent Schwarzschild-coordinate singularity on both $\cH$ and $\cH_{ev}$ 
has disappeared. 
It coincides with the radial Schwarzschild coordinate in both ${\mW}$ and ${\mB}$, hence taking the constant 
value $r_s$ on $\cH_{ev} \cup \cB$; at the same time, the metric singularity, located at $r=0$, corresponds 
to $UV=1$.\\
Let us now focus on the Killing vectors structure.  Per direct inspection of either \eqref{Schw} or 
\eqref{g}, one realizes that there exists a space of Killing
vectors generated both by all the complete Killing fields associated with 
the spherical symmetry -- $\partial_\phi$ for every choice of the polar axis $z$ --
and by a further smooth Killing field $X$. It coincides with $\partial_t$ in both ${\mW}$ and ${\mB}$, 
although
it is timelike and complete in the former static region, while it is spacelike in the latter. Moreover
$X$ becomes light-like and tangent to $\cH$ and  $\cH_{ev}$ (as well as to the whole completion of $\cH_{ev}$
in the Kruskal manifold) while it vanishes exactly on $\cB$, giving rise to the structure of a 
{\em bifurcate Killing horizon} \cite{KW}.
It is finally useful to remark that the coordinates $u$ and $v$
  are respectively well defined on both $\cH_{ev}$ and $\cH^\pm$ where it turns out that:
\begin{equation*}
 X = \mp \partial_u \mbox{ on $\cH^\pm$,} \quad 
 X = \partial_v \mbox{ on $\cH_{ev}$.}
 \end{equation*}
To conclude this short digression on the geometry of Kruskal-Schwarzschild spacetime, we notice that, by 
means of a conformal completion procedure, outlined in Appendix \ref{geometry}, one can coherently 
introduce the notion of future and past  {\bf null infinity}  ${\Im}^\pm$. 
Along the same lines (see again figure 1 and figure 2 in the appendix),
we also shall refer to the formal {\em points at infinity} $i^\pm$, $i^0$, often known as {\bf future}, 
{\bf past} and {\bf spatial infinity} respectively.

\ssb{The Algebra of field observables of the spacetime}\label{observables}
We are interested in the quantisation of the free massless scalar field $\vphi$ \cite{KW,Wald2} on the globally hyperbolic 
spacetime $(\mN,g)$.
The real field $\vphi$ is supposed to be smooth and to satisfy the massless {\bf Klein-Gordon} equation in $(\mN,g)$:
\beq\label{KG}
P_g\vphi =0, \qquad P_g \doteq -\Box_g + \frac{1}{6} R_g\:.
\eeq
Since we would like to use conformal techniques, we have made explicit the conformal coupling 
with the metric,
 even if it has no net effect for the case $\mN=\mM$, since the curvature $R_g$ 
 vanishes therein. Nonetheless, this allows us to make a profitable use of the discussion in Appendix 
 \ref{geometry} when $\mN = \mM$ and  
 $\widetilde{\mM}\supset \mM$. Here $\widetilde{\mM}$ stands for the conformal extension
  (see also figure 2 in the appendix)  of  the previously introduced physical part of Kruskal spacetime 
  ${\mM}$, equipped with the metric $\widetilde{g}$ which coincides with $g/r^2$ in $\mM$. 
 In such case, if the smooth real function $\widetilde{\vphi}$ solves the Klein-Gordon equation in 
$\widetilde{\mM}$  (where now $R_{\widetilde{g}} \neq 0$):
\beq\label{tildeKG}
P_{\widetilde{g}}\widetilde{\vphi} =0, \qquad P_{\widetilde{g}} 
\doteq -\Box_{\widetilde{g}} + \frac{1}{6} R_{\widetilde{g}}\:,
\eeq
$\vphi \doteq \frac{1}{r} \widetilde{\vphi}\spa\rest_{\mN}$ solves (\ref{KG}) in $\mM$. \\
Generally we shall focus our attention to the class $\sS(\mN)$ of real smooth solutions of (\ref{KG})
which have compact support when restricted on a (and thus on every) spacelike smooth Cauchy surface of a 
globally hyperbolic spacetime  $(\mN,g)$. This real vector space becomes a symplectic one $(\sS(\mN),
\sigma_\mN)$ when equipped with the non-degenerate, $\Sigma$-independent, symplectic form
\cite{KW,Wald2,BGP}, for $\vphi_1,\vphi_2 \in \sS(\mN)$,
\beq\label{symp}
\sigma_{\mN}(\vphi_1,\vphi_2)\doteq\int_{\Sigma_\mN} (\vphi_2\nabla_n \vphi_1-\vphi_1\nabla_n \vphi_2) 
d\mu_g(\Sigma_\mN) \:. 
\eeq
Here $\Sigma_\mN$ is any spacelike smooth Cauchy surface of $\mN$ with the metric induced measure 
$\mu_g(\Sigma_\mN)$ and future-directed normal unit vector $n$.\\
Furthermore, for any $\mN' \subset \mN$ such that $(\mN',g\spa\rest_{\mN'})$ is globally hyperbolic,
the following inclusion of symplectic subspaces holds  
$$(\sS({\mN'}),\sigma_{{\mN'}})  \subset (\sS(\mN),\sigma_{\mN})\:.$$ 
Such statement can be proved out of both \eqref{symp} and the independence from the 
used smooth spacelike Cauchy surface. To this avail, it is crucial that every compact portion of a 
spacelike Cauchy surface of $\mN'$ can be viewed as that of a second smooth spacelike, hence acausal, Cauchy 
surface of $\mN$ as shown in  \cite{BS06} (though, for acausality, one should also refer to Lemma 42 in 
Chap. 14 of \cite{ON}).

The quantisation procedure within the algebraic approach goes along the guidelines given in 
\cite{KW,Wald2} as follows: the elementary observables associated with the field $\vphi$ 
are the (self-adjoint) elements of the  {\bf Weyl} ($C^*$-) {\bf algebra} $\cW(\sS(\mN))$ 
\cite{Haag,BR2,KW,Wald2} whose generators will be denoted by $W_\mN(\vphi)$,  $\vphi \in \sS(\mN)$,
 as discussed in the Appendix \ref{algebras}.    

In order to interpret the elements in $\cW(\sS(\mN))$ as {\em local observables} smeared with functions of $C_0^\infty(\mN; \bR)$,
we introduce some further technology. Generally, globally hyperbolicity of the underlying spacetime,
as in the case of $(\mN, g)$, entails the existence of the  {\bf causal propagator}, $E_{P_g}: A_{P_g}-R_{P_
g} :  C_{0}^{\infty}(\mN; \bR) \to \sS(\mN)$  associated to $P_g$ and defined as the difference of the 
advanced and retarded fundamental solution \cite{Wald2,BGP}. Furthermore
$E_{P_g}: C_{0}^{\infty}(\mN; \bR) \to \sS(\mN)$ is linear, surjective with $Ker E_{P_g} = 
P_g(C_0^\infty(\mN; \bR))$ and it is continuous
with respect to the natural topologies of both $C_{0}^{\infty}(\mN; \bR)$ and $C^{\infty}(\mN; \bR)$. 
Finally, given $\psi \in \sS(\mN)$
and any open  neighbourhood $\mN'$ of any fixed smooth spacelike Cauchy surface of $\mN$, there exists $f_\psi
 \in C_{0}^{\infty}(\mN'; \bR)$ with 
$E_{P_g} f_\psi = \psi$. Consequently, $supp\; \psi \subset J^+(supp\; f_\psi; \mN) \cup J^-(supp\; f_\psi; \mN)$.


The standard Hilbert space picture, where the generators $W_\mN(E_{P_g}f)$  are interpreted as exponentials 
of standard field operators, $e^{i\Phi(f)}$, 
can be introduced in the {\em GNS representation}, $(\gH_\omega, \Pi_\omega, \Psi_\omega)$,
of any fixed algebraic state $\omega : \cW(\sS(\mN)) \to \bC$ \cite{Haag,Wald2}, such that
the unitary one-parameter group $\bR \ni t \mapsto \Pi_\omega\left(W_\mN(t E_\mN f)\right)$ is strongly 
continuous.
The {\bf field operators} $\Phi_\omega(f)$ which arise as the self-adjoint generators of those unitary 
one-parameter groups,
$\Pi_\omega\left(W_\mN( E_\mN t f)\right) = \exp\{i t\Phi_\omega(f)\}$,
enjoy all the standard properties of usual quantisation procedure of Klein-Gordon scalar field  based on CCR 
\cite{KW,Wald2}. A different but equivalent definition is presented in the Appendix \ref{algebras}.
A physically important point, which would deserve particular attention, is the choice of physically 
meaningful states, but we shall just come back later to such issue. 

\ssb{Algebras on $\cH$ and $\Im^\pm$} \label{secN}
Let us consider the case $\mN = \mM$, the latter being the physical part of Kruskal spacetime beforehand 
introduced.
 The null $3$-surfaces $\cH$, $\Im^\pm$, as well as, with a certain difference, $\cH_{ev}$ and $\cH^\pm$, 
can be equipped with a Weyl algebra of observables along the guidelines given in 
\cite{DMP2} and references therein. These play a central role in defining  physically interesting states for $\cW(\sS(\mM))$ in the bulk. To keep the paper sufficiently self-contained, we briefly sketch the construction. Let $\cN$ be any $3$-submanifold of a spacetime -- either $(\mM,g)$ 
or its conformal completion $(\widetilde{\mM},\widetilde{g})$ --, whose metric, when  restricted to $\cN$, 
takes the {\bf complete Bondi form}:
\beq 
c_\cN\left(- d\Omega \otimes 
  d\ell - d\ell \otimes d\Omega + h_{\bS^2}(\theta,\phi) \right)
 \eeq 
where $c_\cN$ is a non vanishing constant, while $(\ell, \Omega, \theta,\phi)$ defines a coordinate patch in 
a neighbourhood of
$\cN$ seen as the locus $\Omega=0$ though such that $d\Omega\spa \rest_\cN \neq 0$. Out of this last 
condition we select $\ell\in \bR$ as a complete parameter along the integral lines of $(d\Omega)^a$ and, in 
view of the given hypotheses, $\cN$ turns out to be a null embedded codimension $1$-submanifold 
diffeomorphic to $\bR \times \bS^2$.\footnote{In \cite{DMP,Moretti06,Moretti08,DMP2} it was, more strongly, 
assumed and used the {\bf geodetically complete Bondi form} of the metric, {\it i.e.}, the integral lines of 
$(d\Omega)^a$ forming $\cN$ are complete null geodesics with $\ell \in \bR$ as an affine parameter.  
It happens if and only if, in the considered coordinates, $\partial_\Omega g_{\ell\ell}\spa\rest_\cN =0$ for 
all $\ell \in \bR$ and $(\phi, \theta)\in \bS^2$. This stronger requirement holds here for $\cH$ and ${\Im}^
\pm$.} It is possible to construct a symplectic space  $(\sS(\cN),\si_\cN)$, where 
$\sS(\cN)$ is a real linear space of smooth real-valued functions on $\cN$ which includes
 $C_0^\infty(\cN; \bR)$
and such that the right-hand side of
\beq\label{siN}
\si_\cN (\psi,\psi') \doteq   c_\cN \int_{\cN}   \left(\psi'\frac{\pa \psi}{\pa \ell}  
- \psi\frac{\pa \psi'}{\pa \ell} \right)\; d\ell \wedge  d\bS^2\:, \qquad \psi,\psi' \in \sS(\cN)
\eeq
can be interpreted in the sense of $L^1(\bR \times \bS^2; d\ell \wedge  d\bS^2)$, where $d\bS^2$ is the standard volume form on $\bS^2$. 
Similarly to what it has been done in the bulk, since the only structure of symplectic space is necessary,
one may define the Weyl algebra  
$\cW(\sS({\cN}))$, since the assumption that $C_0^\infty(\cN; \bR) \subset \sS(\cN)$ entails that
$\sigma_\cN$ is non-degenerate, hence $\cW(\sS({\cN}))$ is well-defined.\\
An interpretation of $\si_\cN $ can be given thinking of $\psi,\psi'$ as boundary values of fields  $\vphi,\vphi'\in \sS(\mM)$.
The right hand side of (\ref{siN}) can then be seen as the integral over $\cN$ of the $3$-form $\eta[\vphi,\vphi']$ associated with $\vphi,\vphi'\in \sS(\mK)$
\beq\label{eta}
\eta[\vphi,\vphi'] \doteq  \frac{1}{6} \left(\vphi \nabla^a \vphi' - \vphi' \nabla^a \vphi\right) 
\sqrt{-g}\epsilon_{abcd}dx^b \wedge dx^b \wedge dx^c \:,
\eeq
where $\epsilon_{abcd}$ is  totally antisymmetric  with $\epsilon_{1234} =1$  and where $\psi\doteq \vphi \spa \rest_\cN$,
$\psi'\doteq \vphi' \spa \rest_\cN$. Furthermore, in order to give a sense to the integration of $\eta[\vphi,\vphi']$ over $\cN$, we assume that $\cN$ is positively oriented with respect to its   future-directed normal vector.
The crucial observation  is now that, integrating $\eta[\vphi,\vphi']$ over a spacelike Cauchy surface $\Sigma \subset \mM$, one gets 
exactly the standard symplectic form $\sigma_\mM(\vphi,\vphi')$ in (\ref{symp}) (or that appropriate for the globally hyperbolic 
spacetime containing $\cN$). In view of the validity of the Klein-Gordon equation both for $\vphi$ and 
$\vphi'$, the form $\eta[\vphi,\vphi']$ satisfies $d \eta[\vphi,\vphi']=0$.
Therefore one expects that, as a consequence of Stokes-Poincar\'e theorem it can happen that $\sigma_\mM(\vphi,\vphi')
=\si_\cN (\vphi\spa\rest_\cN,\vphi'\spa\rest_\cN)$.  If this result is valid, it implies the existence of 
an identification of $\cW(\sS(\mM))$ (or some relevant sub algebra) and $\cW(\sS(\cN))$. This is nothing but the idea we want to implement shortly with some generalisations.

In the present case we shall consider the following manifolds  $\cN$ equipped with the Bondi metric and thus the associated 
symplectic spaces $(\sS({\cN}),\sigma_{\cN})$:

(a) $\cH$ with $\ell \doteq U$ where $c_\cN = r_S^2$, $r_S$ being the Schwarzschild radius,  

(b) $\Im^\pm$ with $\ell\doteq u$ or, respectively, $\ell \doteq v$ where $c_\cN= 1$. 

\noindent In the cases (b), the metric restricted to $\cN$ with 
Bondi form is the conformally rescaled and extended Kruskal metric $\widetilde{g}$, with $\widetilde{g}\spa\rest_\mM=
g/r^2$, defined in the conformal completion $\widetilde{\mM}$ of $\mM$, as discussed in the Appendix \ref{geometry}.\\
It is worth stressing that  $\ell$ in Eq. (\ref{siN}) can be replaced, without affecting the left-hand side of (\ref{siN}),
by any other coordinate $\ell' = f(\ell)$, where $f: \bR \to (a,b) \subset \bR$ is any smooth
diffeomorphism.  This allows us to consider the further case of symplectic spaces $(\sS({\cN}),\sigma_{\cN})$ where $\cN$ is: 

(c) $\cH^\pm$
with $\ell\doteq u$ and $c_\cN = r_S^2$,

\noindent independently from the fact that, in the considered coordinates, the metric $g$ over $\cH^\pm$ 
does not take the Bondi form.

\ssb{Injective isometric $*$-homomorphism between the Weyl algebras}
To conclude this section, as promised in the introduction,  we establish the existence of some 
injective (isometric) $*$-homomorphisms which map the Weyl algebras in the bulk into Weyl 
subalgebras defined on appropriate subsets of the piecewise smooth null $3$-surfaces
 $\Im^-\cup \cH $.
To this end we have to specify the definition of $\sS(\cH)$, $\sS(\cH^\pm)$ and 
$\sS(\Im^\pm)$. From now on, referring to the definition of the preferred coordinate $\ell$ as 
pointed out in the above-mentioned list and with the identification of $\cH$, $\cH_{ev}$, $\cH^\pm$ and
$\Im^{\pm}$ with $\bR \times \bS^2$ as appropriate:
\beq\label{SW}
\sS(\cH) \doteq \left\{\psi \in C^\infty(\bR \times \bS^2; \bR)\:\left|
 \: \exists  \exists M_\psi >1,\: C_\psi,C'_\psi  \geq 0
 \mbox{  with  } |\psi(\ell,\theta,\phi)| < \frac{C_\psi}{\ln |\ell|} \right.\right. \:,\nonumber \\  
\left. \left|\partial_\ell\psi(\ell,\theta,\phi)\right|<
\frac{C'_\psi}{|\ell| \ln|\ell|} \quad\mbox{   if } |\ell|>M_\psi\:, (\theta,\phi) \in \bS^2  \right\}\:,
\eeq 
where $\ell = U$ on $\cH$, and
\begin{align}\label{Sscri}
\sS(\Im^\pm) \doteq &\left\{\psi \in C^\infty(\bR \times \bS^2; \bR)\:\left|
 \: \psi(\ell) = 0 \:\mbox{in a neighbourhood of $i^0$ and } 
 \: \exists\exists    C_\psi,C'_\psi  \geq 0
 \mbox{  with  } \nonumber \right.\right.\\
& \left.\left. |\psi(\ell,\theta,\phi)| < \frac{C_\psi}{\sqrt{1+ |\ell|}} \right.\right. \:,
\left. \left|\partial_\ell\psi(\ell,\theta,\phi)\right|<
\frac{C'_\psi}{1+|\ell|} \:,\quad  (\ell, \theta,\phi) \in \bR\times \bS^2  \right\}\:,
\end{align}
where $\ell = u$ on $\Im^+$ or $\ell= v$ on $\Im^-$, and, finally,
\begin{align}\label{SHpm}
\sS(\cH_{ev})\:, \sS(\cH^\pm) \doteq &\left\{\psi \in C^\infty(\bR \times \bS^2; \bR)\:\left| 
 \: \psi(\ell) = 0 \:\mbox{in a neighbourhood of $\cB$ and } \exists\exists   C_\psi,C'_\psi  \geq 0
 \mbox{  with  }\nonumber \right.\right.\\
& \left.\left. |\psi(\ell,\theta,\phi)| < \frac{C_\psi}{1+|\ell|} \right.\right. \:, 
\left. \left|\partial_\ell\psi(\ell,\theta,\phi)\right|<
\frac{C'_\psi}{1+|\ell|}\:,\quad  (\ell, \theta,\phi) \in \bR\times \bS^2  \right\}\:,
\end{align}
where $\ell = v$ on $\cH_{ev}$ and $\ell = u$ on $\cH^\pm$ .\\
It is a trivial task to verify that the above defined sets are real vector spaces; they include 
$C_0^\infty(\bR \times \bS^2; \bR)$ and, if $\psi$ belongs to one of them, $\psi \pa_\ell \psi \in 
L^1(\bR \times \bS^2, d\ell \wedge d\bS^2)$ as requested. Furthermore the above definitions
rely upon the fact that the restrictions of the wavefunctions of $\sS(\mM)$ to the relevant boundaries of 
$\mM$ satisfy the fall-off conditions in (\ref{SW}), (\ref{Sscri}), (\ref{SHpm}) while
approaching $i^\pm$, a fact which will shortly play a crucial role. \\
To go on, notice that, given two real symplectic spaces (with nondegenerate symplectic forms) $(\sS_1, \sigma_1)$ 
and $(\sS_2,\sigma_2)$, we can define the 
direct sum of them, as the real symplectic space $(\sS_1\oplus \sS_1, \sigma_1 \oplus \sigma_2)$, where the nondegenerate
symplectic form $\sigma_1 \oplus \sigma_2 : (\sS_1\oplus \sS_2) \times (\sS_1\oplus \sS_2) \to \bR $ is 
\beq\label{ss}
\sigma_1\oplus \sigma_2 ((f,g), (f',g')) \doteq \sigma_1(f,f') + \sigma_2(g,g')\:, \quad \mbox{for all $f,f' \in \sS_1$
and $g,g' \in \sS_2$.}\eeq
If we focus on the Weyl algebras $\cW(\sS_1)$, $\cW(\sS_2)$, 
$\cW(\sS_1\oplus\sS_2)$, it is natural to identify
the $C^*$-algebra  $\cW(\sS_1\oplus\sS_2)$ with
 $\cW(\sS_1) \otimes \cW(\sS_2)$
 providing, in this way,  
  the algebraic tensor product of the two $C^*$-algebras with a natural $C^*$-norm 
  (there is no canonical $C^*$-norm 
 for the tensor product of two generic $C^*$-algebras). 
  This identification is such that
$W_{\sS_1\oplus \sS_2}((f_1,f_2))$ corresponds to  $W_{\sS_1}(f_1)\otimes W_{\sS_2}(f_2)$ 
for all $f_1\in \sS_1$ and $f_2 \in \sS_2$.

We are now in place to state and to prove the main theorems of this section, making profitable use of the 
results
achieved in \cite{DR05}. Most notably, we are going to show that $\cW(\sS(\mM))$ is isomorphic to a sub 
$C^*$-algebra of $\cW(\sS(\cH))\otimes \cW(\sS(\Im^-))$. 
As a starting point, let us notice that, if $\vphi$ and $\vphi'$ are solutions of the Klein-Gordon equation 
with compact support on any spacelike Cauchy surface $\Sigma$ of $\mM$, the value of $\sigma_\mM(\vphi,\vphi'
)$ is independent on the used $\Sigma$ and, therefore, we can deform it preserving the value of $\sigma_\mM(
\vphi,\vphi')$. A tricky issue arises if one performs a limit deformation where the 
final surface tends to $ \cH \cup \Im^-$ since
\beq\sigma_\mM(\vphi,\vphi') = \sigma_{\cH}\left( \vphi_{\cH},\vphi'_{\cH}\right)
+ \sigma_{\Im^-}\left( \vphi_{\Im^-},\vphi'_{\Im^-}\right)\:, \label{fres}
\eeq
where the arguments of the symplectic forms in the right-hand side (which turns out to belong to the 
appropriate spaces (\ref{SW}), (\ref{Sscri})) are obtained either as restrictions to $\cH$ or as (suitably 
rescaled) limit values towards $\Im^-$ of both $\vphi$ and $\vphi'$. As the map $\vphi \mapsto (\vphi_{\cH},
\vphi_{\Im^-})$ is linear and the sum of the above symplectic forms is the symplectic form $\sigma$ on $\sS 
\doteq \sS(\cH) \oplus \sS(\Im^-)$, this entails that we have built up a symplectomorphism from $\sS(\mM)$ to
$\sS$, $\vphi \mapsto (\vphi_{\cH}, \vphi_{\Im^-})$ which must be injective. 
In view of known theorems \cite{BR2}, this entails the existence of an isometric $*$-homomorphism 
$\imath : \cW(\sS(\mM)) \to 
\cW(\sS(\cH))\otimes \cW(\sS(\Im^-))
$.
Our goal now is to formally state and to prove the result displayed in (\ref{fres}). \\

\teorema \label{Main1} {\em For every $\vphi \in \sS(\mM)$, let us define
$$\vphi_{\Im^-} \doteq \lim_{\to \Im^-} r\vphi\:, \quad \mbox{and}\quad  \vphi_{\cH} \doteq \vphi\rest_{\cH}\:.$$
Then the following facts hold.}

\vskip .2cm

{\em {\bf (a)} 
The linear map
$$\Gamma : \sS(\mM) \ni \vphi \mapsto (\vphi_{\Im^-},\vphi_{\cH})\:,\quad $$
is an injective symplectomorphism of $\sS(\mM)$ into  $\sS(\Im^-) \oplus \sS(\cH)$ 
equipped with the symplectic form, such that, for $\vphi,\vphi' \in \sS(\mM)$:
\beq\label{sigmas}
\sigma_\mM(\vphi,\vphi')
\doteq \sigma_{\Im^-}\left( \vphi_{\Im^-},\vphi'_{\Im^-}\right)
+  \sigma_{\Im_{\cH}}\left( \vphi_{\cH},\vphi'_{\cH}\right).
\eeq
\indent  {\bf (b)} There exists a corresponding injective isometric 
$*$-homomorphism
$$\imath : \cW(\sS(\mM))\spa  \to \cW(\sS(\Im^-))\otimes \cW(\sS(\cH))\:,$$
which is unambiguously individuated by 
$$\imath\left(W_\mM(\vphi)\right) = W_{\Im^-}\left(\vphi_{\Im^-}\right) \otimes W_{\cH}
\left(\vphi_{\cH} \right).
$$}

\noindent{\em Proof}. 
Let us start from point (a). If  $\varphi \in \sS(\mM)$, we can think of it as a restriction to $\mM$ of a 
solution $\varphi'$ of the Klein-Gordon equation in the whole Kruskal manifold.
To this end one should also notice that the initial data of $\varphi$ on a spacelike Cauchy surface of $\mM$
can also be seen as initial data on a spacelike Cauchy surface of the whole Kruskal manifold. This is a
direct application of the results in \cite{BS06} and  \cite{ON}. Therefore $\vphi_{\cH} \doteq \vphi'\rest_{
\cH}$ is well-defined and smooth. Similarly, the functions $\vphi_{\Im^-} \doteq \lim_{\to \Im^-} r\vphi$ are
well defined, smooth and vanish in a neighbourhood of the relevant $i^0$ in view of the following lemma whose
proof is sketched in the Appendix \ref{Appendixproofs}.

\lemma\label{lemma1} {\em If $\vphi \in \sS(\mM)$, $r\vphi$ uniquely extends to a smooth function 
$\widetilde{\vphi}$ defined in $\mM$ joined with open neighbourhoods of $\Im^+$ and $\Im^-$
 included in the conformal extension $\widetilde{\mM}$ of $\mM$ discussed in the Appendix \ref{geometry}.
Furthermore, there are constants $v^{(\vphi)}, u^{(\vphi)} \in (-\infty,\infty)$ such that $\widetilde{\vphi}
$ vanishes in $\mW$ if  $u<u^{(\vphi)}, v> v^{(\vphi)}$ and thus, per continuity, it vanishes in the 
 corresponding limit regions on $\Im^+ \cup \Im^-$.}\\

\noindent Since the map $\Gamma$ is linear by construction, it remains to prove that (i) $\vphi_{\cH} \in 
\sS(\cH)$ and $\vphi_{\Im^\pm}\in \sS(\Im^\pm)$ as defined in (\ref{SW}) and (\ref{Sscri}), and that (ii) 
$\Gamma$ preserves the symplectic forms, {\it i.e.},
\beq\label{sigmas2} 
\sigma_{\mM}(\vphi_1, \vphi_2) = \sigma_{\sS(\cH) \oplus
 \sS(\Im^-)}\left( \Gamma \vphi_1, \Gamma \vphi_2\right)\:. \eeq
Notice that, since $\sigma_{\mM}$ is  nondegenerate, the above identity implies 
that the linear map $\Gamma$ is injective. Let us tackle point (i): since the behaviour 
of  $\vphi_{\Im^-}$ in a neighbourhood of $i^0$ is harmless, we only need to establish the vanishing of both 
$\vphi_{\cH}$ and $\vphi_{\Im^-}$ as they approach $i^-$, with a peeling-off rate consistent with that of
definitions (\ref{SW}) and (\ref{Sscri}). Such a result is a consequence of the following proposition whose 
proof, in Appendix \ref{Appendixproofs}, enjoys a lot from \cite{DR05}.

\proposizione \label{PropDR} {\em  Let us 
 fix $\hat R > r_S$, then the following facts hold:\\
{\bf (a)} If $\vphi \in \sS(\mM)$
and $\widetilde{\vphi}$ extends $r\vphi$ across $\Im^\pm$ as stated in Lemma \ref{lemma1}, 
there exist constants $C_1,C_2\geq 0$  depending on both $\vphi$ and
$C_3,C_4$ 
depending on $\vphi$ and $\hat{R}$, such that the following pointwise bounds hold in both
$\mW\cup \cH_{ev}$ and $\mW\cup \cH^-$:
\beq\label{stime1}
|\vphi|\leq \frac{C_1}{\max\{2,v\}}  \; ,\qquad |X(\vphi)|\leq \frac{C_2}{\max\{2,v\}}\; ,
\eeq
and, respectively,
\beq\label{stime1agg}
|\vphi|\leq \frac{C_1}{\max\{2,-u\}}  \; ,\qquad |X(\vphi)|\leq \frac{C_2}{\max\{2,-u\}}\:.
\eeq
Similarly, if one assumes also  $r \geq \hat R$ and $t > 0$ (including the points on $\Im^+$),
\beq\label{stime2}
|\widetilde{\vphi}|\leq \frac{C_3}{\sqrt{1 +|u|}}  \;, \qquad
|X(\widetilde{\vphi})|\leq \frac{C_4}{1+ |u|}\:, 
\eeq
or, if $r \geq \hat R$ but $t < 0$ (including the points on $\Im^-$),
$$
|\widetilde{\vphi}|\leq \frac{C_3}{\sqrt{1 +|v|}}  \;, \qquad
|X(\widetilde{\vphi})|\leq \frac{C_4}{1+ |v|}. 
$$
$X$ is the smooth Killing vector field on the conformally extended Kruskal spacetime
 with $X =\partial_t$ in $\mW$,
$X = \partial_v$  on $\cH_{ev}$, $X = \partial_u$  on $\cH^-$, $X = \partial_u$ on
 $\Im^+$ and $X = \partial_v$ on
 $\Im^-$.\\ 
 {\bf (b)} If the Cauchy data $(\vphi\spa \rest_{\Sigma}, \nabla_n \vphi\spa \rest_\Sigma)$  
   on $\Sigma\hookrightarrow \mK$ of $\vphi$ tend to $0$ in the sense of the test function (product) topology 
 on $C^\infty_0(\Sigma; \bR)$, then the associated constants $C_i$ tend to $0$, for $i=1,3$.\\
If the Cauchy data $(\vphi'\spa \rest_{\Sigma}, \nabla_n \vphi'\spa \rest_\Sigma)$  on $\Sigma\hookrightarrow \mK$ 
  of $\vphi' \doteq X(\vphi)$ tend to $0$ in the sense of the test function (product) topology 
 on $C^\infty_0(\Sigma; \bR)$, then the associated constants $C_i$ tend to $0$, for $i=2,4$.}\\
 
It is noteworthy to emphasise that, during the final stages of the realization of this paper, a new result on
the peeling-off behaviour of the solutions of the wave equation in Schwarzschild black-hole was made public
\cite{Luk}. Particularly the decay rate on the horizon has been improved; nonetheless, to our purposes, the
original one obtained by Dafermos and Rodnianski suffice.
 
\noindent Since $\vphi$ and $\widetilde{\vphi}$ are smooth, $X(\vphi)=\partial_u\vphi$  on $\cH^-$
and  $X(\widetilde{\vphi})= \partial_v\widetilde\vphi$ on $\Im^-$, it comes out, per direct inspection, that 
$\vphi_{\cH} \in \sS(\cH)$ and $\vphi_{\Im^-}\in \sS(\Im^-)$ since the definitions (\ref{SW}) and 
(\ref{Sscri}) are fulfilled, for $\ell = U$ and $\ell =v$ respectively; furthermore, in view of the last 
statement of the above proposition, it holds $\vphi_{\Im^+}\in \sS(\Im^+)$.\\
In order to conclude, let us finally prove item (ii), that is (\ref{sigmas2}), making use once more of 
Proposition \ref{PropDR}. Let us consider $\vphi,\vphi' \in \sS(\mM)$ and a spacelike Cauchy surface $\Si_\mM$ of $\mM$ so that, 
$$
\sigma_{\mM}(\vphi,\vphi') = 
\int_{\Si_\mM}
\left(\vphi' \nabla_n\vphi -  \vphi \nabla_n\vphi'\right)\:
d\mu_g(\Si_\mM),
$$
where $n$ is the unit normal to the surface $\Si_\mM$ and $\mu_g(\Si_\mM)$ is the metric induced measure on 
$\Si_\mM$ and, in the following, we shall write 
$d\mu_g$ in place of $d\mu_g(\Sigma_\mM)$  and $\Sigma$ in place of $\Sigma_{\mM}$ .
Since both $\vphi$ and $\vphi'$ vanish for sufficiently large $U$, 
we can use the surface $\Sigma$, defined as the locus $t=0$ in $\mW$, 
and, out of the Poincar\'e theorem (employing the $3$-form $\eta$ as discussed in Sec. \ref{secN}), we can 
write
\beq\label{symplecticT}
\sigma_{\mM}(\vphi,\vphi') =
\int_{\Sigma\cap \mW} \vphi' X(\vphi) -  \vphi X(\vphi')  d\mu_g +  
r_S^2 \int_{\cH^+}  \left(
\vphi' \partial_U\vphi -  \vphi \partial_U \vphi'\right)  dU \wedge d\bS^2,
\eeq
where we have used the fact that $\cB \cap \Sigma$ has measure zero.
We shall prove that, if one restricts the integration to $\mW$,
\beq \int_{\Sigma\cap \mW} \vphi' X(\vphi) -  \vphi X(\vphi')  d\mu_g   
= r_S^2 \int_{\cH^-}  \left(
\vphi' \partial_U\vphi -  \vphi \partial_U \vphi'\right)  dU \wedge d\bS^2 \nonumber \\
 + \int_{\Im^-}\left(
\widetilde{\vphi'} \partial_v\widetilde\vphi -  \widetilde\vphi \partial_v\widetilde{\vphi'}\right)du 
\wedge d\bS^2\:. \label{fine1}
\eeq 
Since, with the same procedure, one gets an analogous statement for the portion of the initial integration 
taken in $\mW$, though with the integration in $dU$ extended over $\bR^-$ and the remaining one on $\Im^-$,
this will conclude the proof.\\
In order to prove the identity (\ref{fine1}), we notice at first that:
\begin{gather*}
\int_{\Sigma\cap \mW} \vphi' X(\vphi) -  \vphi X(\vphi')  d\mu_g   =
\int_{[r_S, +\infty) \times\bS^2}   \left.\frac{r^2 \left(
\vphi' X(\vphi) -  \vphi X(\vphi')\right)}{1-2m/r}\right|_{(t=0,r,\theta,\phi)} 
 dr \wedge d\bS^2(\theta,\phi) \:.
 \end{gather*}
Afterwards, we break the integral on the right-hand side into two pieces with respect to the coordinate $r^*$:
\begin{gather*}
 \int_{[r_S, +\infty) \times\bS^2}  \left.\frac{r^2 
 \left(
\vphi' X(\vphi) -  \vphi X(\vphi')\right)}{1-2m/r}\right|_{(t=0,r,\theta,\phi)}   dr \wedge d\bS^2 \nonumber \\
=
\int_{(-\infty, \hat R^*) \times\bS^2}    r^2 \left.\left(
\vphi' X(\vphi) -  \vphi X(\vphi')\right)\right|_{(t=0,r^*,\theta,\phi)} dr^* \wedge d\bS^2 
\nonumber\\
 + \int_{[\hat R^*, +\infty) \times\bS^2} \left.\left(
r\vphi' X(r\vphi) -  r\vphi X(r\vphi')\right)\right|_{(t=0,r^*,\theta,\phi)}  
dr^* \wedge d\bS^2\:.
\end{gather*}
We started assuming $\Sigma$ as the surface $t=0$ in \eqref{symplecticT}; however,
the value of $t$ is immaterial, since we can work,
with a different surface $\Sigma_t$ obtained by evolving $\Sigma$ along the flux of the Killing vector 
$X$. We remind that
$X = \partial_t$ in $\mW$ and $X=0$ exactly on $\cB$, which, as a consequence, is a fixed submanifold 
of the flux. Furthermore we also know that the symplectic form $\sigma_{\mM}(\vphi,\vphi')$ is constructed in
such a way that its value does not change varying $t$, by construction. 
Since $\cB$ is fixed under the flux of $X$, per direct application of 
Stokes-Poincar\'e theorem, one sees that this invariance 
holds also for the integration restricted to $\mW$. In other words, for every $t>0$:
\begin{gather*}
\int_{\Sigma\cap \mW} \sp\sp\sp\vphi' X(\vphi) -  \vphi X(\vphi')  d\mu_g =
\int_{\Sigma_t\cap \mW} \sp\sp\sp\vphi' X(\vphi) -  \vphi X(\vphi')  d\mu_g 
 =
\int_{(u_0(t),+\infty) \times\bS^2}  \sp\sp r^2  
\left.\left(\vphi' X(\vphi) -  \vphi X(\vphi')\right)\right|_{(t,t-u,
\theta,\phi)} \sp du \wedge d\bS^2 \nonumber \\
 + \int_{[v_0(t),+\infty) \times\bS^2}\sp \sp\;
 \left.\left(\widetilde{\vphi'} 
X(\widetilde\vphi) -  \widetilde\vphi X(\widetilde{\vphi'})\right)\right|_{(t,v-t,\theta,\phi)}  
 \sp dv \wedge d\bS^2\nonumber \:,
\end{gather*}
where we have also  changed the variables of integration from $r^*$ either to $v= t+r^*$ or to $u=t-r^*$ and both $u_0(t) \doteq t-\hat R^*$    and  $v_0(t) \doteq t+\hat R^*$ are functions of $t$.
Hence
\beq
\int_{\Sigma\cap \mW} \sp\vphi' X(\vphi) -  \vphi X(\vphi')  d\mu_g 
= \lim_{t\to -\infty}\int_{(u_0(t),+\infty) \times\bS^2}  \sp\sp \left.r^2\left(\vphi' X(\vphi) -  \vphi X(\vphi')\right)
\right|_{(t,t-u,\theta,\phi)} \sp du \wedge d\bS^2 \nonumber \\
 + \lim_{t\to -\infty}\int_{[v_0(t),+\infty) \times\bS^2}\sp \left.\left(
\widetilde{\vphi'} 
X(\widetilde\vphi) -  \widetilde\vphi X(\widetilde{\vphi'})\right)\right|_{(t,v-t,\theta,\phi)} 
  \sp dv \wedge d\bS^2\label{dec}  \:.
\eeq
The former limit should give rise to an integral over $\cH^-$, whereas the latter to an analogous one over 
$\Im^-$. Let us examine them separately and we start from the latter.

To start with we notice that, in view of (a) in Lemma \ref{lemma1}, the integration in $v$ can be performed
in $(-\infty,v_1]$ for some constant $v_1 \in \bR$, without affecting the integral for every $t<0$. 
Therefore 
\begin{gather*}
\lim_{t\to -\infty}\int_{[v_0(t),+\infty)\times\bS^2}\sp\left.\left(\widetilde{\vphi'} X(\widetilde\vphi) - \widetilde
\vphi X(\widetilde{\vphi'})\right)\right|_{(t,v-t,\theta,\phi)}  
 \sp dv \wedge d\bS^2 =\\
\lim_{t\to -\infty} 
 \int_{(-\infty,v_1] \times\bS^2}\sp \chi_{[v_0(t),+\infty)}(v)
 \left.\left(\widetilde{\vphi'} X(\widetilde\vphi) -  \widetilde\vphi X(
 \widetilde{\vphi'})\right)\right|_{(t,v-t,\theta,\phi)}  
 \sp dv \wedge d\bS^2\:,
\end{gather*}
where $\chi_I$ is the characteristic  function of $I\subset \bR$.
In view of the uniform bounds, associated with the constants $C_3$ and $C_4$,
given by  $v$-integrable functions in $(-\infty,v_1]$, as stated in Proposition \ref{PropDR},
 we can now apply Lebesgue's dominated convergence theorem to the limit in the right-hand side:
\beq\label{intone}
\lim_{t\to -\infty}\int_{\bR \times\bS^2}\sp
\chi_{[v_0(t),+\infty)}(v)
\left.\left(\widetilde{\vphi'}
 X(\widetilde\vphi) -  \widetilde\vphi X(\widetilde{\vphi'})\right)\right|_{(t,v-t,\theta,\phi)}  
 \sp dv \wedge d\bS^2 = 
 \int_{\Im^-}\left(
\widetilde{\vphi'} \partial_v\widetilde\vphi -  \widetilde\vphi \partial_v\widetilde{\vphi'}\right)    
dv \wedge d\bS^2\:. 
\eeq

Let us now consider the remaining integral on the right-hand side of (\ref{dec}). 
Fix $u_1 > -\hat R^* \in \bR$ and the following decomposition
\begin{gather*}
\int_{(u_0(t),+\infty)\times\bS^2}  \sp\sp r^2\left.\left(
\vphi' X(\vphi) -  \vphi X(\vphi')\right)\right|_{(t,t-u,\theta,\phi)} \sp du \wedge d\bS^2
=\int_{\Sigma^{(u_1)}_t} \vphi' X(\vphi) -  \vphi X(\vphi')  d\mu_g\:.\nonumber 
\\
+ \int_{(u_0(t),u_1] \times\bS^2}\sp\sp r^2\left.\left(\vphi' X(\vphi) -  \vphi X(\vphi')\right)\right|_{(t,
t-u,\theta,\phi)} \sp\; du \wedge d\bS^2\:.
\end{gather*}
Here we have used the initial expression for the first integral, which is performed 
over the compact subregion $\Sigma^{(u_1)}_t$ of 
$\Sigma_t\cap \mW$ which contains the points with null coordinate $U$ included in $[-\exp\{-u_1/(4m)\}
,0]$. It is noteworthy that such integral is indeed the one of the smooth $3$-form $\eta \doteq \eta [\vphi, 
\vphi']$ defined in (\ref{eta}) and, furthermore, in view of Klein-Gordon equation, $d\eta = 0$. Thus,
by means of an appropriate use of the Stokes-Poincar\'e  theorem, this integral can be re-written as an 
integral of $\eta$ over two regions. The first is a compact subregion of $\cH^+$ which can be constructed as 
the points with coordinate $U\in [U_1,0]$, where $U_1 \doteq -e^{-u_1/(4m)}$; the second, instead  
is the  compact null $3$-surface $S^{(u_1)}_t$ formed by the points in $\mM$ with $U=U_1$ and lying between 
$\Sigma_t$ and $\cH^-$. To summarise:
$$
\int_{\Sigma^{(u_1)}_t} \vphi' X(\vphi) -  \vphi X(\vphi')  d\mu_g = 
\int_{\cH^- \cap \{U_1\leq U \leq 0\}}\eta +  \int_{S^{(u_1)}_t}\eta \:.
$$
If we adopt coordinates $U,V,\theta,\phi$, the direct evaluation of the first integral on the right-hand side
produces: 
$$
\int_{\Sigma^{(u_1)}_t} \vphi' X(\vphi) -  \vphi X(\vphi')  d\mu_g  = 
r_S^2 \int_{\cH^- \cap \{U_1\leq U \leq 0\}}\left( \vphi' \partial_U\vphi -  \vphi \partial_U \vphi' \right) dU \wedge d\bS^2 
+  \int_{S^{(u_1)}_t}\eta \:.  $$ 
We have obtained 
\beq \lim_{t\to -\infty}\int_{(u_1,+\infty)\times\bS^2}  \sp\sp r^2\left.\left(
\vphi' X(\vphi) -  \vphi X(\vphi')\right)\right|_{(t,t-u,\theta,\phi)} \sp\; du \wedge d\bS^2 = 
\nonumber \\
r_S^2 \int_{\cH_R^+ \cap \{U_1\leq U \leq 0\}}\left( \vphi' \partial_U\vphi -  \vphi \partial_U \vphi' \right) dU \wedge d\bS^2 
 + \lim_{t\to -\infty} \int_{S^{(u_1)}_t}\eta 
 \nonumber \\
+ \lim_{t\to -\infty} \int_{(u_0(t),u_1] \times\bS^2}
\sp\sp r^2\left.\left(
\vphi' X(\vphi) -  \vphi X(\vphi')\right)\right|_{(t,t-u,\theta,\phi)} \sp\; du \wedge d\bS^2
\:.\label{agg}
\eeq
If we perform the limit as $t\to -\infty$, one has $\int_{S^{(u_1)}_t}\eta \to 0$, because it is the integral
of a smooth form over a vanishing surface (as $t\to -\infty$), whereas 
\begin{gather*}
 \lim_{t\to -\infty} \int_{(u_0(t),u_1] \times\bS^2}
\sp\sp r^2\left.\left( \vphi' X(\vphi) -  \vphi X(\vphi')\right)\right|_{(t,t-u,\theta,\phi)} \sp\; du \wedge d\bS^2
\\
= \int_{\cH^- \cap \{u_1 \geq u\}}  \chi_{(u_0(t),+\infty)}(u)
r_S^2\left( \vphi' \partial_u \vphi -  \vphi \partial_u\vphi' \right)  du \wedge d\bS^2 \nonumber 
= r^2_S \int_{\cH^- \cap \{U_1\geq U\}}\left( \vphi' \partial_U\vphi -  \vphi \partial_U \vphi' \right) dU 
\wedge d\bS^2,
\end{gather*}
where we stress that the final integrals are evaluated over $\cH^-$ and we have used again Lebesgue's dominated 
convergence theorem thanks to the estimates associated with the constants $C_1$ and $C_2$ in   Proposition 
\ref{PropDR}. Inserting the achieved results in the right-hand side of (\ref{agg}), we find that:
$$
\lim_{t\to -\infty} \int_{\bR \times\bS^2}
\sp\sp \left.r^2\left( \vphi' X(\vphi) -  \vphi X(\vphi')\right)\right|_{(t,t-u,\theta,\phi)} \sp du\; 
\wedge d\bS^2
= r_S^2 \int_{\cH^-}\left( \vphi' \partial_U\vphi -  \vphi \partial_U \vphi' \right) dU \wedge d\bS^2_{
\bS^2}\:.
$$
Such identity, brought in (\ref{dec}), yields, together with (\ref{intone}), (\ref{fine1}), hence concluding 
the proof of (a).\\
Item (b) can be proved as follows. In the following $\sS\doteq \sS(\cH) \oplus \sS(\Im^-)$
and $\sigma$ is the natural symplectic form on such space. Let us consider the closure of the sub $*$-algebra
generated by all the generators $W_{\sS}(\Gamma \vphi) \in \cW(\sS)$ for all $\vphi\in \sS(\mM)$. 
This is still a $C^*$-algebra which, in turn, defines a realization of
 $\cW(\sS(\mM))$ because $\Gamma$ is an isomorphism of the symplectic space 
  $(\sS(\mM), \sigma_{\mM})$
onto the symplectic space $(\Gamma(\sS(\mM)), \sigma\spa\rest_{\Gamma(\sS(\mM)) \times \Gamma(\sS(\mM))})$.
As a consequence of Theorem 5.2.8 in \cite{BR2}, there is a $*$-isomorphism, hence isometric, between 
$\cW(\sS(\mM))$ and the other, just found, realization of the same Weyl algebra,
 unambiguously individuated by the requirement
$\imath_{\mM}(W_{\mM})(\vphi)\doteq W_{\sS}(\Gamma \vphi)$. This isometric $*$-isomorphism individuates an 
injective $*$-homomorphism of $\cW(\sS(\mM))$ into $\cW(\sS,\sigma)\equiv
\cW(\sS(\cH))\otimes \cW(\sS(\Im^-))$.  $\Box$\\ 

\noindent As a byproduct and a straightforward generalisation, the proof of the above theorem also establishes the following:

\teorema \label{Main3}{With the same definitions as in Theorem \ref{Main1} and defining, for $\vphi \in \sS(\mW)$,
$\vphi_{\cH^-} \doteq \lim_{\to \cH^-} \vphi$ and $\vphi_{\cH_{ev}} \doteq \lim_{\to \cH_{ev}} \vphi$, the linear maps 
$$ \Gamma_- : \sS(\mW) 
\ni \vphi \mapsto (\vphi_{\cH^-},\vphi_{\Im^-}) \in \sS(\cH^-) \oplus
 \sS(\Im^-)\:,\quad 
\Gamma_+ : \sS(\mW) 
\ni \vphi \mapsto (\vphi_{\cH_{ev}},\vphi_{\Im^+}) \in \sS(\cH_{ev}) \oplus
 \sS(\Im^+)$$
are well-defined injective symplectomorphisms. As a consequence,
there exists two corresponding injective isometric 
$*$-homomorphisms:
$$\imath^- : \cW(\sS(\mW)) \to \cW(\sS(\cH^-))
\otimes \cW(\sS(\Im^-))\:,\quad \imath^+ : \cW(\sS(\mW)) \to \cW(\sS(\cH_{ev}))
\otimes \cW(\sS(\Im^+))
\:,$$
which are respectively  unambiguously individuated by the requirements for $\vphi\in\sS(\mW)$
$$\quad\imath^-\left(W_{\mW}(\vphi)\right) = W_{\cH^-}
\left(\vphi_{\cH^-} \right)\otimes  W_{\Im^-}\left(\vphi_{\Im^-}\right)\:, \quad 
\quad\imath^+\left(W_{\mW}(\vphi)\right) = W_{\cH_{ev}}
\left(\vphi_{\cH_{ev}} \right)\otimes  W_{\Im^+}\left(\vphi_{\Im^+}\right)
.$$}\\


\noindent Before the conclusion of the present section, we would like to stress that a result similar to 
the one presented in Theorem \ref{Main1} and in Theorem \ref{Main3} can be obtained for the algebra of 
observables defined on the whole Kruskal extension $\mK$ of the Schwarzschild spacetime.
In such case, an injective isometric $*$-homomorphisms
$
\imath_\mK : \cW(\sS(\mK)) \to \cW(\sS(\Im^+_L))
\otimes \cW(\sS(\cH))
\otimes \cW(\sS(\Im^-))
\:
$
can be constructed out of the projection 
$
\Ga_\mK: 
\sS(\mK) 
\ni \vphi \mapsto (\vphi_{\Im^+_L},\vphi_{\cH},\vphi_{\Im^-}) \in \sS(\Im^+_L)\oplus \sS(\cH) \oplus
 \sS(\Im^-)
$
from the requirement
$
\quad\imath_\mK\left(W_{\mK}(\vphi)\right) = W_{\Im^+_L}
\left(\vphi_{\Im^+_L} \right)\otimes  W_{\cH}
\left(\vphi_{\cH} \right)\otimes  W_{\Im^-}\left(\vphi_{\Im^-}\right)
$
where $\Im^+_L$ stands for the future null infinity of the left Schwarzschild wedge in the Kruskal spacetime 
$\mK$. 

\se{Interplay of bulk states and boundary states.}

\ssb{Bulk states induced form boundary states by means of the pullback of $\imath$ and $\imath^-$} 
In this section we construct the mathematical technology to induce algebraic 
states (see Appendix \ref{algebras}) 
on the algebras $\cW(\sS(\mM))$ and $\cW(\sS(\mW))$
 from those defined, respectively, on 
$\cW(\sS(\cH))
\otimes \cW(\sS(\Im^-))$
and $\cW(\sS(\cH^-))
\otimes \cW(\sS(\Im^-))$.
A bit improperly, we shall call {\bf bulk states} those with respect to $\cW(\sS(\mM))$ and
on the other subalgebra defined above while {\bf boundary states} will be called those on 
$\cW(\sS(\cH))\otimes \cW(\sS(\Im^-))$.
To this end, the main tools are Theorem \ref{Main1} and \ref{Main3}. \\
 Let us consider the case of $\cW(\sS(\mM))$ as an example. 
 If the linear functional  $\omega : \cW(\sS(\cH))\otimes \cW(\sS(\Im^-)) \to \bC$ is an algebraic state,
 the 
isometric $*$-homomorphism $\imath$ constructed in Theorem \ref{Main1}
gives rise to $\omega_{\mM}:\cW(\sS(\mM))\to\bC$ defined as
\beq\label{inductedstate}
\omega_{\mM} \doteq \imath^*(\omega)\:,
\qquad\mbox{where}\:\:
\left(\imath^*(\omega)\right)(a) \doteq
\omega \left(\imath(a)  \right)
\:, \quad \mbox{for every $a\in \cW(\sS(\mM))$.}
\eeq
A similar conclusion can be drawn using $\imath^-$ for the corresponding algebra.
The situation will now  be specialised to {\em quasifree states} and, as discussed in Appendix
\ref{algebras}, one of these can be unambiguously defined on $\cW(\sS(\cH))\otimes \cW(\sS(\Im^-))$,
just requiring that
$$
\omega_\mu\left(W_{\cH \cup \Im^-}(\psi)\right) = e^{-\mu(\psi,\psi)/2} \:, \quad 
\mbox{for all $\psi \in \sS(\cH) \oplus \sS(\Im^-)$\:,}
$$
where
$\mu: (\sS(\cH) \oplus \sS(\Im^-)) \times (\sS(\cH) \oplus \sS(\Im^-)) \to \bR$
is a real scalar product satisfying (\ref{sm}). Furthermore the ``quasi-free''-property is stable under
pull-back, {\it i.e.}, if (\ref{inductedstate}) is quasifree, then $\omega_{\mM}$ is such.
Therefore, we can simply turn our attention to quasifree states defined on the boundaries
$\cW(\sS(\Im^\pm))$, $\cW(\sS(\cH))$, and on the possible {\em composition} of such states
in view of the following proposition.

\proposizione\label{propstates} {\em Let $(\sS_1,\sigma_1)$, $(\sS_2,\sigma_2)$ be symplectic spaces and
 $\omega_1$, $\omega_2$ be two quasifree algebraic states on 
$\cW(\sS_1,\sigma_1)$
and $\cW(\sS_2,\sigma_{2})$, induced respectively by the real scalar products $\mu_1 : \sS_1 \times \sS_1 \to \bR$
and $\mu_2: \sS_2 \times \sS_2 \to \bR$. Then the scalar product 
$\mu_1 \oplus \mu_2 : 
(\sS_1\oplus \sS_2) \times (\sS_1\oplus \sS_2) \to \bR$ defined by:
$$\mu_{\sS_1 \oplus \sS_2}((\psi_1,\psi_2), (\psi'_1,\psi'_2)) \doteq 
\mu_1(\psi_1,\psi'_1) + \mu_2(\psi_2,\psi'_2)\:,\quad \mbox{for all
$(\psi_1,\psi_2), (\psi'_1, \psi'_2) \in \sS_1 \oplus \sS_2$}\:,$$
uniquely individuates a quasifree state $\omega_1\otimes \omega_2$ on $\cW(\sS_1)
 \otimes \cW(\sS_2)$ as
$$
\omega_1 \otimes \omega_2\left(W_{\sS_1}(\psi_1)\otimes W_{\sS_2}(\psi_2)\right)
 = e^{-\mu_1 \oplus \mu_2((\psi_1,\psi_2), (\psi_1',\psi'_2))/2} 
\:, \quad \mbox{for all $(\psi_1,\psi_2)\in \sS_1 \oplus \sS_2$}\:.$$}

\noindent{\em Proof}. Only the validity of  (\ref{sm}) for $\mu_1\oplus \mu_2$ has to be proved
with respect to $\sigma_1\oplus \sigma_2$ defined in (\ref{ss}). This fact immediately follows from 
the definition of $\mu_1\oplus \mu_2$ and making use of (2) in remark \ref{remarkstates}. $\Box$\\

\noindent We can iterate the procedure in order to consider the composition of three (or more) states on 
corresponding three (or more) Weyl 
algebras. Hence, in view of the established proposition we may study separately the quasifree states on the 
Weyl algebras $\cW(\sS(\cN))$ associated
to the null surfaces $\cN$  (a)-(c) listed in Sec. \ref{secN}.
 
\ssb{The Kay-Wald quasifree state on $\cW(\cH)$}
We remind the reader that, if $\mu$ individuates a quasifree state over $\cW(\sS,\sigma)$, its 
{\bf two-point function} is defined as 
$\lambda_\mu(\psi_1,\psi_2) \doteq \mu(\psi_1,\psi_2) - \frac{i}{2}\sigma(\psi_1,\psi_2)$ (see Appendix 
\ref{algebras}).
When one focuses on the one-particle space structure $(K_\mu, \sH_\mu)$ (see Appendix \ref{algebras}) one has
$\lambda_\mu (\psi_1,\psi_2) = \langle K_\mu \psi_1, K_\mu \psi_2 \rangle_\mu$, where $\langle \cdot,\cdot  
\rangle_\mu$
is the scalar product in $\sH_\mu$. The two-point function of a quasifree state on a given 
Weyl algebra brings in the same information as the scalar product $\mu$ 
itself since the symplectic form is known {\em a priori}; thus the two-point function
 individuates the state completely.\\ 
In \cite{KW}, some properties are discussed  for a particular state on 
$\cW(\sS(\mK))$, where $\mK$ is the whole Kruskal extension of the Schwarzschild spacetime. 
If existent, such state was proved to be unique with respect to certain algebras of observables 
and to satisfy the KMS property when one works on a suitable algebra of observables in $\mW$. From a physical
perspective, this is nothing but the celebrated Hartle-Hawking state when the background is the whole Kruskal 
spacetime. It is important to remark that, in \cite{KW}, general globally-hyperbolic spacetimes with 
bifurcate Killing horizon are considered, whereas our work only focuses on $\mM$. As an intermediate step, 
Kay and Wald also
showed that the two-point function  $\lambda_{KW}$ of the state has a very particular form when restricted to
the horizon $\cH$, more precisely
\beq\label{lKW}
\lambda_{KW}(\vphi_1,\vphi_2) = \lim_{\epsilon \to 0^+}-\frac{r_S^2}{\pi}
\int_{\bR \times \bR \times \bS^2} \sp\sp\sp\sp\sp 
\frac{\vphi_1\spa\rest_{\cH}\spa(U_1,\theta,\phi)\:\: 
\vphi_2\spa\rest_{\cH}\spa(U_2,\theta,\phi)}{(U_1-U_2 -i\epsilon)^2}\:\:
dU_1 \wedge dU_2 \wedge d\bS^2\:.
\eeq
{\em provided that} $\vphi_1\spa\rest_{\cH},\vphi_2\spa\rest_{\cH} \in C_0^\infty(\bR\times \bS^2; \bR)$.
It is important to stress that the above expression
is valid when $\vphi_1\spa\rest_{\cH}$ and $\vphi_2\spa\rest_{\cH}$ have compact support on $\cH$. 
Actually, the same two-point function was already found both in \cite{Sewell}, while discussing the physical
consequences of the Bisognano-Wichmann theorem, and in \cite{DimockKay}, while analysing the various states in 
the right Schwarzschild wedge $\mW$ of the Kruskal manifold with an {\em $S$-matrix} point of view. In the 
latter paper the two-point function in (\ref{lKW}) was referred to the 
Killing horizon in the {\em two-dimensional Minkowski spacetime rather than Kruskal one}. In such a case 
there are smooth solutions of the Klein-Gordon equation, for $m\geq 0$, and with compactly supported Cauchy 
data, which intersect the horizon in a compact set. These solutions of the {\em characteristic Cauchy 
problem}  can be used in the right-hand side of (\ref{lKW}) when the discussion 
is referred to Minkowski spacetime instead of the Kruskal one.
These ``Minkowskian solutions'', at least in the case $m=0$ where asymptotic completeness 
was proved to hold, are related to the corresponding solutions ({\it i.e.}, $\vphi_1,\vphi_2$) 
in Schwarzschild spacetime by means of a relevant M\o ller operator.                                                                                                             
Unfortunately, in the proper Schwarzschild space, the wavefunctions $\vphi_1$ and $\vphi_2$ with compact support 
on $\cH$ fail to be smooth in general, since they are {\em weak solutions} of the characteristic problem
\cite{DimockKay}, hence they do {\em not} belong to the space $\sS(\mK)$ in general, making difficult the 
direct use of $\lambda_{KW}$.
This is a potential issue in \cite{KW} which has minor consequences for the validity of
  the KMS property discussed below (see also the {\em Note added in proof} in  \cite{KW} for more details).\\
We shall now prove that, actually, such form of the two-point function can
 be extended in order to work on elements of 
$\sS(\cH)$ and, with this extension, it
defines a quasifree state on 
$\cW(\sS(\cH))$. This result is by no means trivial, because the space $\sS(\cH)$  
contains the restrictions to the horizon of the various elements of $\sS(\mK)$, that is all the smooth 
wavefunctions with compact support on spacelike Cauchy surfaces. Our result, which is valid for the 
particular case of the Kruskal spacetime and for $m=0$, is obtained thanks to the achievements recently 
presented in \cite{DR05}. At the same time the space
$\sS(\cH)$ is just the one used in the hypotheses of theorem \ref{Main1}, which assures the existence of the 
$*$-homomorphism
$\imath$. As remarked at the end of the previous section, the procedure can be generalised in order to 
individuate an injective
$*$-homomorphism from the algebra of observables on the whole Kruskal space to the algebras on $\Im_L^+$, 
$\cH$ and $\Im^-$, that is
$\imath_\mK : \cW(\sS(\mK)) \to \cW(\sS(\Im_L^+)) \otimes  \cW(\sS(\cH)) \otimes  \cW(\sS(\Im^-))$.
Therefore, the state on  $\sS(\mK)$  could be used, together with a couple of states on $\cW(\Im^-)$ and on
$\cW(\Im^+_L)$ 
to induce a further one on the whole algebra of observables $\cW(\sS(\mK))$. This should provide an 
{\em existence theorem} 
for the Hartle-Hawking state on the whole Kruskal manifold $\mK$. However we shall not attempt to give such an existence 
proof here and we rather focus attention on another physically interesting state, the so called 
{\em Unruh vacuum} defined only in the 
submanifold $\mM$. Nevertheless, even in this 
case we have to tackle the problem of the extension of the two-point function (\ref{lKW}) to the whole space $\sS(\cH)$. 
  We shall prove the existence of such an extension
  that individuates, moreover, a pure quasifree state on $\cW(\sS(\cH))$, and which
 turns out to be KMS at inverse Hawking's temperature when restricting on a half horizon
  $\cW(\sS(\cH^\pm))$ with respect to the Killing displacements given by $X\spa \rest_{\cH}$.
The way we follow goes on through several steps. 
As a first step we introduce a relevant Hilbert space which we show later to be 
the one-particle space of the quasifree state we wish to define on $\cW(\sS(\cH))$.
The proof of the following proposition is in Appendix \ref{Appendixproofs}.
>From now on,
$$\mF(\psi)
(K,\theta,\phi) \doteq  \int_{\bR}\frac{e^{iKU}}{\sqrt{2\pi}}  \psi(U,\theta,\phi) dU\:,
$$
indicates the $U$-Fourier transform of $\psi$, also in the $L^2$ (Fourier-Plancherel) sense or even in 
distributional sense if appropriate. For all practical purposes, the properties are essentially the same as
for the standard Fourier transform\footnote{For some general properties, see Appendix C of \cite{Moretti08} with the caveat that, in this cited paper, $\mF$ was indicated by $\mF_+$ and the angular coordinates
$(\theta,\phi)$ on the sphere were substituted by the complex ones $(z,\bar{z})$ obtained out of stereographic projection.}.

\proposizione  \label{PropMain4} {\em 
Let  
$\overline{\left(C_0^\infty(\cH; \bC), \lambda_{KW} \right)}$ be the Hilbert completion
of the complex vector space $C_0^\infty(\cH; \bC)$ equipped with the Hermitian scalar product:
\beq\label{lKW2}
\lambda_{KW}(\psi_1,\psi_2) \doteq \lim_{\epsilon \to 0^+} -\frac{r_S^2}{\pi}
\int_{\bR \times \bR \times \bS^2}\spa \frac{\overline{\psi_1(U_1,\theta,\phi)} \psi_2(U_2,\theta,\phi)}
{(U_1-U_2 -i\epsilon)^2}dU_1 \wedge dU_2 \wedge d\bS^2 \:.
\eeq
where $\cH \equiv \bR \times \bS^2$ adopting the coordinate $(U,\theta,\phi)$ over $\cH$.
Denote by 
$\widehat{\psi}_+\doteq\mF(\psi)\spa\rest_{\{K\geq 0,\theta,\phi \in \bS^2\}}$
the restriction to positive values of $K$ of the $U$-Fourier transform of $\psi \in C_0^\infty(\cH; \bC)$.
The following facts hold.

{\bf (a)} The linear map 
$$C_0^\infty(\cH; \bC) \ni \psi \mapsto \widehat{\psi}_+(K,\theta,\phi) \in 
L^2(\bR_+\times \bS^2, 2KdK \wedge r_S^2d\bS^2) \doteq \sH_{\cH}$$ is isometric and uniquely 
extends, by linearity and continuity, to a Hilbert space isomorphism of 
$$
F_{(U)}:  \overline{\left(C_0^\infty(\cH; \bC), \lambda_{KW} \right)} \to
 \sH_{\cH}\:.
$$
 
{\bf (b)} If one switches to $\bR$ in place of $\bC$
$$
 \overline{F_{(U)}\left(C_0^\infty(\cH; \bR) \right)} = \sH_{\cH}\:. 
$$}

\noindent As a second step we should prove that there is a natural way to densely 
embed $\sS(\cH)$ into the Hilbert space 
$\overline{(C_0^\infty(\cH; \bC), \lambda_{KW})}$, that is into $\sH_{\cH}$, as the definition 
of quasifree state requires. However, this is rather delicate because the 
most straightforward way, computing the $U$-Fourier transform of $\psi \in \sS(\cH)$ and 
checking that it belongs to $L^2(\bR_+\times \bS^2, 2KdK \wedge r_S^2d\bS^2) = \sH_{\cH}$,
does not work. The ultimate reason lies in the too slow decay of $\psi$ as $|U| \to +\infty$ obtained in 
\cite{DR05} and embodied 
in the definition of $\sS(\cH)$ itself. As a matter of fact, the idea we intend to exploit is, first, to 
decompose every $\psi \in \sS(\cH)$ as a sum of three functions,
one compactly supported and the remaining ones supported in $\cH^+$ and $\cH^-$ respectively and, then, to 
consider each function separately.
The following proposition, whose proof is in Appendix \ref{Appendixproofs}, analyses the features of 
the last two functions. It also introduces some results, which will be very useful later when dealing 
with the KMS property of the state $\lambda_{KW}$. \\
In the following $H^1(\cH^\pm)_u$ are the Sobolev spaces of the functions $\psi: \bR\times \bS^2 \to \bC$,
 referred to the coordinate $(u,\theta,\phi)\in \bR\times \bS^2$ on $\cH^\pm$, which lie in $L^2(\bR \times 
 \bS^2, du \wedge  d\bS^2)$ together with their first
 (distributional) $u$ derivative. If one follows the same proof as that valid for  
$C_0^\infty(\bR; \bC)$ and  $H^1(\bR)$ along the line of Theorem VIII.6 in \cite{Bre} (employing sequences of
regularising functions which are constant in the angular variables), one establishes  that 
 $C_0^\infty(\cH^\pm; \bC)$ is dense in $H^1(\cH^\pm)_u$.
Every $\psi \in \sS(\cH^\pm)$ is an element of 
 $H^1(\cH^\pm)_u$ as it follows immediately from the definition of $\sS(\cH^\pm)$.

\proposizione \label{propbastarda} {\em The following facts hold, where 
$u\doteq 2r_S \ln(U) \in \bR$ and
$u\doteq -2r_S \ln(-U) \in \bR$ 
are the natural global coordinate covering $\cH^+$ and $\cH^-$, respectively,
while $\mu(k)$ is  the positive measure on $\bR$:
$$
d\mu(k)\doteq 2r^2_S \frac{k e^{2\pi r_S k}}{e^{2\pi r_S k}-e^{-2\pi r_S k}}  dk\:.
$$}

{\em {\bf (a)} 
If $\widetilde{\psi} = (\mF(\psi))(k,\theta,\phi) = \widetilde{\psi}(k,\theta,\phi)$ denotes the
$u$-Fourier  transform of either $\psi \in C_0^\infty(\cH^+; \bC)$ or  $\psi \in C_0^\infty(\cH^-; \bC)$
 the maps 
$$C_0^\infty(\cH^\pm; \bC) \ni \psi \mapsto \widetilde{\psi} \in 
L^2(\bR \times \bS^2, d\mu(k) \wedge d\bS^2)$$ are isometric when $C_0^\infty(\cH^\pm; \bC)$
is equipped with the scalar product $\lambda_{KW}$. It uniquely 
extends, per continuity, to the Hilbert space isomorphisms: 
\beq \label{Fv}
F^{(\pm)}_{(u)} : \overline{C_0^\infty(\cH^\pm; \bC)}\to L^2(\bR \times \bS^2, d\mu(k) \wedge d\bS^2)\:, 
\eeq 
where 
$\overline{C_0^\infty(\cH^\pm; \bC)}$ are viewed as Hilbert subspaces of 
$\overline{\left(C_0^\infty(\cH; \bC), \lambda_{KW} \right)}$\:.}

{\em {\bf (b)} The spaces $\sS(\cH^\pm)$  are naturally identified with
 real subspaces of $\overline{C_0^\infty(\cH; \bC)}$ in view of the following.\\
If either $\{\psi_n\}_{n\in \bN}, \{\psi'_n\}_{n\in \bN} \subset C_0^\infty(\cH^+; \bR)$
or $\{\psi_n\}_{n\in \bN}, \{\psi'_n\}_{n\in \bN} \subset C_0^\infty(\cH^-; \bR)$ and, according to the case,
 both sequences 
$\{\psi_n\}_{n\in \bN}, \{\psi'_n\}_{n\in \bN}$
 converge to the same $\psi \in \sS(\cH^\pm)$ in $H^1(\cH^\pm)$, 
 then  both sequences are of Cauchy type in
  $\overline{\left(C_0^\infty(\cH; \bC), \lambda_{KW} \right)}$ 
  and 
   $\psi_n-\psi'_n \to 0$ in $\overline{\left(C_0^\infty(\cH; \bC), \lambda_{KW} \right)}$. \\
The subsequent identification of $\sS(\cH^\pm)$  with
 real subspaces of $\overline{C_0^\infty(\cH; \bC)}$ is such that:
\beq 
F^{(\pm)}_{(u)}\spa \rest_{\sS(\cH^\pm)} = \mF\spa \rest_{\sS(\cH^\pm)}\:,\label{FF}
\eeq 
where 
 $\mF:  L^2(\bR \times \bS^2, du \wedge d\bS^2) \to L^2(\bR \times \bS^2, dk \wedge d\bS^2)$
stands for the standard $u$-Fourier-Plancherel transform.}\\

\noindent We are finally in place to specify how $\sS(\cH)$ is embedded in $\sH_{\cH}$.
Let us consider a compactly supported smooth function $\chi\in C^\infty(\cH)$, such that 
$\chi=1$ in a neighbourhood of the bifurcation sphere $\cB\in\cH$.
Every $\psi \in \sS(\cH)$ can now be 
decomposed as the sum of three functions:
\beq 
\psi = \psi_- + \psi_0 + \psi_+ \:,
 \quad \mbox{with $\psi_\pm=(1-\chi)  \psi\rest_{\cH^\pm} \in \sS(\cH^\pm)$ and $\psi_0=  \chi\psi \in C_0^\infty(\cH; \bR)$} 
\label{dec0} \;,
 \eeq
Now let us define the map $\sK_{\cH} : \sS(\cH) \to \sH_{\cH} = L^2(\bR_+\times \bS^2, dK \wedge d\bS^2)$
as
\beq\label{mapid}
\sK_{\cH} : \sS(\cH) \ni \psi \mapsto F_{(U)}\left(\psi_-\right)
+ F_{(U)}(\psi_0) +  F_{(U)}\left( \psi_+\right) \in \sH_{\cH}\:,
\eeq
where $F_{(U)}(\psi_\pm)$ makes sense in view of the identification of $\sS(\cH)$ with a real subspace
of  $\overline{\left(C_0^\infty(\cH; \bC), \lambda_{KW} \right)}$ as established in (b) of Proposition 
\ref{propbastarda}. The following proposition yields that $\sK_{\cH}$, in particular, is well-defined 
and injective and, thus, it identifies $\sS(\cH)$ with a subspace of $\sH_{\cH}$. Such identification enjoys 
a nice interplay with the symplectic form $\sigma_{\cH}$.
Furthermore we prove that, $\sK : \sS(\cH) \to \sH_{\cH}$ is continuous if
viewing  $\sS(\cH)$ as a normed space equipped with the norm 
\beq\label{norme}
\| \psi \|_{\cH}^\chi = \| (1-\chi) \psi \|_{H^1(\cH^-)_u} + \| \chi \psi  \|_{H^1(\cH)_U}
+
\|  (1-\chi)\psi \|_{H^1(\cH^+)_u}
\eeq
where
$\|  \cdot  \|_{H^1(\cH^\pm)_u}$ and $\|  \cdot  \|_{H^1(\cH)_U}$ are the norms of the Sobolev spaces $H^1(\cH^\pm)_u$ and
$H^1(\cH)_U$ respectively.\\
Notice that, $\|\cdot\|^\chi_{\cH}$ and $\|\cdot\|^{\chi'}_{\cH}$, defined with respect of different 
decompositions generated by $\chi$ and $\chi'$, are equivalent, in the sense that there are two positive real
numbers $C_1$ and $C_2$ such that
$
C_1 \| \psi \|_{\cH}^\chi 
\leq
\| \psi \|_{\cH}^{\chi'} 
\leq
C_2 \| \psi \|_{\cH}^\chi
$ for all $\psi \in \sS(\cH)$.
The proof of such an  equivalence is based on the decomposition of the various integrals appearing in the mentioned norms with respect to 
both the partitions of the unit $\chi, 1-\chi$ and $\chi', 1-\chi'$.
Afterwards one employs iteratively the triangular  
inequality and the fact that the norms $\| \cdot \|_{H^1(\cH^\pm)_u}$ and 
$\| \cdot\|_{H^1(\cH)_U}$
are equivalent when evaluated on smooth functions whose support is compact and does not include zero, 
because the Jacobian of the change of coordinates in the lone variable $U$ is strictly positive and bounded.
To conclude the proof one should notice that $(\chi-\chi') $ is a compactly supported smooth function on 
the disjoint union of a pair of fixed compact sets
$J\times\bS^2\subset \cH$, that do not contain ${0}$.
Due to such an equivalence, we will often write $\| \psi \|_{\cH}$ in place of $\| \psi \|_{\cH}^\chi$.

\proposizione \label{Propembedding}
{\em The linear map $\sK_{\cH} : \sS(\cH) \to \sH_{\cH}$ in (\ref{mapid}) verifies the following properties:

{\bf (a)} it is independent from the choice of the function $\chi$ used in the 
 decomposition 
(\ref{dec0}) of $\psi \in \sS(\cH)$;

{\bf (b)} it reduces to $F_{(U)}$ when restricting to $C_0^\infty(\cH; \bR)$;

{\bf (c)} it satisfies 
\beq \sigma_{\cH}(\psi,\psi') = -2 Im \langle \sK_{\cH}(\psi), \sK_{\cH}(\psi')
 \rangle_{\sH_{\cH}}\:, \quad \mbox{if $\psi,\psi' \in \sS(\cH)$;}
\label{ultimabastarda1}\eeq

{\bf (d)} it is injective;

{\bf (e)} it holds $\overline{\sK_{\cH}(\sS(\cH))} = \sH_{\cH}$;

{\bf (f)} it is continuous with respect to the norm $\|\cdot\|_{\cH}$ defined in \nref{norme} for every 
choice of the function  $\chi$. Consequently, there exists $C>0$ such that
$$
|\langle \sK_{\cH}(\psi), \sK_{\cH}(\psi')
 \rangle_{\sH_{\cH}}| \leq  C^2 \| \psi \|_{\cH} \cdot \| \psi' \|_{\cH} \;
 \quad \mbox{if $\psi,\psi' \in \sS(\cH)$.}
$$ 
}

\noindent
The proof is in Appendix \ref{Appendixproofs}.
 Collecting all the achievements and presenting some further result,  we can now conclude stating the theorem about the state individuated by
 $\lambda_{KW}$.

\teorema\label{Main4}{The following facts hold referring to $(\sH_{\cH},\sK_{\cH})$.}

{\em  {\bf (a)} The pair $(\sH_{\cH},\sK_{\cH})$ is the one-particle structure for a quasi-free pure state
$\omega_{\cH}$ on  $\cW(\sS(\cH))$ uniquely individuated by the requirement that its two-point function coincides to the right-hand side of 
(\ref{lKW2}) under restriction to $C_0^\infty(\cH; \bR)$.

{\bf (b)} The state $\omega_{\cH}$ is invariant under the natural action of the one-parameter group 
of $*$-automorphisms generated by $X\spa\rest_{\cH}$ and of those generated by the Killing vectors of 
$\bS^2$.

{\bf (c)} The restriction of $\omega_{\cH}$ to $\cW(\sS(\cH^\pm))$ is a quasifree state
$\omega^{\beta_H}_{\cH^\pm}$  individuated by the one particle structure
 $(\sH^{\beta_H}_{\cH^\pm},\sK^{\beta_H}_{\cH^\pm})$ with:
$$
\sH^{\beta_H}_{\cH^\pm} \doteq L^2(\bR\times \bS^2, d\mu(k) \wedge d\bS^2)\quad\mbox{and $\sK^{\beta_H}_{\cH^
\pm} \doteq \mF\spa\rest_{\sS(\cH^\pm)} = F^{(\pm)}_{(u)}\spa\rest_{\sS(\cH^\pm
)}$.}
$$

{\bf (d)} The states $\omega^{\beta_H}_{\cH^\pm}$ satisfy the KMS condition  with respect to one-parameter group 
of $*$-automorphisms generated by, respectively, $\mp X\spa\rest_{\cH}$, with Hawking's inverse temperature
$\beta_H = 4\pi r_S$.

{\bf (e)} If $\{\beta^{(X)}_\tau\}_{\tau \in \bR}$ denotes the pull-back action on $\sS(\cH^-)$ 
of the one-parameter group generated by $X\spa\rest_\cH$, 
that is $(\beta_\tau(\psi))(u,\omega) = \psi(u-\tau,\omega)$, for every $\tau\in \bR$ and every $\psi \in \sS(\cH^-)$
  it holds:
$$
\sK^{\beta_H}_{\cH^-} \beta_\tau^{(X)}(\psi) =  e^{i\tau \hat{k}}\sK^{\beta_H}_{\cH^-} \psi
$$
where $\hat{k}$ is the $k$-multiplicative self-adjoint operator on $L^2(\bR\times \bS^2, d\mu(k) \wedge 
d\bS^2)$. An analogous statement holds for
$\cH^+$.}\\

\noindent{\em Proof}. (a) In view of Proposition \ref{proposition2} (and Lemma \ref{lemma1A}), 
the wanted state is the one uniquely associated with the real scalar product over $\sS(\cH)$
\beq 
\mu_{\cH}(\psi,\psi') \doteq Re \langle \sK_{\cH} \psi, \sK_{\cH}\psi' \rangle_{\sH_{\cH}}\:,
\label{statemuHR}
\eeq
and the one-particle structure is just $( \sH_{\cH},\sK_{\cH})$. 
This holds true provided two conditions are fulfilled, as required in Proposition \ref{proposition2}. 
The first one asks for
\beq
|\sigma_{\cH}(\psi,\psi')|^2 \leq 4 \mu_{\cH}(\psi,\psi)\mu_{\cH}(\psi',\psi')\:.
\label{ultimabastarda}
\eeq
This fact is an immediate consequence of (c) in Proposition \ref{Propembedding}.
The second condition to be satisfied is that $\overline{\sK_{\cH}(\sS(\cH)) + i \sK_{\cH}(\sS(\cH))} = \sH_{
\cH}$ and, actually, a stronger fact holds: $\overline{\sK_{\cH}(\sS(\cH))} = \sH_{\cH}$, because of (e) in 
Proposition \ref{Propembedding}. 
As a consequence, the state $\omega_{\cH}$
 is pure for (d) in Proposition \ref{proposition2}. \\
 (c) We only consider the case of $\cH^+$, the other case being analogous.
  The state  $\omega^{\beta_H}_{\cH^+}$, which is the restriction of $\omega_{\cH}$ to $\cW(\sS(\cH^+))$,
 is by definition completely individuated out of the requirement that
 $$\omega^{\beta_H}_{\cH^+}\left(W_{\cH^+}(\psi) \right) = 
 e^{-\mu_{\cH}(\psi,\psi)/2}
 \quad \mbox{for $\psi \in \sS(\cH^+)$.}$$
 One can also prove the following three facts. (i) If $\psi,\psi' \in \sS(\cH^+)$, then:
 \begin{align*}
 \mu_{\cH}(\psi,\psi') &=  Re \lambda_{KW}(\psi,\psi')
 = Re \langle F^{(+)}_{(u)}\psi, F^{(+)}_{(u)}\psi' \rangle_{\sH^{\beta_H}_{\cH^+}}
 =
Re \langle \widetilde{\psi}, \widetilde{\psi'} \rangle_{L^2(\bR\times \bS^2, d\mu(k)\wedge d\bS^2)}
\nonumber \\ &=
 Re \langle \sK^{\beta_H}_{\cH^+} \psi, \sK^{\beta_H}_{\cH^+}\psi' \rangle_{\sH^{\beta_H}_{\cH^+}}\:, 
 \nonumber
 \end{align*}
 due to (a) and (b) in Proposition \ref{propbastarda}.
 (ii) Condition (\ref{ultimabastarda})  is valid also under restriction to $\sS(\cH^+)$ if one
notices that $\sigma_{\cH^+}= \sigma_{\cH}\spa \rest_{\sS(\cH^+) \times \sS(\cH^+)}$. (iii)
One has  $\overline{\sK^{\beta_H}_{\cH^+}(\sS(\cH^+)) + i \sK^{\beta_H}_{\cH^+}(\sS(\cH^+))} = \sH^{\beta_H}_
{\cH^+}$
 by (a) and (b) of Proposition \ref{propbastarda}, if one bears in mind that $\sS(\cH^+)+ i\sS(\cH^+) 
 \supset C_0^\infty(\cH^+; \bC)$.
 This concludes the proof because (i), (ii) and (iii) entail that $( \sH^{\beta_H}_{\cH^+},\sK^{\beta_H}_{
 \cH^+})$
 is the one-particle structure
 of $\omega^{\beta_H}_{\cH^+}$ in view of  Proposition \ref{proposition2} (and Lemma \ref{lemma1A}).\\
 (b) If $\psi \in \sS(\cH)$, the $1$-parameter group of symplectomorphisms $\beta^{(X)}_\tau$ generated by 
 $X$ 
individuates $\beta^{(X)}_\tau (\psi) \in \sS(\cH)$ such that  $\beta^{(X)}_\tau (\psi)(U,\theta,\phi) =
 (\psi)\left(e^{ \tau/(4m)}U,\theta,\phi\right)$.
This is an obvious consequence of $X= -\partial_u$ on $\cH^+$, $X= \partial_u$ on $\cH^-$ and $X=0$ 
on the bifurcation at $U=0$. Since $\beta^{(X)}$ preserves the symplectic form $\sigma_{\cH}$, 
there must be a representation $\alpha^{(X)}$ of $\beta^{(X)}$,  in terms of $*$-automorphisms of 
$\cW(\sS(\cH))$. We do not need now the explicit form of $\alpha^{(X)}$, rather  let us focus on $\beta^{(X)}
$ again.
If $\psi \in C_0^\infty(\cH; \bR)$, one has immediately, from the definition of $F_{(U)}$, which coincides 
with that of $\sK_{\cH}$ in the considered case, that $\sK_{\cH}(\beta^{(X)}_\tau (\psi))(K,\theta,\phi) = e^
{-\tau/
(4m)}\sK_\cH(\psi)\left(e^{-\tau/(4m)}K,\theta,\phi\right)$. This result generalises to the case where $\psi
\in
\sS(\cH)$ has support in the set $U>0$ (or $U<0)$ as it can be proved along the lines of the proof of (b) 
of Proposition \ref{propbastarda}. Here. if one employs a sequence of smooth functions $\psi_n$ supported in $U>0$ (resp.
$U<0$) which converges to $\psi$ in the Sobolev topology of $H^1(\cH^\pm,du)$ (see the mentioned proof), and 
uses the fact that $\beta^{(X)}_\tau (\psi_n)$ converges to $\beta^{(X)}_\tau (\psi)$ in the same topology. 
Summing up, from definition (\ref{mapid}), one gets that
$\sK_\cH(\beta^{(X)}_\tau (\psi))(K,\theta,\phi) = \left(U^{(X)}_\tau \psi\right)(U,\theta,\phi) \doteq  
e^{-\tau/(4m)}\sK_\cH(\psi)\left(e^{-\tau/(4m)}K, \theta,\phi\right)$
for every $\psi \in \sS(\cH)$ without further restrictions. Since $U^{(X)}_\tau$ is an 
isometry of $L^2(\bR_+\times \bS^2, KdK \wedge d\bS^2)$, in view of the definition of $\omega_{\cH}$ it 
yields that
$\omega_{\cH}(W_{\cH}(\beta^{(X)}_\tau\psi)) = \omega_{\cH}(W_{\cH}(\psi))$ for all $\psi \in 
\sS(\cH)$, and, per continuity and linearity, this suffices to conclude
that $\omega$ is invariant under the action of the group of $*$-automorphisms $\alpha^{(X)}$ induced by $X$. 
The proof for the Killing vectors of $\bS^2$ is similar.\\
 (d) and (e) In $\sS(\cH^-)$, the natural action of the one parameter group of isometries generated by 
 $X\spa\rest_{\cH^-}$
 is $\beta^{(X)}_\tau : \psi \mapsto \beta^{(X)}_\tau(\psi)$ with $\beta^{(X)}_\tau(\psi)(u,\theta,\phi)
 \doteq \psi(u-\tau,\theta,\phi)$, for all
 $u,\tau, \in \bR$, $(\theta,\phi) \in \bS^2$ and for every $\psi \in \sS(\cH^-)$.
 As previously, this is an obvious consequence of $X= \partial_u$ on $\cH^-$.
 Since $\beta^{(X)}$ preserves the symplectic form $\sigma_{\cH^-}$, there must be a representation 
 $\alpha^{(X)}$  of $\beta^{(X)}$,
 in terms of $*$-automorphisms of $\cW(\sS(\cH^-))$. Let us prove that $\alpha^{(X)}$
 is unitarily implemented in the GNS representation of $\omega^{\beta_H}_{\cH^-}$.
To this end, we notice that
 $\beta$ is unitarily implemented 
 in $\sH_{\cH^-}$, the one-particle space of $\omega^{\beta_H}_{\cH^-}$ out of the strongly-continuous
one-parameter group of unitary operators $V_{\tau}$ such that $\left(V_\tau \widetilde{\psi}\right)(k,
\theta,\phi) = e^{i k \tau}\widetilde{\psi}(k,\theta,\phi)$. This describes
 the time displacements with respect to the Killing vector $\partial_u$. 
 Thus the self-adjoint generator of $V$ is
 $h : Dom(\hat{k}) \subset L^2(\bR\times \bS^2, d\mu(k) \wedge d\bS^2) \to L^2(\bR\times \bS^2, d\mu(k) \wedge d\bS^2)$
 with $\hat{k}(\phi)(k,\theta,\phi) = k \phi(k,\theta,\phi)$ and
 $$
 Dom(\hat{k}) \doteq \left\{ \phi \in L^2(\bR\times \bS^2, d\mu(k) \wedge d\bS^2) \:\left| \: 
 \int_{\bR\times \bS^2} |k\phi(k,\theta,\phi)|^2 d\mu(k) \wedge d\bS^2 <+ \infty\right.\right\}\:.
 $$
Per direct inspection, if one employs the found form for $V$ and exploits 
$$
\omega^{\beta_H}_{\cH^-} \left(W_{\cH^-}(\psi)\right) 
= 
e^{-\frac{1}{2}\langle \widetilde{\psi}, \widetilde{\psi}\rangle_{L^2(\bR\times \bS^2, d\mu(k)\wedge
d\bS^2)}}\:,
$$ 
one sees that $\omega^{\beta_H}_{\cH^-}$ is invariant 
under $\alpha^{(X)}$, so that it must admit a unitary implementation \cite{Araki}. 
In order to establish that the $\alpha^{(X)}$-invariant 
 quasifree state $\omega^{\beta_H}_{\cH^-}$ over the Weyl algebra $\cW(\sS(\cH^-))$ 
 is a KMS state with inverse temperature $\beta_H = 4\pi r_S$ with respect to $\alpha^{(X)}$ which, in turn,
is unitarily implemented by $V = \{\exp\{i \tau \hat{k}\}\}_{\tau\in \bR}$ in the one particle space 
$\sH^{\beta_H}_{\cH^-}$, on can use proposition \ref{kkms} in the appendix and prove that 
$\sK^{\beta_H}_{\cH^-} (\sS(\cH^-)) \subset Dom\left( e^{-\frac{1}{2}\beta \hat{k}}\right)$ while 
$\langle e^{i\tau \hat{k}} \sK^{\beta_H}_{\cH^-}  \psi, \sK^{\beta_H}_{\cH^-}  \psi' \rangle 
= \langle e^{-\beta_H \hat{k}/2}\sK^{\beta_H}_{\cH^-} \psi',   e^{-\beta_H \hat{k}/2} e^{i\tau \hat{k}} 
\sK^{\beta_H}_{\cH^-}  \psi \rangle$. Luckily these requirements hold per direct inspection since
$\sK^{\beta_H}_{\cH^-}(\psi) = \widetilde{\psi}\in L^2(\bR \times \bS^2, d\mu(k) \wedge d\bS^2)$. Here we
used the explicit form of the measure $\mu(k)$ and the identity $\overline{\widetilde{\psi}(-
k,\omega)} = 
\widetilde{\psi}(k,\omega)$ if $\psi \in \sS(\cH^-)$ because $\psi$ is real-valued.
 The case of 
$\cH^+$ is strongly analogous, the only difference being $X\spa\rest_{\cH^+} = -\partial_u$.
 $\Box$\\
 
\noindent We conclude stating without proof (straightforward in this case) the following proposition which
concerns the natural $X$-invariant vacuum states  of $\cH^-$  and $\cH_{ev}$(actually, a {\em quasifree regular ground states}
in the sense of \cite{KW}).

\proposizione  \label{PropMain4.5} {\em If $\sK_{\cH^-}: \sS(\cH^-) \to \sH_{\cH^-}\doteq
 L^2(\bR_+ \times \bS^2, 2kdk\wedge d\bS^2)$ 
denotes the standard $u$-Fourier-Plancherel transform, followed by the restriction to $\bR_+\times \bS^2$,
the following facts hold.}

{\em  {\bf (a)} The pair $(\sH_{\cH^-},\sK_{\cH^-})$ is the one-particle structure for a quasi-free pure state
$\omega_{\cH^-}$ on  $\cW(\sS(\cH^-))$.}

{\em  {\bf (b)} The state $\omega_{\cH^-}$ is invariant under the natural action of the one-parameter group 
 of $*$-automorphisms generated by $X\spa\rest_{\cH^-}$ and those generated by the Killing vectors of $\bS^2$.\\
If one replaces the $u$-Fourier-Plancherel transform with the $v$-Fourier-Plancherel one, an analogous 
state $\omega_{\cH_{ev}}$ can be defined, which is invariant under the natural action of the one-parameter 
group of $*$-automorphisms generated by $X\spa\rest_{\cH_{ev}}$ and those generated by the Killing vectors of 
$\bS^2$.}

\ssb{The vacuum state $\omega_{\scrim}$ on $\cW(\Im^-)$}
We now introduce a relevant vacuum state $\omega_{\Im^-}$ on $\cW(\Im^-)$ 
which is invariant with respect to $u$-displacements and under the isometries of $\bS^2$. 
The idea is, in principle, the same as for $\omega_{\cH}$, {\it i.e.}, 
one starts from a two-point function similar to $\lambda_{KW}$, with the important difference that 
the coordinate $U$ is now replaced by $v$. As a starting point we state the following proposition whose 
proof is, {\em mutatis mutandis}, identical to that of proposition \ref{PropMain4}.

\proposizione  \label{PropMain5} {\em Consider the Hilbert completion 
$\overline{\left(C_0^\infty(\Im_R^-; \bC), \lambda_{\Im^-} \right)}$
of the complex vector space $C_0^\infty(\Im^-; \bC)$ equipped with the Hermitian scalar product:
\beq\label{lscri}
\lambda_{\Im^-}(\psi_1,\psi_2) \doteq \lim_{\epsilon \to 0^+} -\frac{1}{\pi}
\int_{\bR \times \bR \times \bS^2}\spa \frac{\overline{\psi_1(v_1,\theta,\phi)} \psi_2(v_2,\theta,\phi)}{
(v_1-v_2 -i\epsilon)^2}dv_1 \wedge dv_2 \wedge d\bS^2 \:,
\eeq
where $\scrim \equiv \bR \times \bS^2$ adopting the coordinate $(v,\theta,\phi)$ over $\scrim$.
The following facts hold.

{\bf (a)} If $\widehat{\psi}_+(k,\theta,\phi) \doteq \mF(\psi)\spa\rest_{\{k\geq 0, (\theta,\phi) \in 
\bS^2\}}(k,\theta,\phi)$ 
denotes the $v$-Fourier transform  of
$\psi \in  C_0^\infty(\scrim; \bC)$
restricted to $k\in  \bR_+$  (see the Appendix C in \cite{Moretti08}), the map 
$$C_0^\infty(\scrim; \bC) \ni \psi \mapsto \widehat{\psi}_+(k,\theta,\phi) \in 
L^2(\bR_+\times \bS^2, 2kdk \wedge d\bS^2) =: \sH_{\scrim}$$ is isometric and it uniquely 
extends, per continuity, to a Hilbert space isomorphism of 
\beq\label{Fu} 
F_{(v)}:  \overline{\left(C_0^\infty(\scrim; \bC), \lambda_{\scrim} \right)} \to
 \sH_{\scrim}\:.
 \eeq
 
{\bf (b)} If one replaces $\bC$ with $\bR$:
 \beq \label{scriaggdens}
 \overline{F_{(v)}\left(C_0^\infty(\scrim; \bR) \right)} = \sH_{\scrim}\:. 
 \eeq}

\noindent We have now to state and to prove the corresponding of the Proposition \ref{Propembedding}, 
which establishes that there exists a state $\omega_\scrim$ which is completely determined by 
$\lambda_\scrim$ and it is such that the one-particle space 
coincides with $\sH_\scrim$. The delicate point is to construct the corresponding of the $\bR$-linear 
map $\sK_{\cH}$, which now has to be thought of as $\sK_{\scrim}: \sS(\scrim) \to \sH_\scrim$. 
Let us notice that $\sK_{\scrim}$ cannot be defined as the 
$v$-Fourier transform (neither the Fourier-Plancherel transform), 
since the elements of $\sS(\scrim)$ do not decay rapidly enough. Similarly to what done before,
 a suitable extension with respect to the topology of 
$\overline{\left(C_0^\infty(\scrim; \bC), \lambda_{\scrim} \right)}$ is necessary. 
To this end, we are going to prove that the real subspace of the functions of $\sS(\scrim)$ supported in the region 
$v>0$ can be naturally identified with a real subspace of $\overline{\left(C_0^\infty(\scrim; \bC), \lambda_{\scrim} \right)}$. 
This is stated in the following proposition whose proof is in the Appendix \ref{Appendixproofs}.
In the following, we pass to the coordinate over $\bR$ defined by $x \doteq \sqrt{v}$ if $v\geq 0$ and $x\doteq -\sqrt{-v}$ if $v\leq0$.
 Then, if we adopt the coordinate $x$ over the factor $\bR$ of $\scrim \equiv \bR \times \bS^2$,
  the Sobolev space $H^1(\scrim)_x$, is that of the functions which belong to
  $L^2(\bR \times \bS^2, dx \wedge d\bS^2)$ with their (distributional) 
  first $x$ derivative. Notice that, in view of the very
  definition of 
  $\sS(\scrim)$, if $\psi$ is supported in the subset of $\scrim$ with $v<0$ ({\it i.e.}, $x<0)$ and $\psi 
  \in \sS(\scrim)$, then 
  $\psi \in H^1(\scrim)_x$.

\proposizione\label{propidscri}
{\em If $\psi \in \sS(\scrim)$ and $\supp(\psi) \subset \bR^*_- \times \bS^2$ (where $\bR^*_-\doteq (-\infty,0)$), 
the following holds.

{\bf (a)}
 Every sequence $\{\psi_n\}_{n\in \bN} \subset C_0^\infty(\bR^*_- \times \bS^2; \bR)$ such 
 $\psi_n \to \psi$ as $n\to +\infty$ in  $H^1(\scrim)_x$
is necessarily of Cauchy type in $\overline{\left(C_0^\infty(\scrim; \bC), \lambda_{\scrim} \right)}$.

{\bf (b)} There is $\{\psi_n\}_{n\in \bN} \subset C_0^\infty(\bR^*_- \times \bS^2; \bR)$ such 
 $\psi_n \to \psi$ as $n\to +\infty$ in $H^1(\scrim)_x$ and, 
 if $\{\psi'_n\}_{n\in \bN} \subset C_0^\infty(\bR^*_- \times \bS^2; \bR)$ converges to the same $\psi$
 in  $H^1(\scrim)_x$, then $\psi'_n-\psi_n\to 0$ in 
 $\overline{\left(C_0^\infty(\scrim; \bC), \lambda_{\scrim} \right)}$.\\
As a consequence every $\psi \in \sS(\scrim)$ with $\supp(\psi) \subset \bR^*_+ \times \bS^2$
 can be naturally identified with a corresponding element of $\overline{\left(C_0^\infty(\scrim; \bC), \lambda_{\scrim} \right)}$,
  which we indicate with the same symbol $\psi$. \\ With this identification it holds
\beq
F_{(v)} \spa\rest_{\sS(\scrim)} = \Theta\cdot \mF\spa\rest_{\sS(\scrim)}\:, \label{FumF}
\eeq
and, for $\psi,\psi'\in \sS(\scrim)$,
\beq
\lambda_{\scrim}(\psi,\psi') = \int_{\bR_+\times \bS^2} \left(\overline{\mF(\psi')}\left( I + C\right)
\mF(\psi)\right)(h,\omega)\: 2hdh\wedge d\bS^2(\omega)\:,\label{antilin}
\eeq
where, $\Theta(h)=0$ if $h\leq 0$ and $\Theta(h)=1$ otherwise. Here
$\mF : L^2(\bR \times \bS^2, dx \wedge d\bS^2) \to L^2(\bR \times \bS^2, dh \wedge d\bS^2)$
is the $x$-Fourier-Plancherel transform ($x\doteq -\sqrt{-v}$ if $v\leq 0$ and $x\doteq \sqrt{v}$ if 
$v\geq 0$) while $C$ stands for the standard complex conjugation.}\\

 \noindent We are in place to define the map $\sK_{\scrim}$ along the lines followed for $\sK_{\cH}$.
 Let $\chi$ be a non-negative  smooth function on $\scrim$ whose support is contained in $\bR^*_-\times\bS^2$, and such that 
$\eta(v,\theta,\phi)=1$ for $v<v_0<0$. 
Consider $\psi \in \sS(\scrim)$
  and decompose it as:
 \beq \psi = \psi_0 + \psi_-\:, \mbox{where $\psi_0 = (1-\eta)\psi$ and 
 $\psi_-=\eta\psi \in \sS(\scrim)$} \label{dec0scri}
 \eeq
Obviously, $\psi_0 \in C_0^\infty(\scrim;\bR)$ and
 $\supp(\psi_-) \subset \bR^*_-\times \bS^2$, where $\bR^*_-$ is referred to the coordinate $v$ on $\bR$. 
 Finally, let us define
 \beq
 \sK_\scri(\psi) \doteq F_{(v)}(\psi_0) +  F_{(v)}(\psi_-)\:, \quad \forall \psi \in \sS(\scrim)\:,
\label{mapidscri}
 \eeq
where, $\psi_-$ in the second term is considered an element of $\overline{\left(C_0^\infty(\scrim; \bC), 
\lambda_{\scrim} \right)}$ in view of Proposition \ref{propidscri}. 
The map $\sK_\scri : \sS(\scrim) \to \sH_\scrim$ is continuous when the domain is equipped 
with the norm 
\beq\label{normeScri}
\| \psi \|_{\scrim}^\eta = \|  \psi_- \|_{H^1(\scrim)_x} + \| \psi_0  \|_{H^1(\scrim)_v}
\eeq
where
$\|  \cdot  \|_{H^1(\scrim)_x}$ and $\|  \cdot  \|_{H^1(\scrim)_v}$ are the norms of the Sobolev spaces 
$H^1(\scrim)_x$ and $H^1(\scrim)_v$ respectively, the latter hence with respect to the $v$-coordinate. 
Let us remark that, as before, different $\eta$ and $\eta'$ produce equivalent norms $\|\cdot\|^\eta_{
\scrim}$ and $\|\cdot\|^{\eta'}_{\scrim}$; for this reason we shall drop the index $\eta$ in $\|\cdot\|^\eta_\scrim$ if 
not strictly necessary. The following proposition states that the definition of $\sK_{\scrim}$, 
given above, is meaningful; its proof, which will be discussed in the Appendix \ref{Appendixproofs}, relies 
on Propositions \ref{PropMain5} and  \ref{propidscri} and it is very similar to that of Proposition 
\ref{Propembedding}.

\proposizione \label{Propembeddingscri}
{\em The linear map $\sK_{\scrim} : \sS(\scrim) \to \sH_{\scrim}$ in (\ref{mapidscri}) enjoys the following properties:

{\bf (a)} it is well-defined, i.e.,  it is independent from the chosen decomposition 
(\ref{dec0scri}) for a fixed $\psi \in \sS(\scrim)$;

{\bf (b)} it reduces to $F_{(v)}$ when restricting to $C_0^\infty(\scrim; \bR)$;

{\bf (c)} it satisfies: 
$$\sigma_{\scrim}(\psi,\psi') = -2 Im \langle \sK_{\scrim}(\psi), \sK_{\scrim}(\psi') \rangle_{\sH_{\scrim}}\:, 
\quad \mbox{if $\psi,\psi' \in \sS(\scrim)$;}
$$

{\bf (d)} it is injective;

{\bf (e)} it holds $\overline{\sK_{\scrim}(\sS(\scrim))} = \sH_{\scrim}$;

{\bf (f)}
it is continuous with respect to the norm $\|\cdot\|_{\scrim}$ defined in \nref{normeScri} for every choice 
of the function $\eta$. Consequently, there exists a constant $C>0$ such that:
$$
|\langle \sK_{\scrim}(\psi), \sK_{\scrim}(\psi') \rangle_{\sH_{\scrim}}|\leq  C^2 \| \psi \|_{\scrim} \cdot \| \psi' \|_{\scrim} \;
 \quad \mbox{if $\psi,\psi' \in \sS(\scrim)$.}
$$ 
}

\noindent We can now define the state $\omega_\scrim$ collecting all the achieved results.

\teorema\label{Main4scri}{The following facts hold referring to $(\sH_{\scrim},\sK_{\scrim})$.}

{\em  {\bf (a)} The pair $(\sH_{\scrim},\sK_{\scrim})$ is the one-particle structure for a quasi-free pure 
state $\omega_{\scrim}$ on  $\cW(\sS(\scrim))$ which is uniquely determined by the requirement that 
its two-point function coincides with the right-hand side of 
(\ref{lscri}) under the restriction to $C_0^\infty(\scrim; \bR)$.

 {\bf (b)} The state $\omega_{\scrim}$ is invariant under the natural action of the one-parameter group 
 of $*$-automorphisms generated both by $X\spa\rest_{\scrim}$ and by the Killing vectors of $\bS^2$.

{\bf (c)} If $\{\beta^{(X)}_\tau\}_{\tau \in \bR}$ denotes the pull-back action on $\sS(\scrim)$ 
of the one-parameter group generated by $X\spa\rest_\scrim$ 
that is $(\beta_\tau(\psi))(v,\omega) = \psi(v-\tau,\omega)$, for every $\tau\in \bR$ and every $\psi \in 
\sS(\scrim)$
  it holds:
$$
\sK_{\scrim} \beta_\tau^{(X)}(\psi) =  e^{i\tau \hat{h}}\sK_{\scrim} \psi
$$
where $\hat{h}$ is the $h$-multiplicative self-adjoint operator on $\sH_\scrim = L^2(\bR\times \bS^2, 2hdh \wedge d\bS^2)$. \\
Analogous statements hold for $\cW(\sS(\Im^{\pm}_{L}))$ and for $\cW(\sS(\Im^{+}))$, hence there exists
the corresponding states $\omega_{\Im^{\pm}_{L}}$ and $\omega_{\Im^+}$ exist}.\\

\noindent{\em Proof}.  The proof of (a) and (b) is essentially identical to that of the 
corresponding items in Theorem \ref{Main4}.
Particularly, the proof of item (b) is a trivial consequence of Lemma \ref{ULTIMOlemma}. $\Box$

\section{The extended Unruh state $\omega_U$.} When a  spherically-symmetric  black hole forms, the metric of
the spacetime outside the event horizon, as well as that inside 
the region containing the singularity away from the collapsing matter, must be of Schwarzschild type 
due to the Birkhoff theorem (see \cite{WR,Wald2} for a more mathematically detailed discussion). A model 
of this spacetime can be realized selecting a relevant subregion of $\mM$ in the Kruskal manifold, 
{\it i.e.},
the so called regions I and II of the Kruskal diagram as depicted in chapter 6.4 of \cite{Wald}. A quantum 
state that accounts for Hawking's radiation in such background was heuristically defined by Unruh in $\mM$, 
who employed a mode 
decomposition approach \cite{Unruh,Candelas,Wald2}. A rigorous, though indirect, definition of $\omega_U$, 
restricted to $\mW$, has been subsequently proposed by Kay and Dimock in terms of an $S$-matrix 
interpretation, though under the assumption of asymptotic completeness, which was proved to hold in the
massless case \cite{DimockKay}. It is
imperative to stress that, in the last cited papers, the restriction to the static region $\mW$ was crucial 
to employ the mathematical techniques used to describe the scattering in stationary spacetimes and, as a
byproduct, the algebras $\cW(\sS(\cH))$ and $\cW(\sS(\Im^-))$ were introduced and used with some differences
with respect to our approach.\\

\ssb{The states $\omega_U$, $\omega_B$ and their basic properties} We are in place to give a rigorous 
definition of the Unruh state by means of the technology previously introduced. Our definition is valid for 
the whole region $\mM$ and it does not require any $S$-matrix 
interpretation, nor formal manipulation of distributional modes as in the more traditional presentations (see 
\cite{Candelas}). Our prescription is a possible rigorous version of Unruh original idea according to which
the state is made of thermal modes propagating in $\mM$ from the white hole and of vacuum modes entering 
$\mM$ from $\scrim$. Together the Unruh state $\omega_U$ on $\cW(\sS(\mM))$ we also define the {\em Boulware 
vacuum},  $\omega_B$ on $\cW(\sS(\mW))$, since it will be useful later.

\definizione\label{omegaHdef} {Consider the states $\omega_\scri$, $\omega_\scrim$,
$\omega_{\cH}$ and $\omega_{\cH_{ev}}$ as in 
Theorem \ref{Main4scri}, Theorem \ref{Main4},  Proposition \ref{PropMain4.5}.
The {\bf Unruh state} is the unique one $\omega_U : \cW(\sS(\mM))\to\bC$ 
such that:
\beq \omega_U\left( W_{\mM}(\vphi)\right) = \omega_{\cH}\left( W_{\cH}(\vphi_{\cH}) \right)
\omega_{\scrim}\left(W_{\scrim}(\vphi_\scrim) \right)\quad \mbox{for all $\vphi \in \sS(\mM)$.}
\label{omegaH}
\eeq
The {\bf Boulware vacuum}  is the unique state $\omega_B : \cW(\sS(\mW)) \to \bC$ 
such that:
\beq \omega_B\left( W_{\mW}(\vphi)\right) = \omega_{\cH_{ev}}\left( W_{\cH_{ev}}(\vphi_{\cH_{ev}}) \right)
\omega_{\scri}\left(W_{\scri}(\vphi_\scri) \right)\quad \mbox{for all $\vphi \in \sS(\mW)$.}
\label{boulware}\eeq
In other words
$
\omega_U\doteq\left(\imath\right)^* \left( \omega_{\cH}\otimes \omega_{\scrim} \right)
$
and
$
\omega_B\doteq\left(\imath^+\right)^* \left( \omega_{\cH^+}\otimes \omega_{\scrim} \right)
$.}\\

\noindent We study now the interplay between $\omega_U$, $\omega_B$
and the action of $X$.
The Killing field $X$ individuates a one-parameter group of (active) symplectomorphisms $\{\beta^{(X)}_t\}_{t
\in\bR}$ on $\sS(\mM)$ which leaves  $\sS(\mM)$ and $\sS(\mW)$ invariant.
As $X$ is defined on the whole manifold $\widetilde{\mM}$, similarly, a one-parameter group of (active) 
symplectomorphisms are induced on $\sS(\Im^\pm)$, $\sS(\cH)$, $\sS(\cH^-)$, $\sS(\cH_{ev})$ and, henceforth, we shall 
use the same symbol $\{\beta^{(X)}_t\}_{t\in \bR}$ for all these groups. In turn, $\{\beta^{(X)}_t\}_{t\in
\bR}$ induces a one-parameter group of $*$-automorphisms, $\{\alpha_t^{(X)}\}_{t\in \bR}$, on $\cW(\mM)$ 
unambiguously individuated by the requirement:
\beq
\alpha^{(X)}_t\left( W_{\mM}(\varphi)\right)\doteq W_{\mM}\left(\beta^{(X)}_t(\varphi)
\right)\:, \quad \mbox{for all $\varphi \in 
\sS(\mM)$.} \label{ad}
\eeq
Whenever $\{\alpha_t^{(X)}\}_{t\in \bR}$ acts on $\cW(\sS(\mM))$ and $\cW(\sS(\mW))$, it leaves these 
algebras fixed and the second one in particular represents the time-evolution, with respect to the 
Schwarzschild time, of the observables therein. Analogous one-parameter groups of $*$-automorphisms, 
indicated with the same symbol, are defined on $\cW(\Im^\pm)$, $\cW(\cH)$, $\cW(\cH^-)$, $\cW(\cH_{ev})$  by $X$. The 
following relations hold true, for all $t\in\bR$ and $\varphi\in\sS(\mM)$:
\beq
 \Gamma\left(\beta^{(X)}_t(\varphi)\right) =  \left(\beta^{(X)}_t(\varphi_{\cH})\:,\: 
  \beta^{(X)}_t(\varphi_{\scrim})\right)\:.\label{cad}
\eeq 
The same result is valid if one replaces $\mM$ with $\mW$, $\cH$ with $\cH^-$ or $\cH_{ev}$ and, in the 
second case, $\scrim$ with $\scri$, so that
$\Gamma$ is substituted by $\Gamma_-$ or $\Gamma_+$ respectively, while $\imath$ by $\imath^-$ or 
$\imath^+$ correspondingly.
The proof is a consequence of the invariance of the Klein-Gordon equation
under $\beta^{(X)}$. Similar identities hold concerning the remaining Killing $\bS^2$-symmetries of both 
$\mM$ and $\mW$. 
 
\proposizione\label{omegaHinvariance} {\em The following facts hold,\\
{\bf (a)} $\omega_U$ and $\omega_B$
are invariant under the action of $\{\alpha_t^{(X)}\}_{t\in \bR}$ and under that of the remaining Killing 
$\bS^2$-symmetries of the metric of $\mM$ and $\mW$ respectively.\\
{\bf (b)} $\omega_B$ is a {\bf regular quasifree ground state}, {\it i.e.}, the unitary one-parameter groups 
which implements $\{\alpha_t\}_{t\in\bR}$ are strongly continuous and the self-adjoint generators have 
positive spectrum with no zero eigenvalues in the one-particle spaces. Hence it coincides to the analogous 
vacuum state defined with respect to the past null boundary of $\mW$, {\it i.e.}, $\omega_B = \left(\imath^-
\right)^*(\omega_{\cH^-}\otimes \omega_{\Im^-})$.}\\

\noindent{\em Proof}. (a) If one bears in mind the same statement for the region $\mW$, the one under
analysis follows from (\ref{ad}), (\ref{cad}) 
together with the definitions (\ref{omegaH}) and (\ref{boulware}). One must also take into account that the 
states 
$\omega_{\cH}$, $\omega_\scrim$, $\omega_{\cH_{ev}}$, $\omega_{\Im^+}$, are invariant under the action of
both $\{\alpha_t^{(X)}\}_{t\in \bR}$ and the remaining Killing symmetries,  as established in 
theorems \ref{Main4}, \ref{Main4scri}  and proposition \ref{PropMain4.5}.\\
(b) By direct inspection one sees that, in the GNS representation space of the quasifree states,
$\omega_B$ and $ \left(\imath^-\right)^*(\omega_{\cH^-}\otimes \omega_{\Im^-})$ are quasifree regular 
ground states with respect to $\{\alpha_t\}_{t\in\bR}$. Thus {\em Kay's uniqueness theorem} \cite{KayU} implies 
that $\omega_B=\left(\imath^-\right)^*(\omega_{\cH^-}\otimes\omega_{\Im^-})$. $\qed$\\

\noindent If $\varphi,\varphi'\in\sS(\mW)$, the function $F_{\varphi,\varphi'}(t)\doteq\omega_{U}
\left(W_{\mW}(\varphi)\alpha_t^{(X)}\left(W_{\mW}(\varphi)\right)\right)$ decomposes in a product 
$$
F_{\varphi,\varphi'}(t) = F^{(\beta_H)}_{\varphi,\varphi'}(t)F^{(\infty)}_{\varphi,\varphi'}(t)\:.
$$
If one refers to the Schwarzschild-time evolution, the first factor fulfils the KMS requirements (see 
definition
\ref{KMSdef}), whereas the second factor enjoys the properties of a ground state two-point function: it can 
be extended to an analytic functions for $Im t >0$ which is continuous and bounded in $Im\; t \geq 0$ and tends 
to $1$ as $\bR\ni t\to\pm\infty$. The term $F^{(\beta_H)}_{\varphi,\varphi'}(t)$, which evaluates only the 
part $\varphi_{\cH^-}$ and $\varphi'_{\cH^-}$ of the wavefunctions, represents the components of the 
wavefunction which brings the thermal radiation entering $\mW$ through the white hole. The latter, which 
evaluates only the components $\varphi_{\Im^-}$ and $\varphi'_{\Im^-}$ of the wavefunctions, represents 
the part of the wavefunction associated with the Boulware vacuum.\\

\subsection{On the Hadamard property}
Let us consider a quasifree  state $\omega$ on the Weyl algebra of the real Klein-Gordon scalar field $\cW(
\mN)$  for a globally hyperbolic spacetime $(\mN, g)$ and let $(\sH_\omega,\sK_\omega)$ be
its one-particle structure which determines the Fock GNS representation $(\gH_\omega, \Pi_\omega, \Psi_\omega)$ of $\omega$.  Finally introduce the field operators ${\Phi}_\omega(f)$ as discussed in sec.  \ref{observables}. 
The {\bf two-point function} of $\omega$ is the 
bilinear form $\lambda : \sS(\mN)\times \sS(\mN) \to \bC$  where  $\lambda_\omega(\psi,\psi') \doteq \langle \sK_\omega \psi, \sK_\omega \psi'
\rangle_{\sH_\omega}$.  Equivalently,  if one follows  Sec. \ref{observables} and the Appendix \ref{algebras},
it turns out that
$$\lambda_\omega (\psi,\psi') =  \langle \Psi_\omega, {\Phi}_\omega(f) {\Phi}_\omega(f') 
\Psi_\omega \rangle\:, \quad \psi = E_{P_g} f\:,\quad \psi' = E_{P_g}f'\:,$$
where the expectation value of the product of two field operators ${\Phi}_\omega(f)$ and ${\Phi}_\omega (f')$ is
computed with respect to the cyclic vector $\Psi_\omega$ of the GNS representation of $\omega$ and where 
$E_{P_g}: C_0^\infty(\mN; \bC) \to \sS(\mN)$
is the causal propagator.  Therefore a {\bf smeared two-point function} can equivalently  be defined as a bilinear map 
 $\Lambda_\omega : C^\infty(\mN; \bR) \times C^\infty(\mN; \bR) \to \bC$ associated with the formal integral kernel $\Lambda_\omega(x,x')$
with
$$  \Lambda_\omega(f,g) \doteq  \int_{\mN\times \mN} \Lambda_\omega(x,x') f(x) g(x') d\mu_g(x) d\mu_g(x')\: \doteq  \langle \Psi_\omega,{\Phi}_\omega(f) {\Phi}_\omega(f') 
\Psi_\omega \rangle \:.$$
Furthermore
$$
 \Lambda_\omega(f,g) =  \lambda_\omega(E_{P_g}f, E_{P_g}g) \quad \mbox{if $f,g \in C^\infty(\mN;\bR)$}\:.
$$
In this framework, the state $\omega$ is said to satisfy the local {\bf Hadamard property}  when, in a geodetically convex neighbourhood of any point 
the two-point (Wightmann) function $\omega(x,x')$ of the state has the structure
$$\Lambda_\omega(x,x') = \frac{\Delta(x,x')}{ \sigma(x,x')} + V(x,x') \ln \sigma(x,x') + w(x,x')\:,$$
where $\Delta(x,x')$ and $V(x,x')$ are determined by the local geometry, $\sigma(x,x')$ is the signed squared geodetical distance of $x$
and $x'$ , while $w$ is a smooth function determining the quasifree state. 
 The precise definition, also at global level and up to the specification of the regularisation procedure 
enclosed in the definition of $\sigma$, was stated in \cite{KW}. 
The knowledge  of the singular part of the two-point function and, thus, of all $n$-point functions
in view of Wick expansion procedure, allows the definition of a suitable renormalisation procedure of 
several physically interesting quantities such as the 
stress energy tensor, to quote just one of the many examples \cite{Wald,Mo03,HW04}. It has thus been the 
starting point 
of a full renormalisation procedure in curved spacetime as well as other very important developments of the 
general theory \cite{BFK, BF00,HW01,BFV03}.
A relevant technical achievement was obtained by Radzikowski \cite{Rada, Radb} who, among other results, 
proved the following: if one refers to the Klein-Gordon scalar field, the global Hadamard condition 
for a quasifree state $\omega$ whose two-point function is a distribution  $\Lambda_\omega \in \mD'(\mN
\times\mN)$, where $(\mN,g)$ 
 is globally hyperbolic and time-oriented, is equivalent to the following 
 constraint on the  {\em wavefront set} \cite{Hormander} of $\Lambda_\omega$. 
 \beq\label{WFgen}
WF(\Lambda_\omega) = \ag (x,y,k_x,k_y)\in T^*(\mN \times \mN)\setminus \{0\} \; |\: \: (x,k_x)\sim (y,-k_y)\;,\; k_x\triangleright 0  \cg\:,
\eeq
that is usually referred to {\bf microlocal spectrum condition} - see \cite{Sanders} for recent developments -. 
One should notice that, above, $0$ denotes the null section of $T^*(\mN \times \mN)$ and $(x,k_x)\sim(z,k_z)$
means that there exists a light-like geodesic 
$\ga$ connecting $x$ to $z$ with $k_x$ and $k_z$ as (co)tangent vectors of $\ga$ respectively at $x$ and
at $z$. Particularly if $x=z$, it must hold that $k_x=k_z$, $k_z$ being of null type. The symbol 
$\triangleright$ indicates that $k_x$ must lie in the future-oriented light cone.  

The aim of this subsection is to prove that the two-point function associated to the state \eqref{omegaH} on $\cW(\mM)$  fulfils the Hadamard property
by means of the microlocal approach based on condition (\ref{WFgen}).
To this avail, the general strategy, we shall follow, consists of combining in a new non trivial way the
results presented in \cite{SV00} and in \cite{Moretti08, DMP3}. 
Since we interpret the two-point function as a map from $C^\infty_0(\mM; \bC) \times C^\infty_0(\mM; \bC) \to \bC$
a useful tool is the map  $\Ga :\sS(\mM)\to\sS(\cH)\oplus\sS(\scrim)$ introduced in the statement of 
Theorem \ref{Main1}. We shall combine it with the causal propagator to obtain  
\beq
(\varphi^f_{\cH},\varphi^f_{\scrim})\doteq\Ga E_{P_g} f\;. \label{BCN}
\eeq
We can now state the following proposition, whose ultimate credit is to allow us to check 
microlocal spectrum condition (\ref{WFgen}) since the two-point function of $\omega_U$ determines a proper 
distribution of $\mD'(\mM\times \mM)$.

\proposizione\label{distrib}{The smeared two-point function $\Lambda_U :C^\infty_0(\mM; \bR) \times C^\infty_0(\mM; \bR) \to\bC$ of the Unruh 
state $\om_U$ can be written as the sum 
\beq \Lambda_U = \Lambda_{\cH}+\Lambda_{\Im^-}, \label{LULHLI}
\eeq
with $\Lambda_{\cH}$ and $\Lambda_{\Im^-}$  defined out of the following relations
for $\la_{\cH}$ and $\la_\scrim$ as  in \eqref{lKW2} and \eqref{lscri}:
$$
\Lambda_{\cH}(f, g)\doteq \lambda_{\cH}(\varphi^f_{\cH}, \varphi^g_{\cH})\;, \qquad   
\Lambda_{\Im^-}(f, g)\doteq \lambda_{\Im^-}(\varphi^f_{\scrim}, \varphi^g_{\scrim})\;, \quad  \mbox{for every $f,g \in 
C^\infty_0(\mM; \bR)$,}
$$
Separately,  $\Lambda_\cH$, $\Lambda_\scrim$ and  
 $\Lambda_U$ individuate  elements of 
$\mD'(\mM\times\mM)$ that we shall indicate with the symbols  $\Lambda_\cH$, $\Lambda_\scrim$ and  
 $\Lambda_U$. These are uniquely individuated by $\bC$-linearity and continuity under the assumption 
 (\ref{LULHLI}) as
\beq\label{twopointseparately}
\Lambda_{\cH}(f \otimes g)\doteq \lambda_{\cH}(\varphi^f_{\cH}, \varphi^g_{\cH})\;, \qquad   
\Lambda_{\Im^-}(f \otimes g)\doteq \lambda_{\Im^-}(\varphi^f_{\scrim}, \varphi^g_{\scrim})\;,   \quad  \mbox{for every $f,g \in 
C^\infty_0(\mM; \bR)$.}
\eeq}

\noindent The proof is in Appendix \ref{Appendixproofs}.

\vskip .2cm

\noindent In the remaining part of this section we shall prove one of the main theorems of this paper, namely
that $\Lambda_U$ satisfies the microlocal spectral condition (\ref{WFgen}) and thus $\omega_U$ is Hadamard.

\vsp 

\teorema\label{maxt}{The two-point function $\Lambda_U\in \mD'(\mM \times \mM)$ associated with the Unruh 
state $\om_U$ satisfies the microlocal spectral condition:
\beq\label{WF2}
WF(\Lambda_U) = \ag (x,y,k_x,k_y)\in T^*(\mM\times \mM) \setminus \{0\},\; (x,k_x)\sim (y,-k_y),\;  k_x 
\triangleright 0   \cg \;,
\eeq
consequently  $\omega_U$ is of Hadamard type.}

\vsp
\noindent{\em Proof}.
As it is often the case with identities of the form \eqref{WF2}, the best approach, to prove them, is to show
that two inclusions $\supset$ and $\subset$ hold separately, hence yielding the desired equality. 
Nonetheless, in this case, we should keep in mind that $\Lambda_U$ is a two-point function, hence it 
satisfies in 
a weak sense the equation of motion \eqref{KG} with respect to $P_g$, a properly supported, homogeneous of 
degree 2, hyperbolic operator of real principal part and
the antisymmetric part of $\Lambda_U$ must correspond to the 
causal propagator $E_{P_g}$ introduced in subsection \ref{observables}. In this framework, all the hypothesis to 
apply the theorem of propagation of singularities (PST), as in Theorem 6.1.1 in \cite{DH}, are met. Hence one 
has all the ingredients necessary to proceed as in the proof of Theorem 5.8 in \cite{SV01}, to conclude that 
the inclusion $\supset$ holds true once $\subset$ has been established.
Therefore, in order to prove (\ref{WF2}), it is enough to establish only the inclusion $\subset$.  
This will be the goal of the remaining part of the proof and we shall divide our reasoning in two
different sequential logical steps. In the first part, below indicated as {\em part 1}, we shall prove 
that the microlocal spectrum condition is fulfilled in the static region $\mW$. In the second,
displayed as {\em part 2}, we apply this result extending it to the full $\mM$, mostly by means of the PST
which strongly constraints the form of $WF(\Lambda_U)$ in the full background. The left-over terms, which are
not fulfilling \eqref{WF2}, are eventually excluded by means of a case-by-case analysis. 

\vskip.3cm

\noindent{\em Part 1}.  In order to establish the validity of the microlocal spectral condition in $\mW$, our overall idea
is to  restrict $\Lambda_U$ to a distribution in $\mD'(\mW\times \mW)$ and to apply/adapt to our case 
the result on the wave-front set of the two-point function of passive quantum states, as devised in 
\cite{SV00}.\\
As a starting point, let us remind that $\mW$ is a static spacetime with respect to 
the Schwarzschild Killing vector $X$, and that the state $\Lambda_U$ is invariant under the associated 
time translation, as established in Proposition \ref{omegaHinvariance}. However, despite this set-up,
 $\Lambda_U$ is not passive in the strict sense given in \cite{SV00} and, hence, we cannot 
directly conclude that the Hadamard property is fulfilled in $\mW$, {\it i.e.}, in other words, Theorem 5.1
in \cite{SV00} does not straightforwardly go through. Nonetheless, luckily enough, a closer look at the proof
of the mentioned statement reveals that it can be repeated slavishly with the due exception of the  
step 2) in which the passivity condition is explicitly employed. Yet, this property is not used to its fullest 
extent and, actually, a weaker one suffice to get the wanted result; in other words, the mentioned ``step
2)'', or more precisely formula (5.2) in the last mentioned paper, can be recast as the following lemma for 
$\Lambda_U$.  

\lemma\label{KV}{The wave front set of the restriction to $\mD(\mW\times \mW)$ of  $\Lambda_U$, satisfies the following inclusion
$$
WF(\Lambda_U\spa \rest_{\mD(\mW\times \mW)})\subset
\ag (x,y,k_x,k_y) \in T^*(\mW\times \mW)\setminus \{0\}, \;  k_x(X)+k_y(X)=0 , \;  k_y(X) \geq 0
\cg \;,
$$
where $X$ is the generator of the Killing time translation.\\
}

\noindent{\it Proof.} As a first step we recall the invariance of $\Lambda_U$, as well as of $\Lambda_\scrim$
and $\Lambda_{\cH}$,  under the action of $X$, an assertion which arises out of part (b) of both Theorem 
\ref{Main4} and \ref{Main4scri}.
Furthermore, out of \eqref{twopointseparately}, it is manifest that both $\Lambda_\scrim$ and $\Lambda_{\cH}$ satisfy 
in a weak sense and in both entries the equation of motion, since they are constructed out of the causal 
propagator \nref{twopointseparately}. Yet their antisymmetric part does not correspond to the causal 
propagator and this lies at the heart of the impossibility to directly apply the proof of theorem 5.1 as it
appears in \cite{SV00}. \\
Nevertheless, if we still indicate by $\beta^{(X)}_t$ ($t\in\bR)$ the pull-back action of one-parameter group
of isometries generated by $X$ on elements in $C^\infty_0(\mW;\bR)$, we can employ 
\eqref{twopointseparately}, as well as the
definition of both $\la_\scrim$ and $\la_{\cH}$, to infer the following: $\Lambda_\scrim$, which we shall 
refer as {\bf vacuum like}, fulfils formula (A1) in \cite{SV00}:
$$
\int\limits_\bR\widehat{f}(t) \Lambda_\scrim(h_1\otimes \beta^{(X)}_t(h_2))dt =0,\quad h_1,\;h_2\in C^\infty_0(\mW;\bR)
$$
for all $\widehat{f}(t)\doteq\int_\bR e^{-ikt} f(k) dk$ such that $f\in C^\infty_0 (\bR^*_-;\bC)$. At the same time 
$\Lambda_{\cH}$ fulfils formula (A2) in the same mentioned paper, which implies that it is {\bf KMS like} at 
inverse temperature $\beta_H$, {\it i.e.},
$$
\int\limits_\bR\widehat{f}(t) \Lambda_{\cH}(h_1\otimes \beta^{(X)}_t(h_2))  dt =
\int\limits_\bR\widehat{f}(t+ i\beta_H) \Lambda_{\cH}(\beta^{(X)}_t(h_2)\otimes  h_1)dt,\quad h_1,\;h_2\in C^\infty_0(
\mW;\bR)\:,
$$
for every  $f\in C^\infty_0(\bR; \bR)$.
The former identity arises out of the Fubini-Tonelli's theorem and of basic properties of the 
Fourier-Plancherel transform. To wit, if one bears in mind the definition of $\Lambda_\scrim$, 
$\omega_\scrim$, the explicit expression of $\sH_{\scrim} =  L^2(\bR_+ \times \bS^2; 2kdk \wedge d\bS^2)$ as
well as part (c) of Theorem  \ref{Main4scri}:
$$ 
\Lambda_\scrim(h_1\otimes \beta^{(X)}_t(h_2)) = \int_{\bS^2} d\bS^2(\omega) \int_0^{+\infty} \overline{\psi_1(k
,\omega)} e^{itk} \psi_2(k,\omega)  2k dk,
$$
for suitable functions $\psi_1$ and $\psi_2 \in L^2(\bR\times \bS^2; 2kdk \wedge d\bS^2)$ which corresponds 
to $h_1$ and $h_2$ . We also stress that the $k$ integration is only extended to the {\em positive real 
axis}, whereas the support of $f$ is contained in $\bR_-$. If one notices that, if $h\in C_0^\infty(\mW;
\bR)$, then $\varphi^h_\cH \in \sS(\cH^-)$, then the second identity follows similarly from Theorem \ref{Main4}. Here
the key ingredients are the definition of $\Lambda_{\cH}$, $\omega^{\beta_H}_{\cH^-}$ and the explicit 
expression of the measure $\mu(k)$ in $\sH^{\beta_H}_{\cH^-}=L^2(\bR \times \bS^2; \mu(k) \wedge d\bS^2)$, 
and point (e)  of Theorem \ref{Main4}.\\
The validity of this pair of identities suffices to establish the statement of Proposition 2.1 in
\cite{SV00}, whose proof can be slavishly repeated with our slightly weaker assumptions, though one should
mind the different conventions in our definition of the Fourier transform. From this point
onwards, one can follow, in our framework and step by step, the calculations leading to the second point in 
the proof of Theorem 
5.1 in \cite{SV00}, which is  nothing but the statement of our lemma. We shall not reproduce all the details here,
since it would lead to no benefit for the reader. $\qed$\\

\noindent Equipped with the proved lemma, and following the remaining steps  of the proof of Theorem 5.1 in \cite{SV00} the last
 statement  in the thesis of Theorem 5.1 in \cite{SV00} can be achieved in our case, too.
As remarked immediately after the proof  of the mentioned theorem in \cite{SV00}, that statement entails the validity of the microlocal 
 spectrum condition for the considered two-point function.  Thus we can claim that

\proposizione\label{propWFW}{The two-point function $\Lambda_U\in\mD'( {\mM\times \mM})$ of the Unruh state, restricted on $C_0^\infty(\mW\times \mW; \bC)$,
satisfies the microlocal spectral condition (\ref{WFgen}) with $\mN= \mW$ and thus $\omega_U\spa\rest_{\cW(\mW)}$ is a Hadamard state.}

\vskip.3cm

\noindent{\em Part 2}.  Our goal is now to establish that the microlocal spectrum condition for $\Lambda_U(x,x')$ 
holds true also considering pairs $(x,x') \in \mM\times \mM$ which do not belong to $\mW\times \mW$. The overall strategy, we shall employ, mainly consists
 of a careful use of the
propagation of singularity theorem which shall allow us to divide our analysis in simpler specific subcases. \\
To this avail, we introduce the following bundle  of null cones  $\mN_g\subset T^*\mM\setminus\left\{0\right\}$
constructed out of the principal symbol of $P_g$, as in \eqref{KG}:
$$
\mN_g \doteq\ag(x,k_x)\in T^*\mM\setminus\left\{0\right\}\;,\;\; g^{\mu\nu}(x)(k_{x})_{\mu}(k_{x})_{\nu}=0
\cg\;.
$$
We define the {\bf bicharacteristic strips} generated by $(x,k_x) \in \mN_g$
$$
B(x,k_x)\doteq\ag(x',k_x')\in \mN_g\; |\; (x',k_{x'})\sim (x,k_x)\cg,
$$
where $\sim$ was introduced in \eqref{WFgen}.  The operator $P_g$ is such that 
we can apply to the weak-bisolution $\Lambda_U$ the theorem of propagation of singularities (PST), as devised in 
Theorem 6.1.1 of \cite{DH}. This guarantees that, on the one hand:
\beq 
WF(\Lambda_U)\subset
\left(\left\{0\right\}\cup\mN_g\right)\times\left(\left\{0\right\}\cup\mN_g\right)\:, \label{PST1}
\eeq
while, on the other hand,
\beq\mbox{if $(x,y,k_x,k_y) \in WF(
\Lambda_U)$  then}\quad 
B(x,k_x)\times B(y,k_{y})\subset WF(\Lambda_U). \label{PST2}
\eeq
A pair of technical results, we shall profitably use in the proof, are given by the following lemma and 
proposition whose proofs can be found in appendix \ref{Appendixproofs}. \\

The proposition characterises the decay property, with respect to $p\in T^*_x\mM$, of the 
distributional Fourier transforms even though one should notice that in
 \cite{Hormander}
the opposite convention concerning the sign in front of $i\langle p,\cdot\rangle$ is adopted:
$$
\varphi^{f_{p}}_\scrim\doteq  \lim_{\to \scrim} E_{P_g} (f e^{i\langle p,\cdot\rangle } ) \;,\qquad    
\varphi^{f_{p}}_\cH\doteq E_{P_g} (f e^{i\langle p,\cdot\rangle } ) \spa \rest_\cH
$$
where we have used the complexified version of causal propagator, which 
enjoys the same causal and topological properties as those of the real one. Henceforth
$\langle \cdot , \cdot \rangle$ denotes  the standard scalar product in $\bR^4$ and $|\cdot|$ the associated
norm, computed after the choice of normal coordinates. From now on we also shall assume to fix a coordinate patch whenever necessary, all the results being 
independent from such a choice, as discussed after Theorem 8.2.4 in \cite{Hormander}.
We remind the reader that, given a function $F: \bR^n \to \bC$, an element $k\in \bR^n \setminus\{0\}$ is said to be 
of {\bf rapid decrease} for $F$ if there exists an open conical set $V_k$, {\it i.e.}, an open set such that,
if $p\in V_k$ then $\lambda p\in V_k$ for all $\lambda >0$, such that, $V_k \ni k$ and, for every $n=1,2,
\ldots$, there exists $C_n \geq 0$ with $|F(p)|\leq C_n/(1+ |p|^{n})$ for all $p \in V_{k}$.

\proposizione\label{RapidDecay}{Let us take $(x,k_x) \in \mN_g$ such that (i)  $x\in\mM\setminus 
\mW$ and (ii) the unique inextensible geodesic $\ga$ (co-)tangent to  $k_x$ at $x$ intersects $\cH$ in a 
point whose $U$ coordinate is nonnegative. Let us also fix $\chi'\in C^\infty(\cH; \bR)$ with $\chi' =1$
in $U\in (-\infty, U_0]$ and $\chi' = 0$ if $U\in [U_1,+\infty)$ for a constant value of $U_0 < U_1 < 0$.\\ 
For any $f\in C^\infty_0(\mM)$ with $f(x)=1$ and  sufficiently small support, 
$k_x$ is a direction of rapid decrease for both
$p\mapsto \|\varphi^{f_{p}}_\scrim\|_\scrim $ and $p\mapsto \|\chi' \varphi^{f_{p}}_{\cH}\|_\cH$.}\\

The pre-announced lemma has a statement which closely mimics an important step in the analysis of the Hadamard
form of two-point functions, first discussed in \cite{SV01}. It establishes the the right-hand side of
(\ref{PST1}) can be further restricted.

\lemma{\label{Nozero} 
Isolated singularities do not enter the wave-front set of $\Lambda_U$, namely
$$
(x,y,k_x,0) \notin WF(\Lambda_U)\;,\quad (x,y,0,k_y) \notin WF(\Lambda_U) \qquad \mbox{if\:\:} x,y \in \mM, k_x \in T_x^*\mM,
k_y \in T_y^*\mM \;.
$$
Thus, as a consequence of  (\ref{PST1}), it holds
$$
WF(\Lambda_U) \subset \mN_g \times \mN_g\:.
$$}

\noindent The next step in our proof consists of the analysis of $WF(\Lambda_U)$, in order to establish the 
validity of \eqref{WF2} with $=$ replaced by $\subset$. We shall tackle the  
cases which are left untreated by the statement of Proposition \ref{propWFW} in particular. As previously discussed,
 this suffices to conclude the proof of the Hadamard property for $\omega_U$.\\
The remaining cases  amount to the points in $WF(\Lambda_U)$ such that, in view of Lemma \ref{Nozero},
 $(x,y,k_x,k_y) \in \mN_g\times \mN_g$  with either $x$, either $y$ or both in $\mM\setminus \mW$.
Therefore, we shall divide the forthcoming analysis in two parts, {\bf case A}, where only one
point is in $\mM\setminus \mW$, and {\bf case B}, where both lie in $\mM\setminus \mW$.

\vskip .2cm

\noindent {\bf Case A}. Let us consider an arbitrary $(x,y,k_x,k_y) \in \mN_g \times \mN_g$ which belongs to 
$WF(\Lambda_U)$ and such that  $x\in\mM\setminus\mW$ 
and $y\in\mW$, the symmetric case being treated analogously. If a representative of the equivalence class 
$B(x,k_x)$ has its basepoint  in $\mW$, (\ref{PST2}) entails that the portion of $B(x,k_x) \times B(y,k_y)$ 
enclosed 
in $T^*(\mW\times\mW)$ must belong to $WF(\Lambda_U\spa\rest_{C_0^\infty(\mW\times \mW;\bC)})$ and, thus, it 
must have 
the shape stated in Proposition \ref{propWFW}. Thanks to the uniqueness of a geodesic which passes through a 
point with a given (co-)tangent vector, it implies that $(x, k_x) \sim (y, -k_y)$ and $k_x\triangleright 0$ as
wanted.\\
Let us consider the remaining subcase where no representative of $B(x,k_x)$ has a basepoint in $\mW$.  
Our goal is to prove that, in this case,  
$(x,y, k_x, k_y) \not \in WF(\Lambda_U)$ for every $k_y$. This will be established showing that 
there are two compactly supported smooth functions $f$ and $g$ with $f(x)=1$ and $g(y)=1$ such that $(k_x,
k_y)$ individuate directions of 
rapid decrease of $(p_x,p_y) \mapsto \Lambda_U((fe^{i\langle p_x, \rangle} \otimes h e^{i\langle p_y, 
\rangle})$.\\
If $B(x,k_k)$ does not meet $\mW$, there must exist 
$(q,k_q)\in B(x,k_x)$, such that $q\in\cH$ and the Kruskal null coordinate $U=U_q$ is nonnegative. Let us
consider, then, the two-point function
$$
\Lambda_U(f\otimes h)= \Lambda_{\cH}(f \otimes h)+\Lambda_\scrim(f \otimes h),\quad f,h\in C^\infty_0(\mM; 
\bR),
$$ 
where $\Lambda_{\cH}$ and $\Lambda_\scrim$ are as in \eqref{twopointseparately}. If the supports of the chosen $f$ 
and $h$ are sufficiently small, we can always engineer a function $\chi\in C^\infty_0(\cH)$ in such a way
that $\chi(U_q,\theta,\phi)=1$ for all $(\theta,\phi)\in\bS^2$
and $\chi=0$ on  $J^-(supp\; h)$ and $\cH$. Furthermore, if we use a coordinate patch which identifies an 
open neighbourhood of $supp(f)$ with $\bR^4$ and we set $\chi' \doteq  1-\chi$, we can arrange
a conical neighbourhood $\Ga_{k_x} \in \bR^4 \setminus \{0\}$ of $k_x$ such that all the bicharacteristics 
$B(s,k_s)$
with $s \in supp(f)$ and $k_s \in \Ga_{k_x}$ do not meet any point of $supp \chi'$ on $\cH$.
If we refer to (\ref{BCN}), we can now divide  $\Lambda_{\cH}(f \otimes h)$ as:
$$
\Lambda_{\cH}(f \otimes h)=\la_{\cH}(\chi \varphi^{f}_{\cH},\varphi_{\cH}^{h})+\la_{\cH}(\chi' \varphi^{f}_{
\cH},\varphi_{\cH}^{h}),
$$
and we separately analyse the behaviour of the following three contributions at large $(k_x,k_y)$ :
\beq\label{distributions}
\la_{\cH}(\chi \varphi^{f_{k_x}}_{\cH},\varphi_{\cH}^{h_{k_y}})
\; , \qquad 
\la_{\cH}(\chi' \varphi^{f_{k_x}}_{\cH},\varphi_{\cH}^{h_{k_y}})
\qquad \text{and} \qquad
\lambda_\scrim(\varphi_\scrim^{f_{k_x}},\varphi_\scrim^{h_{k_y}})\;.
\eeq
Each of these should be seen as the action of a corresponding distribution in $\mD'(\mM\times \mM)$. The 
scenario, we face, is less complicated than it looks at first glance since we know that neither $(x,y,k_x,0)$
nor $(x,y,0,k_y)$ can be contained in $WF(\La_U)$,
as Lemma \ref{Nozero} yields. Hence this implies that, in the splitting we are considering in 
\eqref{distributions}, we can focus only on the points $(x,y,k_x,k_y)$ where both $k_x$ and $k_y$ are not 
zero. If we were able to prove that these points are not
contained in the wave front set of any of the three distributions \eqref{distributions}, we could conclude 
that they cannot be contained in the wave front set of the their sum $\La_U$, because the wave front set of 
the sum of distributions is contained in the union of the wave front set of the single component.
At the same time, the second and third distribution in the right-hand side of
\eqref{distributions} turn out to be dominated by
$ C \|\chi' \varphi^{f_{-k_x}}_{\cH}\|_{\cH} \| \varphi^{h_{k_y}}_{\cH}\|_{\cH}$
and
$C' \|\varphi^{f_{-k_x}}_{\scrim}\|_\scrim \| \varphi^{h_{k_y}}_{\scrim}\|_\scrim$, respectively, $C$ and $C'
$ being
suitable positive constants, whereas $\|\cdot\|_{\cH}$ and $\|\cdot\|_{\scrim}$  stand for the norm
\eqref{norme} and \eqref{normeScri}. This is a by-product of the continuity property presented 
in points (f) of both propositions \ref{Propembedding} and \ref{Propembeddingscri}, here adapted for complex 
functions, too. Furthermore, per Proposition \ref{RapidDecay}, both $\|\chi' \varphi^{f_{k_x}}_{\cH}\|_{\cH}$
and $\|\varphi^{f_{k_x}}_{\scrim}\|_\scrim$ are rapidly decreasing in $k_x \in T^*\mM\setminus\{0\}$ for an 
$f$ with sufficiently small support and if $k_x$ is in a open conical neighbourhood of any null direction. 
The remaining two terms $\|\varphi^{h_{k_y}}_{\cH}\|_{\cH}$ and $\|\varphi^{h_{k_y}}_{\scrim}\|_\scrim$, in 
\eqref{distributions}, can at most grow polynomially in $k_y$. 
The last property can be proved as follows: if
one starts from the bounds for the behaviour of the wavefunctions restricted to on $\cH^-$ and $\scrim$, as
per Proposition \ref{PropDR}, one can estimate the norms 
$\|\varphi^{h_{k_y}}_{\scrim}\|_\scrim$, $\|\varphi^{h_{k_y}}_{\cH}\|_\cH$ embodying the dependence on $k_y$
in the explicit expression of the coefficients  $C_i$ which appear in Proposition \ref{PropDR}.
Then, out of an argument similar to the one exploited in the proof of Proposition \ref{RapidDecay}, for fixed 
$k_y$ and $h$, those coefficients can be bounded by $C \sqrt{|\tilde{E}_5(
\varphi^{h_{k_y}})|}$ as in \nref{E2}, where $\tilde{E}_5(\varphi^{h_{k_y}})$ is the integral of a polynomial
of derivatives of $\varphi^{h_{k_y}}$ on a suitable Cauchy surface $\Sigma\subset\mW$.
Notice that $\varphi^{h_{k_y}}(z) = (E_{P_g}(h_{k_y}))(z)$ is smooth, 
 has compact support when restricted on a Cauchy surface, and together with the compact supports of its derivatives are contained in a 
common compact subset $K\subset \Si$. 
One can exploit the continuity of the causal propagator 
$E_{P_g}$,   
to conclude that,
for every fixed multi-index $\alpha$, $\sup_K| \partial^\alpha E_{P_g}(h_{k_y})|$ is bounded by a corresponding 
polynomial in the absolute values of the components 
of $k_y$. The coefficients are the supremum of derivatives of $h\in C_0^{\infty}(\mM;\bR)$ up to a certain 
order. 
This implies immediately that $\tilde{E}_5(\varphi^{h_{k_y}})$, as well as $\|\varphi^{h_{k_y}}_{\scrim}\|_
\scrim$, $\|\varphi^{h_{k_y}}_{\cH}\|_\cH$ are polynomially bounded in $k_y$, also because the computation of
$\tilde{E}_5(\varphi^{h_{k_y}})$ has to be performed on a compact set $K\subset\Sigma$.\\

We now remind the reader that we have identified $\mK \times \mK$ with $\bR^4\times \bR^4$ 
by means of a suitable pair of coordinate frames. Hence cotangent vectors at different points $x$ and $y$ 
can be thought of as elements of the same $\bR^4$ and, hence, compared. This allows us to define the 
following open cone in $\bR^4$, $\Ga\subset \bR^4\times \bR^4$, {\it i.e.}, with $0<\epsilon< 1$,
\beq\label{cono}
\Ga_{k_x}=\ag (p_x,p_y)\in \bR^4\times \bR^4 \;\left| \; \epsilon |p_x| < |p_y| <\frac{1}{\epsilon}  |p_x| \right. \:,  -p_x \in U_{-k_x}\cg
\eeq
where $U_{k_x}$ is an open cone around the null vector $k_x\neq 0$ where  
$p \mapsto \|\chi' \varphi^{f_{p_x}}_{\cH}\|^\eta_{\cH^-}$
and $p\mapsto \|\varphi^{f_{p_x}}_{\scrim}\|_\scrim$ decrease rapidly.
Hence, per construction, for any  direction $(k_x,k_y)$ with both $k_x \neq 0$ and $k_y\neq 0$ 
of null type, there is a cone $\Ga_{k_x}$ containing it.
Moreover  all the directions contained in $\Ga_{k_x}$  are of rapid decrease for both
$\la_{\cH}(\chi \varphi^{f_{k_x}}_{\cH},\varphi_{\cH}^{h_{k_y}})$ and 
$\la_{\cH}(\chi' \varphi^{f_{k_x}}_{\cH},\varphi_{\cH}^{h_{k_y}})$
because, just in view of the shape of $\Gamma_{k_x}$, the rapid decrease of $\|\chi' \varphi^{f_{-p_x}}_{\cH}
\|_{\cH}$ and $\|\varphi^{f_{-p_x}}_{\scrim}\|_\scrim$  controls the polynomial growth in $|p_y|$ of $\|
\varphi^{h_{p_y}}_{\cH}\|_{\cH}$ and $\|\varphi^{h_{p_y}}_{\scrim}\|_\scrim$ respectively.\\
We are thus left off only with the first term in \eqref{distributions} and, also in this case, if the support
of $f$ and $h$ are chosen sufficiently small, $\la_{\cH}(\chi \varphi^{f_{k_x}}_{\cH},\varphi_{\cH}^{
h_{k_y}})$ can be shown to be rapidly decreasing in both $k_x$ and $k_y$. 
To this end, let us thus choose $\chi''\in C^\infty(\cH; \bR
)$ such that both $\chi''(p)=1$ for every $p$ in $\supp(\varphi^{h_{k_y}}_{\cH})$ (also for every $k_y$) and
$\chi''\cap \chi =\emptyset$. We can write
$$
\la_{\cH}(\chi \varphi^{f_{k_x}}_{\cH},\varphi_{\cH}^{
h_{k_y}}) = \int_{\cH\times \cH}  \sp\sp \chi(x') \left(E_{P_g}(f_{k_x})\right)(x')\: T(x',y') \chi''(y') 
\varphi^{h_{k_y}}_{\cH}(y')\quad dU_{x'}d\bS^2(\theta_{x'},\phi_{x'})\:dU_{y'}d\bS^2(\theta_{y'},\phi_{y'})
$$
Theorem 8.2.14 of \cite{Hormander} guarantees us that 
$$
(x',y',k_{x'},k_{y'})\not\in WF\left((T\chi'')\circ (\chi E_{P_g}\spa \rest_{\cH})\right)\;\qquad \forall (y',k_{y'})\in T^*\mM,
$$ 
where $T$ is the integral kernel of $\la_{\cH}$ seen as a distribution on $\mD'(\cH\times \cH)$,
while $\circ$ stands for the composition on $\cH$  {\em with $E_{P_g}$ on the left of $T$}. Finally $E_{P_g}\spa \rest_{\cH}$ means that the left entry of the 
causal propagator has been restricted on the horizon $\cH$, an allowed operation thanks to theorem 8.2.4 in 
\cite{Hormander}. One can convince himself, out of a direct construction, that the set of normals associated to
the map embedding $\cH$ in $\mW$ does not intersect the wave front set of $E_{P_g}$. 
The integral kernel of $(\chi T\chi'')(x',y')$, with the entry  $x'$  restricted on the support of $\chi$
and the entry $y'$  restricted on that of $\chi''$, moreover, is always smooth and, if one keeps $x'$ fixed, 
it is dominated by a smooth function whose $H^1$-norm in $y'$ is, uniformly in $x'$, finite. This also yields
that, the $H^1(\cH)_U$-norm of $\|(T \chi''  )\circ \chi E_{P_g} f_{k_x}\|_{H^1(\cH)_U}$ is dominated by the product of two 
integrals one over $x'$ and one over $y'$. The presence of the compactly supported function $\chi$ and the
absence of points of the form $(x,y,k_x,0)$ and $(y,x,0,k_y)$ in $WF(E_{P_g})$ assures that the integral
kernel of $\chi T\chi''$ is rapidly decreasing in $k_x$. 
Summing up we have that 
\beq\label{aggI}
|\la_{\cH}(\chi \varphi^{f_{k_x}}_{\cH},\varphi_{\cH}^{h_{k_y}})| \leq  C \|\left((T\chi'' )\circ (\chi E_{P_g})\right)
(f_{-k_x})\|_{H^1(\cH)_U} \:\: \| \varphi_{\cH}^{h_{k_y}} \|_{\cH} \;,
\eeq
where the second norm in the right-hand side is given in \eqref{norme}. This bound proves that, for
a fixed $k_y$, $k_x \to \la_{\cH}(\chi \varphi^{f_{k_x}}_{\cH},\varphi_{\cH}^{h_{k_y}})$
is rapidly decreasing.\\
To conclude, if we look again at \eqref{aggI} and if we introduce a cone as in \eqref{cono}, 
out of Lemma \ref{Nozero}, we can control the (at most) polynomial growth of $\| \varphi_{\cH}^{h_{k_y}} 
\|_{\cH}$ using the rapidly decreasing map $k_x \mapsto  \|\left((T\chi'' )\circ (\chi E_{P_g})\right)
(f_{-k_x})\|_{H^1(\cH)_U} $. Hence we establish that $(k_x,k_y)$ is a direction of fast decreasing of $\la_{
\cH}(\chi \varphi^{f_{k_x}}_{\cH},\varphi_{\cH}^{h_{k_y}})$.\\
\vskip .2cm

\noindent {\bf Case B}. We shall now tackle the case in which we consider an arbitrary but fixed $(x,y,k_x,
k_y)\in \mN_g\times \mN_g$, with both $x$ and $y$ lying $\mM\setminus \mW$.\\
If one assumes that $(x,y,k_x,
k_y) \in WF(\Lambda_U)$ we have to prove that both $(x,k_x)\sim(y,-k_y)$ and  $k_x\triangleright 
0$ have to be valid. If $B(x,k_x)$ and $B(y,k_y)$ are such that both admit representatives in $\mW$, 
we make use of both (\ref{PST2}) and of the fact that elements in the wavefront set of the restriction of 
$\Lambda_U$ to $\mW$ fulfils  $(x',k'_{x'})\sim(y',-k'_{y'})$ and $k'_{x'}\triangleright 0$. Hence one extends
this property  to $(x, y, k_x, k_y)$ following the same reasoning  as the one at the beginning of the {\em 
Case A}. If, instead, only one representative, either of $B(x,k_x)$ or of $B(y,k_y)$ lies in $\mW
$, then we fall back in {\em Case A} studied above again thanks to (\ref{PST2}).
Thus, we need only to establish the wanted behaviour
of the wave front set when it is possible to find representatives of both $B(x,k_x)$ and $B(y,k_y)$ which
intersect $\cH$ at a nonnegative value of $U$.  We shall follow a procedure
similar to the one already employed in \cite{Moretti08}.\\
In this framework, let us consider the following decomposition of 
$\Lambda_U(\varphi^{f_{k_x}}\otimes \varphi^{h_{k_y}})$:
$$\la_U(\varphi^{f_{k_x}},\varphi^{h_{k_y}})=\la_{\cH}(\varphi^{f_{k_x}},
\varphi^{h_{k_y}})+\la_\scrim(\varphi^{f_{k_x}},\varphi^{h_{k_y}}),
$$
where $f, h\in C^\infty_0(\mM)$ and they attain the value $1$ respectively at the point $x$ and $y$.\\
As before, we start decomposing the first term in the preceding expression by means of a partition of 
unit $\chi, \chi'$ on $\cH$, where $\chi, \chi' \in C^\infty_0(\cH)$ satisfy 
$\chi+\chi'=1  :\cH \to \bR$. We obtain
\begin{gather}
\la_{\cH}(f_{k_x},h_{k_y}) =
\la_{\cH}(\chi \varphi^{f_{k_x}}_{\cH}, \chi \varphi^{h_{k_y}}_{\cH})+
\la_{\cH}(\chi' \varphi^{f_{k_x}}_{\cH}, \chi \varphi^{h_{k_y}}_{\cH})+\notag\\
\la_{\cH}(\chi\varphi^{f_{k_x}}_{\cH}, \chi' \varphi^{h_{k_y}}_{\cH})+
\la_{\cH}(\chi'\varphi^{f_{k_x}}_{\cH}, \chi'\varphi^{h_{k_y}}_{\cH}).\label{decomp}
\end{gather}
Furthermore, the above functions $\chi, \chi' $ can be engineered in such a way that the inextensible null geodesics 
$\ga_x$ and $\ga_y$, which starts respectively at $x$ and $y$ with  cotangent vectors $k_x$ and $k_y$,
intersect $\cH$ in  $u_x$ and $u_y$ (possibly $u_x=u_y)$, respectively, included in  two corresponding open 
neighbourhoods $O_x$ and $O_y$ (possibly $O_x=O_y$) where $\chi'$ vanishes.
Let us start from the first term in the right hand side of \eqref{decomp} and, particularly, we 
shall focus on the wave front set of the unique extension of  $f\otimes g \mapsto \la_{\cH}(\chi \varphi^{f}_
{\cH},\chi\varphi^{h}_{\cH})$ to a distribution in $\mD'(\mM\times \mM)$. If we indicate as $T$ the 
integral kernel of $\la_{\cH}$, interpreted as distribution of $\mD'(\cH\times\cH)$, 
we notice that, as an element in $\mD'(\mM\times \mM)$, $\la_{\cH}$ can be 
written as:
$$\la_{\cH}(\chi \varphi^{f}_{\cH}, \chi \varphi^{h}_{\cH})\doteq\;\chi T\chi\left(E_{P_g}\rest_{\cH}\otimes  
E_{P_g}\rest_{\cH}\at f\otimes h\ct\right),$$
where $E_{P_g}\rest_{\cH}$ is the causal propagator with one entry restricted on the horizon $\cH$ and $\chi T 
\chi\in\cE'(\cH\times\cH)$. Thanks to the insertion of the compactly supported smooth functions $\chi$, 
and with the knowledge that $WF(E_{P_g}\otimes E_{P_g})_{\cH\times\cH}=\emptyset$ (see \cite{Moretti08}), we can make
sense of the previous expression as an application of Theorem 8.2.13 in \cite{Hormander}, of which we also
employ the notation. The wave front set of $T$ has been already explicitly written in Lemma 4.4 of \cite{Moretti08} and, hence,
still Theorem 8.2.13 in \cite{Hormander} guarantees us that if $(x,y,k_x,k_y)$ is contained in the wave 
front set of the resulting distribution then $(x,k_x)\sim(y,-k_y)$  and $k_x\triangleright 0$ hold.\\
If we come back to the remaining terms in \eqref{decomp}, it is possible to show that all of them, together 
with $\lambda_\scrim$ are rapidly decreasing in both $k_x$ and $k_y$, provided that $f$ and $h$ have 
sufficiently small support. Hence they give no contribution to $WF(\Lambda_U)$.\\
Here we analyse in details only the second term in \nref{decomp} since the others can be 
treated exactly in the same way. To start with,  notice that, due to  (f) in proposition \ref{Propembedding} $|\la_{\cH}(\chi' \varphi^{f_{k_x}}_{\cH
}, \chi \varphi^{h_{k_y}}_{\cH})|$ is bounded by $C \|\chi' \varphi^{f_{k_x}}_{\cH}\|_{\cH^-
} \|\chi \varphi^{h_{k_y}}_{\cH}\|_{\cH}$, where $\|\cdot\|_{\cH}$ is the norm introduced in 
\eqref{norme} and $C>0$ is a constant.
Due to Proposition \ref{RapidDecay}, $\|\chi' \varphi^{f_{k_x}}_{\cH}\|_{\cH}$ is rapidly decreasing in
$k_x$ for some $f$ with sufficiently small support.
Finally,  the rapid decrease of $\|\chi' \varphi^{f_{-k_x}}_{\cH}\|_{\cH}$ can control the at-most 
polynomial growth of $\|\chi \varphi^{h_{k_y}}_{\cH}\|_{\cH}$, as discussed above in the analysis of 
\eqref{distributions} using the fact that $(x,y,k_x,0)$ and $(x,y,0,k_y)$ cannot be contained in the wave 
front set of $\Lambda_U$; this leads to the construction of the open cone $\Gamma_{k_x}$.\\
If we collect all the pieces of information we have got about the shape of $\Lambda_U$, we can state that: 
$$
WF(\Lambda_U) \subset  \ag (x,y,k_x,k_y)\in T^*(\mM\times \mM)\setminus \{0\},\; (x,k_x)\sim (y,-k_y),\;  k_x 
\triangleright 0   \cg
$$
and this concludes the proof. $\qed$

\vskip .3cm

To conclude this section we have an important remark on the physical
interpretation of $\omega_U$, that arises if one combines the results presented above, on the Hadamard property fulfilled by $\om_U$
in the {\em full region} $\mM$,
together with the
known achievements due to Fredenhagen and Haag
\cite{Fredenhagen} (see also the discussion Sec VIII.3 in \cite{Haag}).
In such paper the authors showed that,
whenever a state $\om'$ is vacuum like far away from the black hole,
its two-point function $\La'(x,x')$ tends to zero when the spatial separation between $x$ and $x'$ tends to infinity
and whenever it is of Hadamard form in a neighborhood of $\cH_{ev}$, then, towards future infinity, the Hawking radiation appears.
More precisely, if $h$ is a compactly supported smooth function supported far away from the black hole, they show that, for positive large values of $T$, the expectation value of
$\La'(\beta_T^{X}(h),\beta_T^{X}(h))$, interpreted as the response of a detector,
is composed by two contributions. One relates the signals
received by outward directed detectors (looking away from the collapsed
star) and such contribution is completely due to the Boulware vacuum.
The other one takes, instead, the approximated form
\beq
\sum_{l,m} \int_{\bR} d\epsilon \frac{|D_\ell(\epsilon)|^2}{\epsilon
(e^{\epsilon \beta_H} -1)} |\widetilde{h}_{lm}(\epsilon)|^2  \label{LAST}
\eeq
valid at positive large $T$ and assuming that the support of $h$ stays in a region $r>> R >0$.
Above $|D_\ell(\epsilon)|^2$ is the (gravitational) barrier penetration factor at energy $\epsilon$.
Furthermore $|\widetilde{h}_{lm}(\epsilon)|^2$ is the sensitivity of the
detector to quanta of energy $\epsilon$ and to angular momentum individuated by the quantum numbers $l$ and $m$, found employing an approximated mode decomposition as displayed in equation
(VIII.3.45) of  \cite{Haag}.  Formula \eqref{LAST} 
shows that the asymptotic counting rate is the one produced by an outgoing flux of radiation at temperature $1/\beta_H$ modified by the barrier penetration effect.\\
We stress that, as $\omega_U$ is of Hadamard form on $\mM$,  the Fredenhagen-Haag's result can be applied to the Unruh state, proving that it describes the appearance of the Hawking radiation near $\Im^+$ as described in \eqref{LAST}.
Here, the splitting of $\La_U(\beta_T^{(X)}(h),\beta_T^{(X)}(h))$  in the two contributions
$\La_{\cH}(\beta_T^{(X)}(h),\beta_T^{(X)}(h))$, due to $\cH$ and  $\La_{\scrim}(\beta_T^{(X)}(h),\beta_T^{(X)}(h))$ due to $\scrim$
is already embodied in the very construction of $\omega_U$ as
$\omega_\cH\otimes \omega_\scrim$.  Furthermore, they coincide separately with the two terms surviving in the limit $T\to+\infty$ according to the analysis given in \cite{Fredenhagen}. This last extent can be shown
by invariance of $\omega_U$ under $\beta^{(X)}$, hence
moving towards the past
the Cauchy surface used in \cite{Fredenhagen},  instead of
moving the support of $h$ towards the future.
In the limit $T\to+\infty$, the contribution due to $\cH$, gives rise to the expression (4.13)  in \cite{Fredenhagen} in particular.
Finally, one can follow almost slavishly the very steps of \cite{Fredenhagen}, which are based on the asymptotic behaviour of the solutions of the Klein-Gordon equation in $\mW$, to get (\ref{LAST}).\\
As a last comment, we emphasize that this result is stable under perturbations of the state $\om_U$ that involve only modification of $\omega_{\cH}$ in such a way that its integral kernel, seen as a bilinear map on $\sS(\cH)$, differs from the one of $\la_\cH$ in \eqref{lKW2} by a smooth function on $\cH\times\cH$, integrable
in the product measure $(dU \wedge d\bS^2)\wedge (dU' \wedge d\bS'^2)$.
Let us indicate by $\om''$ the perturbed state and by $\La''$ its two-point function, which is supposed to be a well defined distribution in $\mD'(\mM\times\mM)$; per direct application of Lebesgue's dominated convergence theorem, one finds that the contribution due to the perturbation vanishes at large $T$ under the action of $\beta^{(X)}_T$, so that:
$$
\lim_{T\to\infty} \La''(\beta^{(X)}_T f\otimes \beta^{(X)}_T h)=\La_U(f\otimes h) \;. \qquad f, h\in C^\infty_0(\mW)
$$
This is tantamount to claim that, for large positive values of $T$,
$\om''$ tends weakly to the Unruh state. In other words, far in the future, the effects seen in $\om''$
coincide with those shown in $\om_U$, hence the Hawking radiation also appears in the perturbed state.


\se{Conclusions}

In this paper we employed a bulk-to-boundary reconstruction procedure to rigorously and unambiguously 
construct and characterise on $\mM$ ({\it i.e.}, the static joined with  the black hole region of Schwarzschild 
spacetime, event horizon included) the so-called Unruh state $\om_U$. Such state plays the role of natural candidate to be used in the
quantum description of the radiation arising during a stellar collapse. Furthermore we proved that $\om_U$ 
fulfils the so-called Hadamard condition, hence it can be considered a genuine ground state for a massless 
scalar field theory living on the considered background. Overall, the achieved result can be seen as a novel
combination of earlier approaches \cite{DMP,Moretti08, DMP2, DMP3} with the theorems proved in \cite{SV00} as well as
with the powerful results obtained  by Rodnianski and Dafermos in \cite{DR03, DR07, DR05}.

Therefore we can safely claim that it is now possible to employ the Unruh vacuum in order to use the 
analysis in \cite{Fredenhagen} as a starting point, to study quantum effects such as the role of the back 
reaction of Hawking's radiation, a phenomenon which was almost always discarded as negligible. 

At the same time it would be certainly interesting to try to enhance the results of this paper since, as one
can readily infer from the main body of the manuscript, $\om_U$ has been here constructed only on $\mM$.
It is worth stressing that  it is, however, possible to extend $\omega_U$ to the whole Kruskal manifold 
following our induction procedure, 
defining a further part of the state  on $\Im_L^+$. The obtained state on the whole Weyl algebra $\cW(\mK)$ 
would be invariant under the group of Killing isometries generated by $X$ and without zero modes, if one
refers to the one-parameter group of isometries. The problem with this extension is related with the Hadamard 
constraint. Indeed, we do not expect that this extension is of Hadamard form on $\cH$, due to a theoretical 
obstruction beyond Candelas' remarks \cite{Candelas}. In view of the uniqueness and KMS-property theorem proved in \cite{KW} 
for a large class of spacetimes including the Kruskal one,
the validity of the Hadamard property on the whole spacetime together with the invariance under $X$ and the 
absence of zero-modes 
imply that the state is unique on a certain enlarged algebra of observables $\mA$ on $\mK$. Furthermore
it coincides with a KMS state with respect to the Killing 
vector $X$ at the Hawking temperature, {\em i.e.} it must be the Hartle-Hawking state for a certain 
subalgebra of observables $\mA_{0I} \subset \mA$ supported in the wedge $\mW$. 
These algebras are obtained out of a two steps approach. At first one enlarges $\cW(\mK)$ to $\mA$, whose 
Weyl generators are smeared with both
the standard solutions of KG equation with compactly supported Cauchy data in $\mK$ and a certain class 
of weak solutions of the same equation. Afterwards one restricts this enlarged algebra to a certain 
subalgebra of observables $\mA_{0I}$ supported in $\mW$, in a suitable sense related with the properties of 
the supports of the smearing distributions across the Killing horizon. 
With respect to our state we know that the KMS property is not verified in a neighbourhood of $\Im^-$, so we 
do not expect that any extension of that state satisfies there the KMS property. Nonetheless the issue is not 
completely clear since the extension we are discussing and the failure 
of the KMS condition are both referred to $\cW(\mW)$
rather than $\mA_{0I}$. Hence further investigations in such direction would be desirable.\\ 
 A further and certainly enticing possible line
of research consists of using the very same approach discussed in this paper in order both to rigorously define the very
Hartle-Hawking state and to prove its Hadamard property; although, from a physical perspective, this is
certainly a very interesting problem, from a mathematical perspective, it amounts to an enhancement of the
peeling behaviours for the solutions of the Klein-Gordon equation discussed by Rodnianski and Dafermos, also
beyond what recently achieved in \cite{Luk}.
Although there is no proof that the obtained ones are sharp conditions, the high degree of mathematical
specialisation, needed to obtain the present results, certainly makes the proposed programme a challenging
line of research, which we hope to tackle in future papers.\\
As an overall final remark, it is important stress that all our results are only valid for the massless case,
since the massive one suffers of a potential sever obstruction which is the same as the one pointed out in
\cite{DMP}. To wit it appears impossible to directly project on null infinity a solution of the massive wave 
equation and, hence, the problem must be circumvented with alternative means, as it has been done, for
example, in Minkowski spacetime in \cite{Da08a}. A potential solution of this puzzle in the Schwarzschild
background would be certainly desirable.

\section*{Acknowledgements.} The work of C.D. is supported by the von Humboldt Foundation and that of N.P. 
has been supported by the German DFG Research Program SFB 676. We are grateful to S. Hollands for the useful 
discussions and for having pointed out \cite{DR05}. We also deeply acknowledge B.S. Kay for the useful 
clarifications of the various equivalent definitions of KMS states as well as for having pointed out some 
relevant references.
 
\appendix

 
\se{Further details on the geometric setup.}\label{geometry}
In this paper, the extension of the underlying background to include null infinities as well as a region  
beyond them, plays a pivotal role and we shall now dwell into a few more details. To this end, one purely
follows \cite{SW83} and rescales the global metric 
$g$ in (\ref{g}) by a factor $1/r^2$ after which one can notice that the obtained manifold $(\mM, g/r^2)$ 
admits a smooth 
larger extension $(\widetilde{\mM},\widetilde{g})$. We have to notice that, in this case, the singularity 
present at $r=0$
in $(\mM,g)$ is pushed at infinity in the sense that the non-null geodesics
 takes an infinite amount of affine parameter to reach a point situated at $r=0$.
The extension of $(\widetilde{\mM},\widetilde{g})$ obtained in this way does not cover the 
sets indicated as $i^\pm$ and $i_0$ in figure 1 and figure 2, though it includes the boundaries 
${\Im}^{\pm}$, called {\bf future} and {\bf past null infinity} respectively. These represent subsets of 
$\widetilde{\mM}$ which are null $3$-submanifold of $\widetilde{\mM}$ formally localised at $r=+\infty$.

Let us now examine the form of the (rescaled or not) extended metric
restricted to the
Killing horizon $\cH$ as well as to the null infinities ${\Im}^\pm$.
Per direct inspection and without rescaling the metric, one finds that,
if one fixes $\Omega \doteq 2V/e$, which vanishes on $\cH$,
$$ g\spa\rest_{\cH} = r_S^2\left( - dU \otimes  d\Omega - d\Omega
\otimes d U + h_{\bS^2}(\theta,\phi)\right)\:. $$
In this case $U\in \bR $ is the complete affine parameter of the null
$\widetilde{g}$-geodesics
 generating $\cH$ and $\cH$ itself is obtained setting $V=0$.
 This form of the metric,
 up to the constant factor $r_S^2$,
  is called  {\bf geodetically complete Bondi form}.\\
The same structure occurs on $\Im^+$, formally individuated by $v=
+\infty$ and
on $\Im^-$, formally individuated by $u= -\infty$, where the metric
$\tilde{g} = \Omega^2 g = g/r^2$
has still a {\bf geodetically complete Bondi form}, namely
$$ 
\widetilde{g}\spa\rest_{\Im^+} =  d\Omega \otimes
 du + du \otimes d\Omega + h_{\bS^2}(\theta,\phi)\:,
$$
where $\Omega \doteq 1/r$ individuates $\scri$ for $\Omega=0$. Similarly
$$
\tilde{g}\spa\rest_{\Im^-} = - d\Omega \otimes dv - dv \otimes d\Omega
+ h_{\bS^2}(\theta,\phi)\:,
$$
where $\Omega \doteq 1/r$ individuates $\Im^-$ for $\Omega =0$.
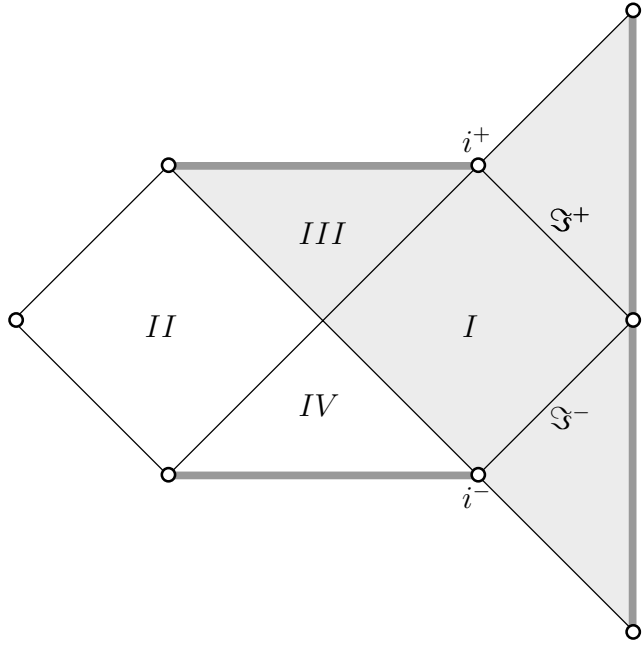
\begin{figure}
\centering
     \begin{picture}(150,200)(0,0) 
       \put(0,0){\includegraphics[height=8.5cm]{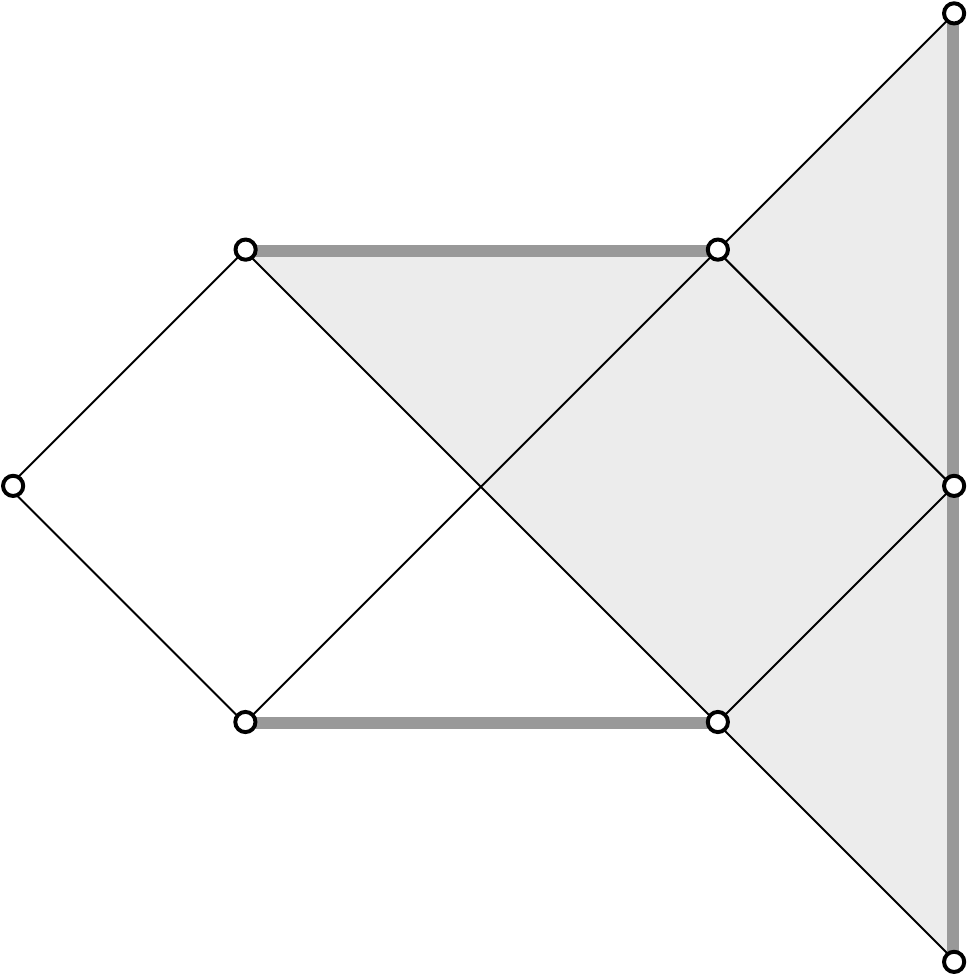}}
           \put(205,80){\large $\scrim$}
       \put(205,155){\large $\scri$}
        \put(205,155){\large $\scri$}
       \put(172,50){\large $i^-$}          
       \put(172,185){\large $i^+$}
       \put(172,115){\large $I$}
        \put(52,115){\large $II$}
          \put(110,150){\large $III$}
         \put(110,85){\large $IV$}
     \end{picture}
       \caption{The Kruskal spacetime $\mK$ is the union of the open regions $I$,$II$,$III$, and $IV$ including their common boundaries.
$\mM$ is the union of $I$ and $III$ including the common boundary $\cH_{ev}$.
The conformal extension $\widetilde{\mM}$ of $\mM$ beyond $\scri$ and $\scrim$ is the gray region. The thick lines denote the metric singularities at $r=0$.}
    \end{figure}
In both cases the coordinates
$u$ and $v$ are well defined and they coincide with 
the complete affine parameters of the null $\widetilde{g}$-geodesics forming $\Im^+$
and $\Im^-$ respectively.\\
With respect to Killing symmetries, we notice that the $g$-Killing vector $X$ is also a Killing vector for 
$\widetilde{g}$ and it extends to a $\widetilde{g}$-Killing vector $X$ defined on $\widetilde{\mM}$.
Particularly, in $\partial \mM$ it satisfies
$$X=\partial_u\;\;\textrm{on}\;\Im^+,\qquad X=\partial_v\;\;\textrm{on}\;\Im^-.$$

\section{Weyl algebras, quasifree states, KMS condition.} \label{algebras}
 A $C^*$-algebra $\cW(\sS)$ is called {\bf Weyl algebra} associated 
with a (real) symplectic space $(\sS,\sigma)$ (the symplectic form $\sigma$ being nondegenerate) 
if it contains a class of non-vanishing elements $W(\psi)$ for all
 $\psi \in \sS$, called {\bf Weyl generators}, 
which satisfy {\bf Weyl relations}\footnote{Notice that in \cite{KW} a different convention for the sign of 
$\sigma$ in (W2) is employed.}:
$$(W1)\quad\quad W(-\psi)= W(\psi)^*\:,\quad\quad\quad\quad (W2)\quad\quad W(\psi)W(\psi') =
 e^{i\sigma(\psi,\psi')/2} W(\psi+\psi') \:.$$
$\cW(\sS)$ coincides with the closure of the $*$-algebra (finitely) generated
 by Weyl generators. 
As a consequence of (W1) and (W2), one gets: $W(0) = \bI$ (the unit element), 
$W(\psi)^*= W(\psi)^{-1}$, 
$||W(\psi)||=1$  and, out of the non degenerateness of $\sigma$, $W(\psi)=W(\psi')$ 
iff $\psi=\psi'$ .\\ 
 $\cW(\sS)$ is {\em uniquely} determined by $(\sS,\sigma)$  (theorem 5.2.8 in \cite{BR2}): 
Two different realizations admit a unique $*$-isomorphism which transform the former into the latter,
preserving Weyl generators, and the norm on $\cW(\sS)$ is unique, since
$*$ isomorphisms of $C^*$-algebras are isometric. This result implies that every GNS $*$-representation of a 
Weyl algebra is always faithful
and isometric. It is also worth mentioning that, per construction, any GNS $*$-representation of a Weyl 
algebra is such that 
the generators are always represented by unitary operators, but it is not the case for other
$*$-representations in Hilbert spaces.\\
$\cW(\sS)$ can always be realized in terms of bounded operators on $\ell^2(\sS)$, viewing $\sS$ as a
Abelian group and defining the generators as
$(W(\psi)F)(\psi')\doteq e^{-i\sigma(\psi,\psi')/2}F(\psi+\psi')$ for every $F\in \ell^2(\sS)$.
In this realization (and thus in every realization) it turns out that the generators $W(\psi)$ are 
{\em linearly independent}. 
 A state $\omega$ on $\cW(\sS)$, with GNS triple $(\gH_\omega, \Pi_\omega, \Omega_\omega)$, is called
{\bf regular} if the maps $\bR\ni t\mapsto \Pi_\omega(W(t\psi))$ are strongly continuous. 
In general,  strong continuity of the unitary group implementing a $*$-automorphism representation
$\beta$ of a topological group $G \ni g \mapsto \beta_g$ for a $\beta$-invariant 
state  $\omega$ on a Weyl algebra 
$\cW(\sS)$, 
is equivalent to $\lim_{g\to \bI} \omega(W(-\psi)\beta_g W(\psi)) = 1$ for all $\psi \in \sS$.
The proof follows immediately from  
$||\Pi_\omega\left(\beta_{g'}  W(\psi)\right)\Omega_\omega - \Pi_\omega\left(\beta_{g}  W(\psi)\right)\Omega_\omega||^2
= 2- \omega\left(W(-\psi)\beta_{g'^{-1}g}W(\psi)\right) -  
\omega\left(W(-\psi)\beta_{g^{-1}g'}W(\psi)\right)$
 and $\overline{\Pi_\omega(\cW(\sS))\Omega_\omega} = \gH_\omega$.\\
If $\omega$ is regular, in accordance with Stone theorem, one can write 
$\Pi_\omega(W(\psi)) = e^{i\sigma(\psi,\Phi_\omega)}$,  
 $\sigma(\psi,\Phi_\omega)$ being the (self-adjoint) {\bf field operator symplectically-smeared} with $\psi$.\\
When $\cW(\sS)= \cW(\sS(\mN))$ is the Weyl algebra on the space of Klein-Gordon equation solutions as in  Sec. \ref{observables},
the field operator $\Phi_\omega(f)$ introduced in that section, smeared with smooth compactly supported 
functions $f \in C_0^\infty(\mN;\bR)$, is related with $\sigma(\psi,\Phi_\omega)$ by 
$$
\Phi_\omega(f) \doteq \sigma(E_{P_g}(f),\Phi_\omega)\quad \mbox{for all $f \in C_0^\infty(\mN;\bR)$,}
$$
where we exploit the notations used in Sec. \ref{observables}.
In this way, the field operators enter the theory in the Weyl algebra scenario.
At a formal level, Stone theorem together with (W2) imply both $\bR$-linearity and the standard CCR:
$$
(L)\quad \sigma(a\psi + b \psi', \Phi_\omega) = a \sigma(\psi,\Phi_\omega) + b \sigma(\psi',\Phi_\omega)\:, \quad\:\: 
(CCR) \quad \mbox{$[$}\sigma(\psi,\Phi_\omega), \sigma(\psi',\Phi_\omega)\mbox{$]$} = -i\sigma(\psi,\psi')I\:,$$  
for $a,b\in \bR$ and $\psi,\psi' \in \sS$. 
Actually (L) and (CCR) hold rigorously in an invariant dense set of analytic vectors
by Lemma 5.2.12 in \cite{BR2} (it holds if $\omega$ is quasifree by proposition 
\ref{proposition2}).\\
In the standard approach of QFT, based on bosonic real scalar field operators $\Phi$, {\em either a vector or
a density matrix} state are {\em quasifree} 
if the associated $n$-point functions satisfy
 (i) $\langle\sigma(\psi,\Phi) \rangle =0$ for all $\psi\in \sS$ and (ii)
 the  $n$-point functions  $\langle  \sigma(\psi_1,\Phi)\cdots \sigma(\psi_n,\Phi)\rangle$
are determined from the functions $\langle \sigma(\psi_i,\Phi)\sigma(\psi_j,\Phi) \rangle$, with
$i,j=1,2,\cdots, n$, using standard Wick's expansion. 
A technically different but substantially equivalent definition, completely based on the Weyl algebra
 was presented in \cite{KW}. It relies on the following three observations:
(a) if one works formally with (i) and (ii), one finds that it holds
$\langle  e^{i\sigma(\psi,\Phi)} \rangle =
 e^{-\langle \sigma(\psi,\Phi)\sigma(\psi,\Phi) \rangle/2}$.
In turn, at least formally, such identity determines the $n$-point functions by Stone theorem and (W2).
(b) From (CCR) it holds $\langle \sigma(\psi,\Phi)\sigma(\psi',\Phi)\rangle = \mu(\psi,\psi') - (i/2) \sigma(\psi,\psi')$,
where $\mu(\psi,\psi')$ is the symmetrised  two-point function 
$(1/2)(\langle \sigma(\psi,\Phi)\sigma(\psi',\Phi)\rangle + \langle  \sigma(\psi',\Phi)\sigma(\psi,\Phi)
\rangle)$
 which defines a symmetric positive-semidefined bilinear form on $\sS$.
 (c) $\langle A^\dagger A\rangle\geq 0$
for  elements $A\doteq [e^{i\sigma(\psi,\Phi)} -I]+i[e^{i\sigma(\psi,\Phi)}-I]$ 
entails: 
\beq
|\sigma(\psi,\psi')|^2 \leq 4\:\mu(\psi,\psi)\mu(\psi',\psi')\:, 
\quad\quad \mbox{for every $\psi,\psi' \in \sS$}\label{sm}\:,
\eeq
which, in turn,  implies that {\em $\mu$ is strictly positive defined} because $\sigma$ is non degenerate. 
If one reverses the procedure, the general definition of quasifree states on Weyl algebras is the following.

\definizione \label{defquasifree}
{\em Let $\cW(\sS)$ be  a Weyl algebra and $\mu$ a real scalar product on $\sS$ 
satisfying (\ref{sm}).
A state $\omega_\mu$ on $\cW(\sS)$ is called the {\bf quasifree state}  associated with $\mu$
if
$$\omega_\mu(W(\psi)) \doteq e^{-\mu(\psi,\psi)/2} \:, \quad \mbox{for all $\psi\in \sS$.}$$} 
\noindent The following technical lemma is useful to illustrate the GNS triple of a quasifree state as 
established in the subsequent theorem. 
The last statement in the lemma arises out of the Cauchy-Schwarz inequality and
the remaining part out of Proposition 3.1 in \cite{KW}.

\lemma \label{lemma1A}
{\em Let $\sS$ be a real symplectic space with $\sigma$ non degenerate and $\mu$  
 a real scalar product on $\sS$ fulfilling (\ref{sm}).
There exists a complex Hilbert space $\sH_\mu$
and a map $K_\mu: \sS \to \sH_\mu$ with: 

(i)  $K_\mu$ is $\bR$-linear with dense complexified range, i.e. $\overline{K_\mu(\sS) + i K_\mu(\sS)}= \sH_\mu$,

(ii) for all $\psi,\psi' \in \sS$,
$\langle K_\mu\psi , K_\mu\psi'\rangle= \mu(\psi,\psi') - (i/2) \sigma(\psi,\psi')$.\\
Conversely, if the pair $(\sH,K)$ satisfies (i) and 
$\sigma(\psi,\psi')= -2 Im \langle K\psi , K\psi'\rangle_\sH$, with $\psi,\psi' \in \sS$, 
the unique real scalar product $\mu$ on  $\sS$ satisfying (ii) verifies  (\ref{sm}).} \\

\noindent   An existence theorem for quasifree states can be proved using the lemma above with the following proposition
relying on Lemma A.2, Proposition 3.1 and a comment on p.77 in \cite{KW}).

\proposizione\label{proposition2}
{\em For every $\mu$ as in definition \ref{defquasifree} the following hold.\\
{\bf (a)} There exists a unique quasifree state $\omega_\mu$ 
associated with $\mu$ and it is  
regular.\\   
{\bf (b)} The GNS triple $(\gH_{\omega_\mu}, \Pi_{\omega_\mu}, \Omega_{\omega_\mu})$
   is determined as follows with respect to  $(\sH_\mu,K_\mu)$ as in 
  lemma \ref{lemma1A}. (i) $\gH_{\omega_\mu}$ is the symmetric Fock space with  
  one-particle space $\sH_\mu$. (ii) The cyclic vector   $\Omega_{\omega_\mu}$ is the vacuum vector of $\gH_\omega$. 
   (iii) $\Pi_{\omega_\mu}$ is  determined by
$\Pi_{\omega_\mu}(W(\psi)) = e^{i\overline{\sigma(\psi,\Phi_{\omega_\mu})}}$, the bar denoting the
closure, where\footnote{The field operator $\Phi(f)$, with $f$ in the complex Hilbert space $\gh$, 
used in \cite{BR2} in propositions 5.2.3 and 5.2.4 is related to $\sigma(\psi,\Phi)$ by means of
$\sigma(\psi,\Phi)= \sqrt{2} \Phi(iK_\mu \psi)$ assuming $\sH\doteq \gh$.}
$$
\sigma(\psi,\Phi_{\omega_\mu}) \doteq  ia(K_\mu\psi) -ia^\dagger(K_\mu\psi)\:, \quad \mbox{for all $\psi\in
\sS$} $$
$a(\phi)$ and $a^\dagger(\phi)$, $\phi\in \sH_\mu$, being the usual annihilation (antilinear in $\phi$) 
and creation operators defined 
in the dense linear manifold spanned by the states with finite number of particles.\\
{\bf (c)} A pair $(\sH,K)\neq (\sH_\mu,K_\mu)$ satisfies (i) and (ii)  in lemma \ref{lemma1A} 
for $\mu$, thus determining the same quasifree state $\omega_\mu$, if and only if there is a unitary operator
$U: \sH_\mu\to \sH$ such that $UK_\mu=K$.\\
{\bf (d)} $\omega_\mu$ is pure, i.e., its GNS representation is irreducible
if and only if $\overline{K_\mu(\sS)} = \sH_\mu$.
In turn, this is equivalent  to
$4\mu(\psi',\psi') =  \sup_{\psi\in \sS\setminus\{0\}} |\sigma(\psi,\psi')|/\mu(\psi,\psi)$ for every
$\psi'\in \sS$.}

\remark \label{remarkstates}\\
{\bf (1)} $K_\mu$ is always injective due to (ii) and non degenerateness of 
$\sigma$.  \\
{\bf (2)} Consider the real Hilbert space obtained by taking the completion of
 $\sS$ with respect to $\mu$. 
The requirement (\ref{sm}) is equivalent to the fact that there is 
 is a bounded operator $S$ everywhere defined over the mentioned Hilbert space,  with $S=-S^*$, $||S|| \leq 1$ and
 such that $\frac{1}{2}\sigma(\psi,\psi') = \mu(\psi, S \psi')$, for all $\psi,\psi' \in \sS$.\\
{\bf (3)} The pair $(\sH_\mu,K_\mu)$ is called the {\bf one-particle structure} of the quasifree state $\omega_\mu$.\\

\noindent Let us pass to discuss the KMS condition \cite{hug,Haag,BR2}. KMS state are the algebraic 
counterpart, for infinitely extended systems, of thermal states of standard statistic mechanics.
 There are several different equivalent definitions of KMS states, see  \cite{BR2} for a list of various equivalent definitions.
While bearing in mind Definition 5.3.1 and Proposition 5.3.7 in \cite{BR2}, we adopt the following one:\\

\definizione\label{KMSdef} A state $\omega$ on a $C^*$-algebra $\mA$ is said to be a  {\bf KMS state at 
inverse temperature $\beta\in \bR$}  with respect to a one-parameter group of 
$*$-automorphisms $\{ \alpha_t\}_{t\in \bR}$ which represents, from the algebraic point of view, some notion 
of time-evolution if, for every pair $A,B \in \mA$, and with respect to the function
$\bR \ni t \mapsto \omega\left(A\alpha_t(B)\right)=: F^{(\omega)}_{A,B}(t)$, the following facts hold.\\
(a) $F^{(\omega)}_{A,B}$ extends to a continuous complex function $F^{(\omega)}_{A,B} = F^{(\omega)}_{A,B}(z)$
 with domain  $$\overline{D_\beta} \doteq \{z \in \bC\:|\: 
0 \leq Im z \leq \beta\}\quad \mbox{if $\beta\geq 0$, or}\quad
\overline{D_\beta} \doteq \{z \in \bC\:|\: 
\beta \leq Im z \leq 0\} \quad \mbox{if $\beta\leq 0$,}
$$
(b) $F^{(\omega)}_{A,B} = F^{(\omega)}_{A,B}(z)$ is analytic in the interior of $\overline{D_\beta}$;\\
(c) it holds, and this identity is -- a bit improperly -- called the {\bf KMS condition}:
$$F^{(\omega)}_{A,B}(t+i\beta) = \omega\left(\alpha_t(B)A \right)\:, \quad \mbox{for all $t \in \bR$.}$$
With the given definition, an $\{ \alpha_t\}_{t\in \bR}$-KMS state $\omega$ turns out to be invariant 
under $\{ \alpha_t\}_{t\in \bR}$ \cite{BR2};
 the function 
$\overline{D_\beta} \ni z \mapsto 
F^{(\omega)}_{A,B}(z)$ is uniquely determined by its restriction
to real values of $z$ (by the ``edge of the wedge theorem'') and $\sup_{\overline{D_\beta} } |F^{(\omega)}_{A,B}|
= \sup_{\partial D_\beta} |F^{(\omega)}_{A,B}|$ (by the ``three lines theorem'')
\cite{BR2}.\\
 Equivalent definitions of KMS states are obtained by the following propositions, 
the second for quasifree states,
due to Kay \cite{Kay12,KW} and relying upon earlier results by
Hugenholtz \cite{hug}. We  sketch  the proofs since they are very spread in the literature.
 
\proposizione\label{KMSkay}
{\em An algebraic state $\omega$, on the $C^*$-algebra $\mA$, which is invariant under the one-parameter group of 
$*$-automorphisms $\{ \alpha_t\}_{t\in \bR}$ is a KMS state at the inverse temperature $\beta\in \bR$ if and only if its GNS triple
$(\cH_\omega, \Pi_\omega, \Omega_\omega)$ satisfies the following three requirements. \\
(1) The unique unitary group $\bR \ni t \mapsto U_t$
which leaves $\Omega_\omega$ invariant and implements $\{ \alpha_t\}_{t\in \bR}$ -- i.e.
$\Pi_\omega\left(\alpha_{t}(A)\right) = U_t \Pi_\omega(A) U^*_t\quad \mbox{for all $A\in \mA$ and $t\in \bR$}$ --
 is strongly continuous, so that $U_t = e^{itH}$
 for some self-adjoint operator $H$ on $\cH_\omega$.\\ (2) $\Pi_\omega\left(\mA\right) \Omega_\omega\subset Dom\left(e^{-\beta H/2}\right)$. \\
 (3) There exists an antilinear operator $J: \cH_\omega \to \cH_\omega$ with $JJ=I$ such that:
$$
Je^{-it H} = e^{-it H}J \quad \mbox{for all $t\in \bR$,  and}\quad
 e^{-\beta H/2} \Pi_\omega(A)\Omega_\omega = J \Pi_\omega(A^*)\Omega_\omega \quad \mbox{for all $A\in \mA$.}
$$}

\noindent {\em Proof}. A $\{ \alpha_t\}_{t\in \bR}$-KMS state  with inverse temperature $\beta$ 
is $\{ \alpha_t\}_{t\in \bR}$-invariant and fulfils 
the conditions (1), (2) and (3) due to Theorem 6.1 in \cite{hug}.
Conversely, consider an $\{ \alpha_t\}_{t\in \bR}$-invariant state $\omega$ on $\mA$ which fulfils
 the conditions (1), (2) and (3).
When $A$ and $B$ 
are entire analytic elements of $\mA$ (see \cite{BR2}), $\bR \ni t \mapsto F^{(\omega)}_{A,B}(t)$ uniquely 
extends to an analytic function on the whole $\bC$ and thus (a) and (b) in def. \ref{KMSdef} are true.
(1), (2), (3) and $e^{z H} \Omega_\omega = \Omega_\omega$, for all 
$z\in \overline{D_\beta}$ (following from (2) and (3)) also entail (c):
$$\omega(\alpha_t(B) A) = \langle\Omega_\omega,\: U_t \Pi_\omega(B) U^*_t \Pi_\omega(A) \Omega_\omega \rangle 
= \langle  \Pi_\omega(B^*)\Omega_\omega,\: U_t^* \Pi_\omega(A) \Omega_\omega\rangle
= \langle  J U_t^* \Pi_\omega(A)\Omega_\omega  ,\: J\Pi_\omega(B^*) \Omega_\omega\rangle$$
$$= \langle   U_t^* e^{-\beta H/2}\Pi_\omega(A^*) \Omega_\omega ,\: e^{-\beta H/2} \Pi_\omega(B) \Omega_\omega\rangle
= \langle\Omega_\omega,\: \Pi_\omega(A) e^{i(t+i\beta)H} \Pi_\omega(B)e^{-i(t+i\beta)H}  \Omega_\omega\rangle
= F_{A,B}^{(\omega)}(t+i\beta)\:.$$
The validity of conditions (a), (b) and (c) for  entire analytic elements $A,B \in \mA$ implies the validity  
for all $A,B\in \mA$, as established in \cite{BR2} (compare Definition 5.3.1 and Proposition 5.3.7 therein). 
$\Box$

\proposizione \label{kkms}
{\em 
 Consider a quasifree algebraic state $\omega_\mu$ on the Weyl-algebra $\cW(\sS)$, with
one-particle structure $(\sH_\mu,\sK_\mu)$. Assume that (i) $\omega_\mu$ is invariant under
the one-parameter group of $∗$-automorphisms
$\{ \alpha_t\}_{t\in \bR}$ and that (ii) $\{ \alpha_t\}_{t\in \bR}$ is implemented by a strongly continuous
unitary one-parameter group $\{U_t\}_{t\in \bR}$ in the GNS Hilbert Fock space,
leaving fixed the vacuum vector, and obtained by tensorialization of a
unitary one-parameter group $\{V_t = e^{i \tau h} \}_{t\in \bR}$ in $\sH_\mu$ .
The following facts are equivalent. 

{\bf (a)} $\omega_\mu$ is a KMS state at the inverse temperature $\beta \in \bR$ with respect to $\{ \alpha_t\}_{t\in \bR}$.

{\bf (b)} There is an anti-unitary operator
$j: \sH_\mu \to \sH_\mu$ with $jj = I$ and the following facts hold:\\
(i) $\sK_\mu (\sS) \subset Dom\left( e^{-\frac{1}{2}\beta h}\right)$,
 (ii) $[j, V_t] =0$ for all $t\in \bR$, 
(iii) $e^{-\frac{1}{2}\beta h} \sK_\mu  \psi = -j \sK_\mu \psi$ for all $\psi \in \sS(\cH_\mu)$.

{\bf (c)}  $\sK_\mu (\sS) \subset Dom\left( e^{-\frac{1}{2}\beta h}\right)$ and
$\langle e^{-it h} x, y \rangle = \langle e^{-\beta h/2} y , e^{-it h} e^{-\beta h/2} x\rangle$
 if $x,y \in \sK_\mu(\sS)$ and  $t\in \bR$.}\\

\noindent{\em Proof}. (a) is equivalent to (b) as proved on pages 80-81 in \cite{KW}. 
(b) entails (c) straightforwardly. If one assumes (c) and exploits (i) of lemma \ref{lemma1A}, $j:\sH_\mu\to 
\sH_\mu$, which fulfils (b), is completely individuated by continuity
and anti-linearity under the request that $j \sK_\mu \psi = -e^{-\frac{1}{2}\beta h}\sK_\mu\psi$ when $\psi 
\in \sS$. $\Box$
 
\section{Proofs of some propositions.} \label{Appendixproofs}

\noindent{\bf Proof of Lemma \ref{lemma1}}.
As in Appendix \ref{geometry}, let us consider the conformal extension $(\widetilde{\mM},\widetilde{g})$ 
of the spacetime $(\mM,g)$ determined in \cite{SW83} where $\widetilde{g} = g/r^2$ in $\mM$ (see figure 2).
In view of the previously illustrated properties of $E_{P_g}$, if $\varphi \in \sS(\mM)$, there is a smooth function $f_\varphi$ with support contained in 
$\mM$ and such that $\varphi = E_{P_g}f_\varphi$, and $supp \varphi \in  J^+(supp f_\varphi ; \mM) \cup  
J^-(supp f_\varphi ; \mM)$. Since
$J^\pm(supp f_\varphi ; \mM)  \subset J^\pm(supp f_\varphi ; \widetilde{\mM})$,
 the very structure of $\widetilde{\mM}$  (see figure 2) guarantees
  that, if the smooth extension $\widetilde{\varphi}$ of $r\varphi$ in a neighbourhood 
 of $\Im^\pm\subset \widetilde{\mM}$ exists, it  must have support  bounded 
 by  constants $v^{(\vphi)}, u^{(\vphi)} \in (-\infty,\infty)$. Here we adopt the relevant null coordinates 
 in the considered neighbourhood: $(\Omega, u, \theta, \phi)$
 or  $(\Omega, v, \theta, \phi)$ respectively, where $\Omega = 1/r$ in $\mW$.  Furthermore, in view of the 
 shape of 
 $J^\pm(supp f_\varphi ; \widetilde{\mM})$, the analogous property  holds true for the support of $\varphi$ 
 in  $\mW$.
 The existence of $\widetilde{\varphi}$ can be established examining the various possible cases. 
To start with, let us assume that $supp f_\vphi \subset \mW$.
 Let $p\in \mW$ be in the chronological past of $supp f_{\varphi}$ 
sufficiently close to $i^-$. Afterwards, let us consider a second point $q$ beyond $\Im^+$, though
sufficiently close to $\Im^+$ so that the closure of
$\mN_{p,q} \doteq I^+(p;\widetilde{\mM}) \cap I^-(q; \widetilde{\mM})$ does not meet the timelike 
singularity in the conformal extension of $\mM$ on the 
right of $\Im^+$. Let us consider $\mN_{p,q} $ as a spacetime equipped with the metric $\tilde{g}$.
It is globally hyperbolic since, per direct inspection, one verifies that the diamonds $J^+(r; \mN_{p, q} )
\cap J^-(s; \mN_{p,q} )$ are empty or compact 
for $r,s \in \mN_{p, q}$ while the spacetime itself is 
causal. Hence  $E_{P_{\widetilde{g}}}$ is well defined and individuates 
 a solution $\widetilde{\varphi} \doteq E_{P_{\widetilde{g}}} f_\varphi$ of the Klein-Gordon equation 
 associated with $P_{\widetilde{g}}$, as in (\ref{tildeKG}), with $\widetilde{g} = g/r^2$. 
Thanks to the properties of the Klein-Gordon equation under conformal rescaling \cite{Wald}, one has 
$\widetilde{\varphi}  = r \varphi$ in  $\mM$ because $\widetilde{g} = g/r^2$ therein. If we keep $p$ fixed 
while moving $q$ in a parallel way to $\Im^+$ towards $i^+$, one obtains an increasing 
 class of larger globally-hyperbolic spacetimes $\mN_{p, q}$
 and, correspondingly, a class of analogous extensions $\widetilde{\varphi}$ on corresponding $\mN_{p, q}$. 
Furthermore, if one considers two of these 
 extensions, they coincide in the intersections of their domains (see figure 3).
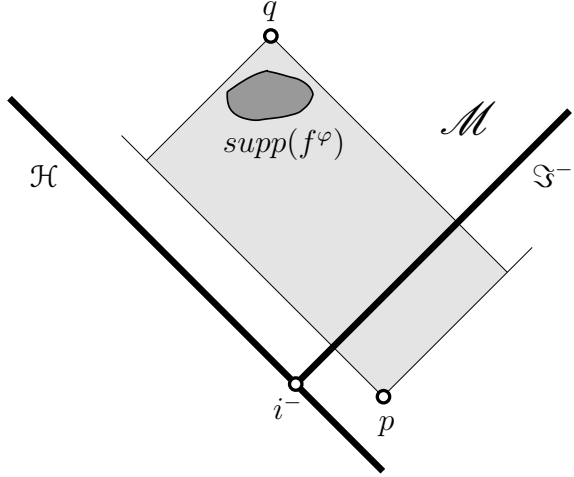
\begin{figure}
\centering
     \begin{picture}(150,170)(0,0) 
       \put(0,0){\includegraphics[height=6cm]{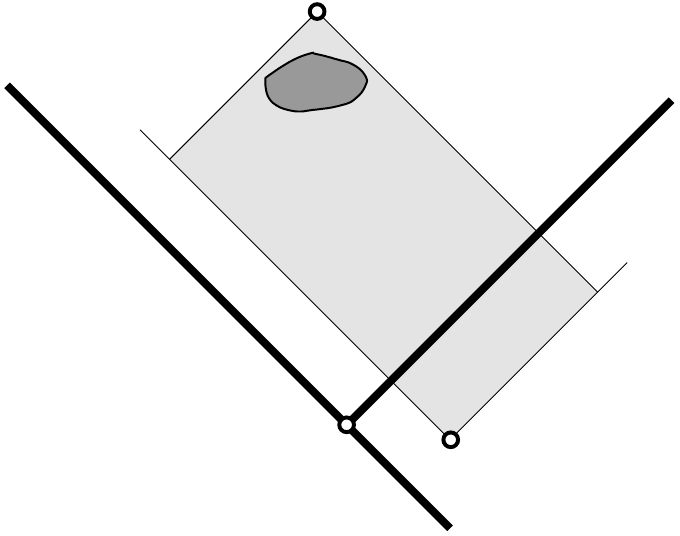}}
       \put(102,22){\large $i^-$}
       \put(98,175){\large $q$}
        \put(83,123){\large $supp(f^\varphi)$}
       \put(142,18){\large $p$}
       \put(165,130){\LARGE $\mM$}
       \put(200,110){\large $\scrim$}
       \put(10,110){\large $\cH$}          
       \end{picture}
     \caption{The gray region indicates the globally hyperbolic subspacetime $\mN_{p,q}$ of $\widetilde{\mM}$, the point $p$ eventually tends  to $i^-$. }
\end{figure}
In order to draw these conclusions, we exploited of the uniqueness of the solution of a Cauchy problem as
well as the property according to which any compact portion of a spacelike Cauchy surface of a globally 
hyperbolic spacetime can be extended  to a smooth spacelike Cauchy surface of any larger globally hyperbolic 
spacetime \cite{BS06}. Hence the initial data can be read on the larger spacelike Cauchy surface, also thanks
to its acausal structure (Lemma 42 from Chap. 14 in \cite{ON}). Accordingly, a smooth extension of $r\varphi$
turns out to be defined in a neighbourhood of $\Im^+$ and an almost slavish procedure yields the analogous 
extension on $\Im^-$.  Let us now suppose that $supp f_{\varphi} \subset \mB$. In such case $\varphi$ cannot
reach $\Im^+$ and, thus, the only extension  of $r\varphi$ concerns $\Im^-$.
The employed procedure is similar to the one above, though the class of globally hyperbolic spacetime is 
constructed as follows. Let us take a point $p$ beyond $\Im^-$ sufficiently close to $i^-$ and let us 
consider the intersection $\mN_p \doteq I^+(p;\widetilde{\mM})\cap I^-(\mM;\widetilde{\mM})$. If one moves 
$p$ parallelly to $\Im^-$ drawing closer to $i^-$, one obtains an increasing class of globally hyperbolic
spacetimes. equipped with the metric $\widetilde{g} = g/r^2$, and, correspondingly, a class of solutions of
the rescaled Klein-Gordon equation. These define the smooth extension $\widetilde{\varphi}$ of $r\varphi$ in
an open neighbourhood of $\Im^-$.  
Let us now consider the case where $supp f_{\varphi}$ is concentrated in an arbitrarily shrunk open 
neighbourhood of $\cH_{ev}$. While the behaviour of $\varphi$ in a neighbourhood of $\Im^-$, mimics the
previously examined one, that around $\Im^+$ deserves a closer look mostly with reference to the construction
of the relevant globally hyperbolic spacetimes. To this end, let us fix a point $p\in \mW$ in the 
chronological past of $supp f_{\varphi}$ sufficiently close to $\cH$. Afterwards, let us
consider a smooth spacelike surface $\Sigma$ in the chronological future of $supp f_{\varphi}$, 
which lies in the past of $i^+$ in $\widetilde{\mM}$ and it intersects $\Im^+$ for some $u=u_\Sigma$.
The relevant class of globally hyperbolic spacetimes is now made of the sets
$\mN_{p, \Sigma} \doteq  J^-(\Sigma; \widetilde{\mM}) \cap J^+(p; \widetilde{\mM})$ when $\Sigma$ moves towards $i^+$.
It remains to consider the case where $supp f_{\varphi}$ intersect $\cH_{ev}$,  but it is not confined in a small 
neighbourhood of $\cH_{ev}$. In this case, if one takes into account the linearity of both the causal 
propagator and $P_{\widetilde{g}}$, we can reduce ourselves to a combination  of the three above considered 
cases. If one decomposes the constant function $1$ in $\mW$ as the sum of three non-negative smooth functions
$1= f_1+f_2+f_3$, with $f_1$ supported in $\mB$, $f_2$ supported in $\mW$ and $f_3$ supported in an 
arbitrarily shrunk open neighbourhood of $\cH_{ev}$, we have $f_\varphi = f_\varphi\cdot f_1+f_\varphi\cdot
f_2+f_\varphi\cdot f_3$. If we fix $r\varphi_i\doteq r E_{P_g} (f_\varphi \cdot f_i)$, $i=1,2,3$,  each 
wavefunction can be treated separately as discussed above, hence yielding corresponding extensions 
$\widetilde{\varphi}_i$ to $\Im^+$. The sum of these extensions is, per construction, the wanted one
$\widetilde{\varphi}$ of $r\varphi$. The same procedure applies to the case of $\Im^-$. $\Box$\\

\noindent {\bf Proof of Proposition \ref{PropDR}.}
(a) We consider the proof for the case of $t>0$, {\it i.e.}, the behaviour of the wavefunctions about
 $\cH_{ev}$ and $\Im^+$ only), the remaining case being
then an immediate consequence of the symmetry $X \to -X$ of the Kruskal geometry.\\ 
To start with, it is worth noticing that each of our coordinates $u,v$ amounts to twice the corresponding 
one defined in \cite{DR05} and the difference of our $r^*$ and that defined in \cite{DR05} is 
$3m +2m\ln m$.\\
The bounds concerning the constants $C_1$ and $C_3$ are proved in Theorem  1.1
of \cite{DR05}. Here, sufficiently regular solutions of the massless Klein-Gordon equation are considered
and initial data are assigned on a smooth complete spacelike Cauchy surfaces of the full Kruskal extension of
$\mM$ which is asymptotically flat at spatial infinity. Furthermore it is imperative that the said data 
vanish fast enough at space infinity. In our case these requirements are 
fulfilled because the elements of $\sS(\mM)$ are smooth and have compact support on every smooth 
spacelike, hence acausal, Cauchy surface of $\mM$; therefore we can employ the results in \cite{BS06} to view 
these as Cauchy data on a smooth spacelike Cauchy surface of the full Kruskal extension. 
The bound which concerns $C_2$ has the same proof as that for $C_1$ because, when $\vphi\in\sS(\mM)$, 
$X(\vphi)\in\sS(\mM)$, $X$ itself being a smooth Killing vector field. To conclude the proof, it is enough 
to show the last bound, related with the constant $C_4$. To this end, let us fix $\vphi \in \sS(\mM)$ and 
re-define, if necessary, the origin of the killing time $t$ in $\mW$ in order that $u^{(\vphi)} \geq 2$, 
where $u^{(\vphi)}$ is the constant defined in Lemma \ref{lemma1}. Now 
 we focus  on the proof contained in sec. 13.2 of \cite{DR05} and particularly on the part called ``{\em 
 decay in $r\geq \hat{R}$''} which concerns the bound associated with $C_3$.  We want to 
adapt such proof to our case, replacing the solution $\phi$ there considered with our 
$X(\vphi)$, so that also $r\phi$ is replaced by  $rX(\vphi) = \widetilde{X}(r\vphi)$ in $\mW$. Furthermore it
smoothly extends to $X(\widetilde{\vphi})$ on $\mW \cup \Im^+$. It is remarkable that it suffices to prove 
the bound in the region $\{r>\hat R\}\cap  \{t>0\}$ in $\mW$, since it would then hold on $\Im^+$ per 
continuity.\\
One should notice that only the region $\{r \geq \hat R\}  \cap \{t>0\} \cap \{ u \geq 2\}$ has to be 
considered. Indeed, in the set $\{ u < 2\} \cap \{v > v^{(\vphi)}_0\}$ for some $v^{(\vphi)}_0 \in \bR$, $X(
\widetilde{\vphi})$ vanishes due to Lemma \ref{lemma1}. Hence $X(\widetilde{\vphi})$ vanishes in $\{r\geq\hat
R\}  \cap \{t>0\}\cap \{ u < 2\} \cap \{v > v^{(\vphi)}_0\}$, trivially satisfying the wanted bound.
The region individuated by $\{r \geq \hat R\}  \cap \{t\geq 0\} \cap  \{ u \leq 2\}
\cap \{v\leq v^{(\vphi)}_0\}$ is, moreover, compact thus $X(\widetilde{\vphi})$ is bounded 
therein and it also satisfies the looked-for bound.\\
In the region  $\{r \geq \hat R\}  \cap \{t>0\} \cap \{ u \geq 2\}$, along the lines of p.916-917 in
\cite{DR05}, though with $\phi$ replaced by $\varphi' \doteq {\widetilde X}(\vphi)$,  we achieve, out of a 
Sobolev inequality on the sphere
 $$r^2 \left|\varphi'(u,v,\theta,\phi)\right|^2 \leq C \int_{\bS^2} r^2  \left|\varphi'\right|^2
d\bS^2 + C \int_{\bS^2} \left|r \spa \not{\nabla} \varphi'\right|^2  r^2 d\bS^2 + 
 C \int_{\bS^2}\left|r^2\not{\nabla} \spa \not{\nabla} \varphi'|^2\right) r^2 d\bS^2\:,$$
where $\not{\nabla}$ denotes the covariant derivative with respect to metric induced on the sphere of radius 
$r$, while $d\bS^2$ is the volume form on the unit sphere. If the squared angular momentum operator is 
denoted as $\Omega^2\doteq r^2\not{\nabla}\spa\not{\nabla}$ the above inequality can be re-written as:
\beq \label{inter}
r^2 \left|\varphi'(u,v,\theta,\phi)\right|^2 \leq C \int_{\bS^2}   \left|\Omega^0 \varphi'\right|^2
r^2 d\bS^2+ C \int_{\bS^2}
  \left|\Omega^0 \varphi'\right| \left| \Omega^1 \varphi'\right|  r^2 d\bS^2 + 
   C \int_{\bS^2}\left|  \Omega^2 \varphi'\right|^2 r^2 d\bS^2\:.\eeq
To conclude it is sufficient to prove that, for $k=0,1,2$ and if $r\geq \hat R$, $u\geq 2$, $t>0$:
\beq\label{middle}
 \int_{\bS^2}   \left|\Omega^k \varphi'\right|^2
r^2 d\bS^2 \leq B_k/u^2,
\eeq
for some constants $B_k \geq 0$. Let us notice that, in view of Cauchy-Schwartz inequality, the second 
integral in the right hand side of (\ref{inter}) is bounded by the product of the square root of the 
integrals with $k=0$ and $k=1$ in the left-hand side of (\ref{middle}). If we follow \cite{DR05} and if we 
pass to the coordinates 
$(t,r^*, \theta,\phi)$ (see Section \ref{geometry}), and for some constant $D\geq 0$:
\beq\label{inter2}
\int_{\bS^2}   \left|\Omega^k \varphi'\right|^2
r^2(t,r^*,\theta,\phi) d\bS^2\leq \int_{\bS^2}   \left|\Omega^k \varphi'\right|^2
r^2(t,\tilde{r}^*,\theta,\phi) d\bS^2 \nonumber \\
+ D \int_{\tilde{r}^*}^{r^*}\int_{\bS^2}
|\partial_\rho \Omega^k \varphi'||\Omega^k\varphi'| r^2(t,\rho,\theta,\phi) d\rho d\bS^2
+ D \int_{\tilde{r}^*}^{r^*}\int_{\bS^2}  |\Omega^k \varphi'|^2 r(t,\rho,\theta,\phi) d\rho d\bS^2\:.
\eeq
If we stick to \cite{DR05}, the parameter $\tilde{r}^*\geq \hat{R}^*$ can be fixed so that the first integral
in the right-hand side satisfies
\beq\label{bastarda0}
\int_{\bS^2}   \left|\Omega^k \varphi'\right|^2
r^2(t,\tilde{r}^*,\theta,\phi) d\bS^2 \leq \bar{E}_2/t^{2} \leq \bar{E}_2/ u^2,
\eeq
where the constant $\bar{E}_2$ was defined in \cite{DR05} and it depends 
on $\Omega^k \varphi'\in \sS(\mM)$. Here we have also used the fact that
$u = t - r^* \geq 2$ with $r^*>0$ and $t>0$; hence that $u \leq t$.
>From now on, our procedure departs form that followed in \cite{DR05}. With respect to the third
integral in the right-hand side of (\ref{inter2}), it can be re-written
$$
\int_{\tilde{r}^*}^{r^*}\int_{\bS^2} |\partial_t\Omega^k \vphi|^2 r(t,\rho,\theta,\phi) d\rho d\bS^2
\leq const. \int_{\tilde{r}^*}^{r^*}\int_{\bS^2} |\partial_t\Omega^k \vphi|^2 r^2(t,\rho,\theta,\phi) d\rho 
d\bS^2\leq F(\cS),
$$
where $\cS$ is the achronal hypersurface individuated by the fixed time $t$, the interval  $[\tilde{r}^*, 
r^*]$ and the coordinates $(\theta,\phi)$ which vary over $\bS^2$. $F(\cS)$ is then the flux of energy through $\cS$ associated with the Klein-Gordon 
field $\Omega^k \vphi$. Theorem 1.1 in \cite{DR05} assures now that, for some constant $C'$, which depends on 
$\vphi$,
$$
F(\cS) \leq C'/v_+(\cS)^2 + C'/u_+(\cS)^2\:,
$$ 
where $v_+(\cS) = \max\{\inf_\cS v, 2\}$ and $u_+(\cS) = \max\{\inf_\cS u, 2\}$.
In our case, per construction, we have
$\max\{\inf_\cS v, 2\}\geq t+\hat{R}^*$ and $\max\{\inf_\cS u, 2\} = u(t,r^*,\phi,\theta)$.
For $t>0$, $r\geq \hat R$, $u>2$, one can conclude:
$$
 \int_{\tilde{r}^*}^{r^*}\int_{\bS^2} |\Omega^k \varphi'|^2 r(t,\rho,\theta,\phi) d\rho d\bS^2 \leq 
 \frac{C'}{(t + \hat{R}^*)^2} + \frac{C'}{u^2} =
  \frac{C'}{(u + r^*+ \hat{R}^*)^2} + \frac{C'}{u^2} \leq 
  \frac{2C'}{u^2}\:,
$$ 
and thus
\beq\label{bastarda1}
 \int_{\tilde{r}^*}^{r^*}\int_{\bS^2} |\Omega^k \varphi'|^2 r(t,\rho,\theta,\phi) d\rho d\bS^2 \leq
 const. \int_{\tilde{r}^*}^{r^*}\int_{\bS^2} |\Omega^k \varphi'|^2 r(t,\rho,\theta,\phi)^2 d\rho d\bS^2 \leq  
  \frac{K}{u^2}.
\eeq
Let us finally consider the second integral in the right-hand side of (\ref{inter2}). We notice
that
$$
\int_{\tilde{r}^*}^{r^*}\int_{\bS^2} |\partial_\rho \Omega^k \varphi'|^2 r(t,\rho,\theta,\phi)^2 d\rho d\bS^2
\leq F'(\cS),
$$
where $F'(\cS)$ is the flux of energy through $\cS$ associated with the Klein-Gordon 
field $\Omega^k \varphi'$. If we deal with it as before, we obtain the bound
\beq\label{bastarda2}
 \int_{\tilde{r}^*}^{r^*}\int_{\bS^2} |\partial_\rho \Omega^k \varphi'|^2 r(t,\rho,\theta,\phi)^2 d\rho 
 d\bS^2 \leq  
  \frac{K'}{u^2}\:.
\eeq
The Cauchy-Schwartz inequality, together with (\ref{bastarda1}) and (\ref{bastarda2}), leads to
\beq\label{bastarda3}
\int_{\tilde{r}^*}^{r^*}\int_{\bS^2}
|\partial_\rho \Omega^k \varphi'||\Omega^k\varphi'| r(t,\rho,\theta,\phi)^2 d\rho d\bS^2 \leq  
\frac{K''}{u^2}\:.
\eeq
If one puts all together in the right-hand side of (\ref{inter2}), the bounds (\ref{bastarda0}), 
(\ref{bastarda1}) and (\ref{bastarda3}) yield  (\ref{middle}). \\
(b) Let us fix $\Sigma$ as any smooth spacelike Cauchy surface of $\mM$. Notice that if the sequence of 
initial data converge to zero in the test function topology 
on $\Sigma$, there is a compact set $C\subset\Sigma$ which, per definition, contains all the supports of the 
initial data of the sequence.
 In view of \cite{BS06}, we can construct  a smooth spacelike Cauchy surface 
$\Sigma'$ of the complete Kruskal manifold $\mK$, which includes that compact. Thus, the sequence of initial
data tends to $0$
in the test function topology of $\Sigma'$  as well.  Such data on $\Sigma$ can be read on $\Sigma'$
since the supports of the solutions cannot further intersect $\Sigma'$ as it is acausal.
 From standard results of continuous dependence from compactly-supported initial data of the smooth solutions
of hyperbolic equations in globally hyperbolic spacetimes (see Theorem 3.2.12 in \cite{BGP}), if the initial
data on a fixed spacelike Cauchy surface $\Sigma'$ tend to $0$ in the test function topology, then also the 
solution tends to $0$ in the topology of $C^\infty(\mK; \bR)$.
At the same time, as one can prove out of standard results on the topology of causal sets ({\it e.g.}, see 
\cite{Wald} and particularly theorems 8.3.11 and 8.3.12 in combination with the fact that the open double 
cones form a base of the topology) $J^+(C; \mK) \cup J^-(C; \mK)$ has compact intersection 
with every spacelike Cauchy surface of $\mK$, since $C$ is compact in $\Sigma'$. So all initial data on 
$\Sigma''$ of the considered sequence of solutions are contained in a compact, too.
 From these results we conclude that, if the initial data tend to $0$ in the test function topology on 
$\Sigma'$, the associated solution, whenever restricted on any other Cauchy surface $\Sigma''\subset\mK$
yields, per restriction, new initial data, which also tend to $0$.
For convenience, we fix $\Sigma''$ as an extension of the spacelike Cauchy surface of $\mW$ 
(whose closure intersects $\cB$)  individuated in $\mW$ as the locus 
$t=1$.  If we refer to (a), one sees that the coefficients $C_i$ are obtained as the product of universal 
constants and  integrals of derivatives of the compactly supported 
Cauchy data of both $\vphi$, and, where appropriate, $X(\vphi)$ over $\Sigma'' \cap \overline{\mW}$.
This is explained in Theorem 1.1, Theorem 7.1 as well as in the formulae appearing in sec. 4 of \cite{DR05}, 
though one should reformulate them with respect both to $r^*$ and to the global coordinates $U$ and $V$ 
instead of $u$, $v$ and $r$). From these formulas it follows immediately that the constants $C_i$
vanish provided that the Cauchy data tend to $0$ in the test function topology on $\Sigma''$, and this 
requirement is valid in our hypotheses. $\Box$\\


\noindent {\bf Proof of Proposition \ref{PropMain4}}. 
(a) To start with, we notice that, by direct inspection, as shown in \cite{KW} and \cite{Moretti08},
though the angular coordinates $(\theta,\phi)$ substituted by the complex ones $(z,\bar{z})$ obtained out of 
stereographic projection, it turns out that:
\begin{gather*}
\langle \widehat{\psi}_+, \widehat{\psi'}_+ \rangle_{\sH_{\cH}} \doteq
\int_{\bR_+\times \bS^2} \overline{\widehat{\psi}_+}(K, \theta,\phi) \widehat{\psi'}_+(K,\theta,\phi)\: 
2K dK \wedge r^2_S d\bS^2\nonumber \\
= \lim_{\epsilon \to 0^+} -\frac{r_S^2}{\pi}
\int_{\bR \times \bR \times \bS^2}\spa \frac{\overline{\psi(U_1,\theta,\phi)} \psi'(U_2,\omega)}
{(U_1-U_2 -i\epsilon)^2}
dU_1 \wedge dU_2 \wedge d\bS^2,
\end{gather*}
for $\psi,\psi' \in C_0^\infty(\cH; \bC)$.\\
As a consequence we have obtained that the map 
$M: C_0^\infty(\cH; \bC) \ni \psi \mapsto \widehat{\psi}_+(K,\omega) \in  \sH_{\cH}$
is isometric and thus, per continuity, it uniquely extends to  a Hilbert space isomorphism $F_{(U)}$ of 
$\overline{\left(C_0^\infty(\cH;\bC), \lambda_{KW} \right)}$ onto the closed Hilbert space
 $\overline{M(C_0^\infty(\cH; \bC))} \subset \sH_{\cH}$. To conclude the proof of the first statement in (a), 
 it is enough to establish
that $\overline{M(C_0^\infty(\cH; \bC))} = \sH_{\cH}$. 
This immediately follows from the two lemma proved below. 

\lemma \label{Lemma1TeoMain4} {\em $\overline{M(C_0^\infty(\cH; \bC))}$
includes the space $\mS_0$ whose elements $f = f(K,\omega)$ are the restrictions to $\bR_+\times \bS^2$
 of the  functions in $\mS(\bR\times \bS^2)$  
and they vanish in a neighbourhood of $K=0$ depending on $f$.}\\
 
 \noindent Above and henceforth $\mS(\bR\times \bS^2)$ denotes  the  complex {\bf Schwartz space}
on $\bR \times \bS^2$, {\em i.e.} the space of complex-valued smooth functions on $\bR\times \bS^2$ which 
vanish, with all their $K$-derivatives of every order,  as $|K|\to +\infty$ 
uniformly in the angles and faster than every inverse power of $|K|$. This space can be equipped with the 
usual topology induced by seminorms (see Appendix C of \cite{Moretti08}).

\lemma \label{Lemma2TeoMain4}  {\em  $\mS_0$ is dense in  $\sH_{\cH}$.}\\

\noindent Concerning (b), we notice that, if $f\in \mS_0$,  $if \in \mS_0$ and that
both  vanish in a neighbourhood of 
$K=0$. Therefore, it is possible to arrange  
two {\em real} functions in $\mS(\cH)$, $g_1$ and $g_2$ such that $\widehat{g_1}_+ = f$ and $\widehat{g_2}_+ = if$.
With the same proof of Lemma \ref{Lemma1TeoMain4} one can establish that $g_i$ are the the limits, in the 
topology of $\lambda_{KW}$, of sequences $\{f_{(i) n}\} \subset C_0^\infty(\cH; \bR)$. We have obtained that
every complex element of the dense subspace $\mS_0 \subset \sH_{\cH}$ is the limit of elements of 
$F_{(U)}\left(C_0^\infty(\cH; \bR)\right)$.
$\Box$\\

 \noindent {\bf Proof of Lemma \ref{Lemma1TeoMain4}}. 
Let us take $f \in \mS_0$. As a consequence, it can be written as the restriction to $\bR_+\times \bS^2$ of 
$F \in \mS(\bR\times \bS^2)$. In turn, $F= \mF_+(g)$ for some $g \in  \mS(\bR\times \bS^2)$, since the 
Fourier transform
is bijective from   $\mS(\bR\times \bS^2)$ onto $\mS(\bR\times \bS^2)$
(see Appendix C of \cite{Moretti08}). Since $C_0^\infty(\bR \times \bS^2; \bC)$
is dense in $\mS(\bR\times \bS^2)$ in the topology of the latter, there is a sequence $\{g_n\} \subset
C_0^\infty(\bR \times \bS^2;\bC)$ with $g_n \to g$ in the sense of ${\mS(\bR\times \bS^2)}$. Since the 
Fourier transform is continuous with respect to that topology, we conclude that
 $\mF_+(g_n) \to F$ in the sense of ${\mS(\bR\times \bS^2)}$. By direct inspection one finds that the achieved result
 implies that $\mF_+(g_n)\rest_{\bR_+\times \bS^2} \to F\rest_{\bR_+\times \bS^2}$ in the topology of every 
 $L^2(\bR_+\times \bS^2, c K^n dK \wedge r_S^2 d\bS^2)$ for every power $n=0,1,2,\ldots$ and $c>0$. 
Particularly it happens
 for $n=c =2$. We have found that, for every $f \in \mS_0$, there is a sequence in $M(C_0^\infty(\bR \times \bS^2;\bC))$
 which tends to $f$ in the topology of $\sH_{\cH_R}$ and thus $\mS_0 \subset \overline{M(C_0^\infty(\bR \times \bS^2;\bC))}$.
 $\Box$\\
 
\noindent {\bf Proof of Lemma \ref{Lemma2TeoMain4}.} In this proof $\bR^*_+ \doteq (0,+\infty)$ and 
$\bN^* = \{1,2,\ldots\}$. A well-known result is that
$C_0^\infty((a,b); \bC)$ is dense in $L^2((a,b), dx)$ so that, particularly, $C_0^\infty((1/n,n); \bC)$ is 
dense in $L^2((1/n,n),dx)$, and, thus, if we introduce the new variable $K = \sqrt{x}$, it turns out that the
space $C^\infty_0((1/\sqrt{n},\sqrt{n}); \bC)$ is dense in $L^2((1/\sqrt{n},\sqrt{n}),2KdK)$.\\ Since, in the 
sense of the Hilbertian direct sum, $\oplus_{n\in \bN} L^2((1/\sqrt{n},\sqrt{n}),2KdK) = L^2(\bR^*_+, 2KdK)$ 
(for instance making use of Lebesgue's dominated 
 convergence theorem), we conclude that $C_0^\infty(\bR_+^*; \bC) = \cup_{n\in \bN^*}C_0^\infty((1/n,n);\bC)$
is dense in  $L^2(\bR^*_+, 2KdK)= L^2(\bR_+, 2KdK)$ and, thus, 
there must exist a Hilbert base $\{f_n\}_{n\in \bN} \subset C_0^\infty(\bR_+^*; \bC)$.    \\
By standard theorems on Hilbert spaces with product measure, we know that 
a Hilbert base of the space $L^2(\bR_+ \times \bS^2, 2KdK\wedge r^2_Sd\bS^2)$ is
$\{f_{n} Y_{m}\}_{n,m \in \bN}$, provided that $\{Y_m\}_{m\in \bN}$ and $\{f_n\}_{n\in\bN}$ are respectively
one for $L^2(\bS^2,r^2_Sd\bS^2)$ and for $L^2(\bR_+, 2KdK)$. The elements $Y_m$ can be chosen as harmonic 
functions so that they are smooth and compactly supported. Therefore, if  $\{f_n\}_{n\in \bN}\subset C_0^
\infty((0,+\infty); \bC)$, it holds that  $\{f_{n} Y_{m}\}_{n,m \in \bN} \subset 
C_0^\infty(\bR^*_+\times \bS^2; \bC)$ and, thus, trivially, the space $C_0^\infty(\bR^*_+\times \bS^2;\bC)$ 
is dense in $L^2(\bR_+ \times \bS^2, 2KdK\wedge r^2_Sd\bS^2)$. Since it holds $C_0^\infty(\bR^*_+\times\bS^2;
\bC)\subset\mS_0$, the achieved result proves the thesis. $\Box$\\


\noindent{\bf Proof of Proposition \ref{propbastarda}.} We only consider the case of $\cH^+$, the proof for 
$\cH^-$ being identical.\\
(a) If $\psi_1, \psi_2 \in C_0^\infty(\cH^+; \bC)$, then:
\beq\label{lKWs} 
\lambda_{KW}(\psi_1,\psi_2) =
\lim_{\epsilon \to 0^+} -\frac{1}{4 \pi}
\int_{\bR \times \bR \times \bS^2}\spa \frac{\overline{\psi_1(u_1,\theta,\phi)} \psi_2(u_2,\theta,\phi)}{
\left[\sinh\left(\frac{u_1-u_2}{4r_S}\right) -i\epsilon\right]^2}
du_1 \wedge du_2 \wedge d\bS^2\:.
\eeq 
That identity follows from the expression of $\lambda_{KW}$ given in (\ref{lKW}),
passing to the coordinates $u_1,u_2$ and making an appropriate use of
Sokhotskys formula $1/(x-i0^+)^2 = 1/x^2 - i \delta'(x)$ (where $1/x^2$ is the 
derivative of the distribution $-1/x$ interpreted in the sense of the principal value). 
Actually, passing to coordinates $u_1, u_2$ from the
inital ones $U_1, U_2$, a bounded strictly positive factor arises in front
of $\epsilon$, but it can safely be replaced by the costant $1$, as it can be
easily proved (especially taking into account that the used test
functions have compact support).  \\
In spite of the different 
relation between the coordinate $U$ and $u$, the same result arises referring to $\cH^-$ instead
of $\cH^+$. We notice that the $u$-Fourier 
transform of the distribution $-\frac{1}{4\pi}\frac{1}{\left[\sinh\left(\frac{u}{4r_S}\right)-i0^+\right]^2}$
turns out to be just $\frac{1}{\sqrt{2\pi}}\frac{d\mu(k)}{dk}$. Hence, the limit as $\epsilon \to 0^+$ of the 
integral in the right-hand side of (\ref{lKWs})
can be interpreted as the $L^2(\bR\times \bS^2, dv\wedge d\bS^2)$ scalar product of $\psi_1$ and the the 
$L^2(\bR\times \bS^2, dk\wedge d\bS^2)$  function obtained by the $u$-convolution of the Schwartz 
distribution  $const. /\left[\sinh\left(\frac{u}{4r_S}\right) -i0^+\right]$
with the compactly-supported function $\psi_2$. The convolution makes sense if one interprets $\psi_2$ as a 
distribution with compact support; it produces a distribution which is the antitransform of
$\widetilde{\psi_2} d\mu/dk$ which, in turn, belongs to the Schwartz space by construction. Hence, up to an
antitransformation, the said convolution has to be an element of $L^2(\bR\times \bS^2, du\wedge d\bS^2)$ as 
previously stated. In this sense we can apply first the convolution theorem for Fourier transforms and, 
afterwards, the fact that the Fourier transform is an isometry, achieving: 
$$ \lim_{\epsilon \to 0^+} -\frac{1}{4 \pi}
\int_{\bR \times \bR \times \bS^2}\spa \frac{\overline{\psi_1(u_1,\theta,\phi)} \psi_2(u_2,\theta,\phi)}{
\left[\sinh\left(\frac{u_1-u_2}{4r_S}\right) -i\epsilon\right]^2}
du_1 \wedge du_2\wedge d\bS^2 = \int_{\bR\times \bS^2} \overline{\widetilde{\psi}(k,\theta,\phi)}
 \widetilde{\psi}(k,\theta,\phi) \frac{d\mu}{dk} dk\wedge d\bS^2 \:,$$
which implies that the map $C_0^\infty(\cH^+;\bC) \ni \psi \mapsto \widetilde{\psi} \in L^2(\bR\times \bS^2, 
d\mu(k)\wedge d\bS^2)$ is isometric, when the domain is equipped with the scalar product $\lambda_{KW}$.
The fact that this map extends to a Hilbert space isomorphism $F^{(+)}_{(u)}:\overline{C_0^\infty(\cH^+;\bC)}
\to L^2(\bR\times \bS^2, d\mu(k)\wedge d\bS^2)$ is very similar to the proof of the analogue for $F_{(U)}$
and the details are left to the reader.\\
(b) Let us indicate by $\widetilde{\psi}\equiv\mF(\psi)$ the Fourier-Plancherel transform of $\psi$, computed
with respect to the coordinate $u$. Per definition, if $\psi \in \sS(\cH^+)$, one has $\psi,
\partial_u\psi \in L^2(\bR\times \bS^2, du\wedge  d\bS^2)$, so that $\psi$ belongs to the Sobolev space $H^1(\cH^+)_u$ and,
equivalently,
$\widetilde{\psi} \in L^2(\bR\times \bS^2, dk\wedge  d\bS^2) \cap L^2(\bR\times \bS^2, k^2dk\wedge  d\bS^2)$.
The last inclusions also implies that $\widetilde{\psi}$ belongs to $L^2(\bR\times \bS^1,|k|dk\wedge d\bS^2)$
and  $L^2(\bR\times \bS^1, d\mu\wedge  d\bS^2)$. 
Since  $C_0^\infty(\cH^+; \bC)$ is dense in $H^1(\cH^+)_u$, if $\psi \in \sS(\cH^+)$, there is a sequence of functions 
$\psi_n \in C_0^\infty(\cH^+; \bR)$ with $F^{(+)}_{(u)}(\psi_n) = \mF(\psi_n) \to \widetilde{\psi}$, in the 
topology of both
$L^2(\bR\times \bS^2, dk\wedge  d\bS^2)$ and $L^2(\bR\times \bS^2, k^2dk\wedge d\bS^2)$. In turn this implies 
the convergence  in the topology of $L^2(\bR\times \bS^2, d\mu\wedge  d\bS^2)$. Since $L^2(\bR\times \bS^1,
d\mu\wedge d\bS^2)$ is isometric  to $\overline{C_0^\infty(\cH^+; \bC)}$, the sequence $\{\psi_n\}$ is of 
Cauchy type in $\overline{(C_0^\infty(\cH; \bC),\lambda_{KW})}$. For the same reason, 
any other $\{\psi'_n\} \in C_0^\infty(\cH^+; \bR)$ which converges 
to the same $\psi$, is such that $\psi_n -\psi'_n \to 0$ in $\overline{(C_0^\infty(\cH; \bC),\lambda_{KW})}$.
Therefore $\psi$ is naturally identified with an element of 
$\overline{C_0^\infty(\cH^+; \bC)}$, which we shall denote with the same symbol  $\psi$. With this 
identification, for $\psi \in \sS(\cH^+)$,
the fact that  $F^{(+)}_{(u)}(\psi_n) \to \widetilde{\psi} = \mF(\psi)$ in the topology of 
$L^2(\bR\times \bS^2, kdk\wedge  d\bS^2)$ implies that $F^{(+)}_{(u)}(\psi) = \mF(\psi)$ by continuity of $F_{(u)}$.
$\Box$\\

\noindent {\bf Proof of Proposition \ref{Propembedding}.}
The  map  $\sK_{\cH}$ is per construction linear. Let us prove that (a) is valid, {\it i.e.}, 
$\sK_{\cH}$ does not depend on the particular decomposition 
(\ref{dec0}) for a fixed $\psi \in \sS(\cH)$. Consider a different analogous decomposition
$\psi = \psi'_- + \psi'_0 + \psi'_+$. We have that the two definitions of $\sK_{\cH}\psi$ coincides because 
their difference is:
\begin{align*}
& F_{(U)}(\psi_-) - F_{(U)}(\psi'_-) + F_{(U)}(\psi_0) -F_{(U)}(\psi'_0)+  F_{(U)}(\psi_+)- F_{(U)}(\psi'_+)\nonumber \\
&= F_{(U)}(\psi_-- \psi'_-) + F_{(U)}(\psi_0 -\psi'_0)+  F_{(U)}(\psi_+- \psi'_+)
= \widehat{\psi_-- \psi'_-} + \widehat{\psi_0-\psi'_0} + \widehat{\psi_+- \psi'_+}\nonumber\\
& = \widehat{\psi-\psi} =0 \:,
\end{align*}
Here we have used the fact that, per construction, 
 $\psi_\pm -\psi'_\pm$ and $\psi_0-\psi_0$ belongs to $C_0^\infty(\cH; \bR)$
 and thus $F_{(U)}$, 
acting on each of them, produces the standard $U$-Fourier transform indicated by 
$\widehat{\cdot}$.\\
(b) The statement is valid per definition of $\sK_{\cH}$. Let us thus prove (c).
>From now on we write $\sigma$ instead of $\sigma_{\cH}$.
Let us take $\psi,\psi' \in \sS(\cH)$ and decompose them as 
$\psi= \psi_1+\psi_2+\psi_3$ and as $\psi'= \psi'_1+\psi'_2+\psi'_3$
 where $\psi_1,\psi'_1 \in \sS(\cH^+)$, $\psi_2,\psi'_2 \in C_0^\infty(\cH; \bR)$ and
$\psi_3,\psi'_3 \in \sS(\cH_R^-)$. In this way we have:
\begin{gather*}\sigma(\psi,\psi') = \:\sigma(\psi_1,\psi'_1)+\sigma(\psi_2,\psi'_2)+\sigma(\psi_3,\psi'_3) +
\sigma(\psi_1,\psi'_2)+\sigma(\psi_1,\psi'_3)\nonumber\\  +  
\sigma(\psi_2,\psi'_1)+\sigma(\psi_2,\psi'_3) + \sigma(\psi_3,\psi'_1)+\sigma(\psi_3,\psi'_2)\:.\end{gather*}
Let us examine each term separately. Consider $\sigma(\psi_1,\psi'_1)$. 
>From now on $\widetilde{\psi}\equiv \mF(\psi)$ is the Fourier-Plancherel transform of $\psi$, computed with 
respect to the coordinate $u$.
Notice that  $\widetilde{\psi}_1(-k,\theta,\phi) = \overline{\widetilde{\psi}_1(k, \theta,\phi)}$
since $\psi_1$ and $\psi'_1$ are real. By direct inspection, if one uses these ingredients and the 
definition of $d\mu(k)$, one gets immediately the first identity:
 \begin{gather} \sigma(\psi_1,\psi'_1) = 
 - 2 Im \langle \widetilde{\psi}_1, \widetilde{\psi'}_1\rangle_{L^2(\bR\times \bS^1, d\mu\wedge d\bS^2)}=
 \nonumber \\ 
-= 2 Im \langle F_{(U)}\circ (F^{(+)}_{(u)})^{-1}(\widetilde{\psi_1}),  F_{(u)}\circ (F^{(+)}_{(u)})^{-1}(\widetilde{\psi'_1})
\rangle_{L^2(\bR_+\times \bS^1, dK\wedge d\bS^2)}
=- 2 Im \langle \sK_{\cH} \psi_1, \sK_{\cH} \psi'_1\rangle_{\sH_{\cH}}\:, \label{p1}
\end{gather}
The second identity arises form the fact that $ F_{(U)}\circ (F^{(+)}_{(u)})^{-1}$ is an isometry as follows
from (b) in Proposition \ref{propbastarda} and (a) in Proposition \ref{PropMain4}.
 The last identity is nothing but the definition
of $\sK_{\cH}$. With the same procedure we similarly have
$$
 \sigma(\psi_3,\psi'_3) = - 2 Im \langle \sK_{\cH} \psi_1, \sK_{\cH} \psi'_1\rangle_{\sH_{\cH}}\:. 
$$
If we refer to $\sigma(\psi_2,\psi'_2)$, we can employ the coordinate $U$ taking into 
account that the support of those smooth functions is compact when referred to the coordinates $(U,\theta,
\phi)$ over $\cH$.\\ Hence, $\psi_2,\psi'_2, \partial_U\psi_2,\partial_U\psi'_2 \in L^2(\bR \times \bS^2, 
dU \wedge d\bS^2)$ so that, at a level of $U$-Fourier transforms, it holds
$\widehat{\psi}_{2+},\widehat{\psi'}_{2+} \in L^2(\bR_+ \times \bS^2, dK \wedge d\bS^2)\cap 
L^2(\bR_+ \times \bS^2, K dK \wedge d\bS^2)$. Finally, in the considered case, directly by the definition,
$\sK_{\cH} \psi'_2 = \widehat{\psi'}_{2+}$ and $\sK_{\cH} \psi_2 = \widehat{\psi}_{2+}$.
If one uses the fact that  $\widehat{\psi}_2(-K,\theta,\phi)=\overline{\widehat{\psi}_2(K, \theta,\phi)}$
since $\psi_2$ and $\psi'_2$ are real, one straightforwardly achieves the first identity:
\begin{gather*} \sigma(\psi_2,\psi'_2) = 
- 2 Im \langle \widehat{\psi}_{2+}, \widehat{\psi'}_{2+} \rangle_{L^2(\bR_+\times\bS^2, 2K dK\wedge d\bS^2)} 
= \nonumber \\
= - 2 Im \langle  F_{(U)}{\psi}_{2+}, F_{(U)}{\psi'}_{2+} \rangle_{L^2(\bR_+\times\bS^2, 2K dK\wedge d\bS^2)}
= - 2 Im \langle \sK_{\cH} \psi_2, \sK_{\cH} \psi'_2\rangle_{\sH_{\cH}}\:, 
 \end{gather*}
The remaining identities follow from the definition of $F_{(U)}$ and $\sK_{\cH}$. As a further step we 
notice that
$$
\sigma(\psi_1,\psi'_3) = 0 = 
- 2 Im \langle \sK_{\cH} \psi_1, \sK_{\cH} \psi'_3\rangle_{\sH_{\cH}}\quad \sigma(\psi_3,\psi'_1) = 0 = - 2
Im \langle \sK_{\cH} \psi_3, \sK_{\cH} \psi'_1\rangle_{\sH_{\cH}}\:. 
$$
Let us focus on the first identity, the second being analogous; it holds true
because the functions have disjoint supports, whereas  $\langle \sK_{\cH} \psi_3, \sK_{\cH} \psi'_1\rangle_{\sH_{\cH}}=0$
since, per direct application of (b) in Proposition \ref{propbastarda},
$\psi_1 \in \sS(\cH^+)$ is the limit of a sequence of real smooth functions $f^{(1)}_n$ with support in 
$\cH^+$ whereas $\psi'_3 \in \sS(\cH^+)$ is the limit of a sequence of real smooth functions $f^{(3)}_n$
with support in $\cH^-$. Hence
\begin{align}&Im \langle \sK_{\cH} f^{(1)}_n, \sK_{\cH} f^{(2)}_m\rangle_{\sH_{\cH}}
= Im \lambda_{KW}(f^{(1)}_m, f^{(2)}_n) \nonumber \\ &= 
 -\frac{r_S^2}{\pi}
Im \int_{\bR \times \bR \times \bS^2}\spa \frac{{f^{(1)}_n(U_1,\theta,\phi)} f^{(2)}_m(U_2,\theta,\phi)}{(U_1-U_2 -i0^+)^2}
dU_1 \wedge dU_2 \wedge d\bS^2(\theta,\phi)\nonumber \\&= 
-\frac{r_S^2}{\pi}
Im \int_{\bR \times \bR \times \bS^2}\spa \frac{\partial_{U_1}{f^{(1)}_n(U_1,\theta,\phi)} f^{(2)}_m(U_2,\theta,\phi)}{U_1-U_2 -i0^+}
dU_1 \wedge dU_2 \wedge d\bS^2(\theta,\phi)\nonumber\\
&= -r_S^2
 \int_{\bR \times \bR \times \bS^2}\spa \partial_{U_1}{f^{(1)}_n(U_1,\theta,\phi)} f^{(2)}_m(U_2,\theta,\phi) \delta(U_1 -U_2)
dU_1 \wedge dU_2 \wedge d\bS^2(\theta,\phi) =0, \nonumber
\end{align}
since $f^{(1)}_n$ and $f^{(2)}_m$ have disjoint support. Let us examine the term $\sigma(\psi_1,\psi'_2)$: 
in this case we decompose $\psi_1 = f_1 + g_1$ where $f_1 \in C_0^\infty(\cH^+; \bR)$ and $g_1 \in \sS(\cH^+)$, but
$\supp(g_1) \cap \supp(\psi'_2) = \emptyset$. We have:
$$
\sigma(\psi_1,\psi'_2) = \sigma(f_1,\psi'_2) + \sigma(g_1,\psi'_2)\:.
$$
At the end of this proof we shall also prove that:
\beq
\sigma(\psi_1,\psi'_2) = 0 = - 2 Im \langle \sK_{\cH} \psi_1, \sK_{\cH} \psi'_2\rangle_{\sH_{\cH}}
\label{laaast}\:.\eeq
Conversely
$\sigma(f_1,\psi'_2) = - 2 Im \langle \sK_{\cH} f_1, \sK_{\cH} \psi'_2\rangle_{\sH_{\cH}}$, exactly as in the 
case $\sigma(\psi_2,\psi'_2)$ examined above. If we sum up, per $\bR$-linearity:
$$
\sigma(\psi_1,\psi'_2) = - 2 Im \langle \sK_{\cH} \psi_1, \sK_{\cH} \psi'_2\rangle_{\sH_{\cH}}\:.
$$
With an analogous procedure we also achieve:
$$
\sigma(\psi_2,\psi'_1) = - 2 Im \langle \sK_{\cH} \psi_1, \sK_{\cH} \psi'_2\rangle_{\sH_{\cH}}
\quad \sigma(\psi_2,\psi'_3) = - 2 Im \langle \sK_{\cH} \psi_2, \sK_{\cH} \psi'_3\rangle_{\sH_{\cH}},
$$
and
\beq
 \sigma(\psi_3,\psi'_2) = - 2 Im \langle \sK_{\cH} \psi_3, \sK_{\cH} \psi'_2\rangle_{\sH_{\cH}}
\label{p7}\:.
\eeq
The identities (\ref{p1})-(\ref{p7}), per $\bR$-linearity, yield the thesis:
\beq
 \sigma(\psi,\psi') = - 2 Im \langle \sK_{\cH} \psi, \sK_{\cH} \psi'\rangle_{\sH_{\cH}}
\nonumber\:.
\eeq
The proof ends provided we demonstrate (\ref{laaast}). We only sketch the argument leaving the details 
to the reader. The proof is based on the following result. If $\psi \in \sS(\cH)$ and $T\in \bR$, let us denote by 
$\psi_T \in \sS(\cH)$ the function such that $\psi_T(U,\theta,\phi) \doteq \psi(U-T,\theta,\phi)$. It is possible to prove that
\beq\label{notte}
\left(\sK_{\cH}(\psi_T)\right)(K,\theta,\phi)=e^{-iKT}\left(\sK_{\cH}(\psi)\right)(K,\theta,\phi)\:.
\eeq
 The proof of (\ref{notte}) is straightforward when $\psi \in C_0^\infty(\cH; \bR)$, since, in such case 
 $\sK_{\cH}$ is the positive frequency part of the $U$-Fourier transform of $\psi$. If $\psi \not\in 
 C_0^\infty(\cH; \bR)$, we can decompose it as $\psi_- + \psi_0 +\psi_+$, as in the definition of $\sK_{\cH}$, 
 fixing $\psi_-$ and $\psi_+$ in order that
 $(\psi_{\pm})_T$ are still supported in $(-\infty,0)$ and $(0,+\infty)$ respectively if $|T'| \leq T$.
If one uses the fact, which can be proved by inspection, that -- up to a re-definition of the 
 initially taken $\psi_n$ -- $C_0(\cH^+; \bR) \ni (\psi_{n})_T \to (\psi_+)_T$ in $H^1(\cH^+)_u$
 if $C_0(\cH^+; \bR) \ni \psi_n \to \psi_+$, one gets that (\ref{notte}) is valid for $\psi_+$.
The very same argument applies also $\psi_-$. The very definition of $\sK_{\cH}$ entails  the validity of 
(\ref{notte}) for every $\psi \in \sS(\cH)$, which, in turn, yields (\ref{laaast}) immediately, because, in 
the examined case,
$$ 
\sigma(\psi_1,\psi'_2) = 0 = - 2 Im \langle \sK_{\cH} \psi_1, \sK_{\cH} \psi'_2\rangle_{\sH_{\cH}}.
$$
The left hand side vanishes as $\psi_1,\psi'_2$ have disjoint supports, whereas the right-hand side can be 
re-written as:
\begin{gather*}
- 2 Im \int_{\bR \times \bS^2} \overline{e^{-iTK} \left(\sK_{\cH} g_1\right)(K,\theta,\phi)}
 e^{-iTK} \left(\sK_{\cH} \psi'_2\right)(K,\theta,\phi)\: 2KdK \wedge d\bS^2(\theta,\phi) 
 =\\ 
 -2 Im \left\langle \sK_{\cH}((g_1)_T), \sK_{\cH}((\psi'_2)_T) \right\rangle\:. 
\end{gather*}
 Such term is also vanishing, because we can fix $T$ so that $\supp((g_1)_T) \subset \cH^-$ and
 $\supp ((\psi'_2)_T) \subset \cH^+$, hence reducing to the case $\sigma(\psi_1,\psi_3')=0 =
- 2 Im \langle \sK_{\cH} \psi_1, \sK_{\cH} \psi'_3\rangle_{\sH_{\cH}}$ examined beforehand. \\
(d) is a trivial consequence of (c) : if $\sK_{\cH} \psi =0$, then
$Im \langle \sK_{\cH} \psi, \sK_{\cH} \psi'\rangle =0$ and thus $\sigma_{\cH}(\psi,\psi')=0$ for every 
$\psi' \in \sS(\cH)$. Since $\sigma_{\cH}$ is nondegenerate, it implies $\psi=0$. 
Let us prove (e). As $C_0^\infty(\cH;\bR)) \subset \sS(\cH)$, 
$$
\sH_{\cH} = \overline{F_{(U)}(C_0^\infty(\cH;\bR))}=
\overline{\sK_{\cH}(C_0^\infty(\cH;\bR))} \subset  \overline{\sK_{\cH}(\sS(\cH))}\subset 
 \sH_{\cH}\quad \mbox{and thus
$\overline{\sK_{\cH}(\sS(\cH))}   = \sH_{\cH}$.}$$
The first identity arises out of (a) in Proposition \ref{PropMain4}, the second out of (b) in 
Proposition \ref{Propembedding}.\\
We can now conclude proving (f). The continuity of $\sK_\cH$ with respect to the considered norm holds for 
the following reason. If $\{\psi_n\}_{n\in \bN} \subset \sS(\cH)$ and $||\psi_n||_\cH^\chi \to 0$, then, 
if we decompose $\psi_n = \psi_{0n}+ \psi_{+n}+ \psi_{-n}$, separately,
$\psi_{0 n}$ and $\psi_{\pm n} \to 0$ in the respective Sobolev topologies. In turn $\sK_\cH(\psi_{0n}) 
= F_U(\psi_{0n})\to 0$ because the Sobolev topology is stronger than that of $L^2(\bR \times \bS^2; dU\wedge 
d\bS^2)$ and $\sK_\cH(\psi_{\pm n})\to 0$ for (b) in Proposition \ref{propbastarda}.
Per definition of $\sK_\cH$, it hence holds $\sK_\cH(\psi_{n}) \to 0$. Thus the linear map $\sK_\cH:\sS(\cH) 
\to sH_\cH$ is continuous it being continuous in $0$. Particularly, we conclude, that there exists $C_\chi 
>0$ (the value $0$ is not allowed since $\sK_\cH$ cannot be null function) with $||\sK_\cH(\psi)||_{\sH_\cH} 
\leq  C_\chi ||\psi||_\cH^\chi$ for every $\psi \in \sS(\cH)$. The Cauchy-Schwartz inequality implies the 
one displayed in (f). $\Box$\\

\noindent {\bf Proof of Proposition \ref{propidscri}.} Let us define $v=x^2$ if $x\geq 0$
and $v= -x^2$ if $v<0$ .Per direct inspection one sees that, if
$\psi,\psi'\in C^\infty_0(\bR^*_-\times \bS^2; \bR)$, 
\begin{gather} 
\int_{\bR^2\times \bS^2} \frac{\psi(v,\theta,\phi)\psi'(v',\theta,\phi)}{(v-v'-i0^+)^2} dv
\wedge dv'\wedge d\bS^2(\theta,\phi)  = 
\int_{\bR^2\times \bS^2} \frac{\psi(v(x),\theta,\phi)\psi'(v(x'),\theta,\phi)}{(x'-x-i0^+)^2} dx\wedge dx' 
\wedge d\bS^2(\theta,\phi) \notag \\
+\int_{\bR^2\times \bS^2} \frac{\psi(v(x),\theta,\phi)\psi'(v(-x'),\theta,\phi)}{(x'-x-i0^+)^2} dx\wedge dx' 
\wedge d\bS^2(\theta,\phi)\label{AGG}\:.
\end{gather}
At a level of $x$-Fourier transform, let us denote as $\dot{\psi}= \dot{\psi}(h,\theta,\phi)$, with
$h\in \bR$ and $(\theta,\phi) \in \bS^2$ the $x$-Fourier transform of $\psi(v(x))$. Let us also define
$\dot{\psi}_+ \doteq \dot{\psi}\spa \rest_{\bR_+ \times \bS^2}$,
then, out of the fact that if $\phi$ is real valued, as it happens for $\psi$ 
and $\psi'$, then $\overline{\dot{\phi}_+(h,\theta,\phi)}$ is the $x$ Fourier transform of $x \mapsto 
\phi(-x,\theta,\phi)$), \eqref{AGG} can be re-written as
\begin{gather}
\lambda_{\scri}\left(\psi, \psi'\right) = \int_{\bR_+\times \bS^2}  \overline{\dot{\psi'}_+(h,\theta,\phi)}
\dot{\psi}_+(h,\theta,\phi) 2hdh \wedge d\bS^2 +\notag\\
 \int_{\bR_+\times \bS^2}  \overline{\dot{\psi'}_+(h,\theta,\phi)}\left(C\dot{\psi}_+\right)(h,\theta,\phi) 
 2hdh \wedge d\bS^2\:,\label{ddd}
 \end{gather}
where the operator $C : L^2(\bR_+\times \bS^2, 2hdh) \to L^2(\bR_+\times \bS^2, 2hdh)
$ is anti-unitary and it is nothing but the complex conjugation. Now let us take $\psi \in \sS(\scrim)$ which
is completely supported in $\bR_+\times \bS^2$. Per definition of $\sS(\scrim)$, the function $\psi=\psi(v(x)
)$ and its $x$-derivative belong to $L^2(\bR\times \bS^2, dx \wedge d\bS^2)$ and, thus, $\psi$ belongs to 
$H^1(\scri)_x$. A sequence of functions $\psi_n \in C^\infty(\bR^*_-\times \bS^2; \bR)$ which converges to 
$\psi$ in $H^1(\scrim)_x$ can be constructed as $\psi_n = \chi_n \cdot \psi$, where
$\chi_n(x) \doteq \chi(x/n)\geq 0$ with $\chi(x) = 1$ if $x\in (-1, +\infty)$ and $\chi(x) = 0$ for $x\leq -2$.
Per direct inspection and thanks to Lebesgue's dominated convergence theorem, one achieves that 
$C^\infty(\bR^*_-\times\bS^2;\bR)\ni\psi_n\to\psi$
in $H^1(\scrim)_x$ as $n\to +\infty$. Consequently $\dot{\psi_n}\to \dot{\psi}$ both in $L^2(\bR\times \bS^2, dh)$
and in $L^2(\bR_+\times \bS^2, h^2dh)$. Therefore $\dot{\psi_n}_+\to \dot{\psi}_+$ in the topology of 
$L^2(\bR_+\times \bS^2, 2hdh)$. Finally, in view of (\ref{ddd}) and on account of the continuity of $C$, the 
sequence $\{\psi_n\}_{n\in \bN}$ is of Cauchy type with respect to $\lambda_{\scrim}$. The same argument 
shows that, if $C^\infty(\bR^*_-\times \bS^2; \bR) \ni \psi'_n \to \psi$
in $H^1(\scrim)$ as $n\to +\infty$, then $\lambda_{\scrim}(\psi_n-\psi'_n, \psi_n-\psi'_n) \to 0$ as $n\to+
\infty$. Hence (\ref{FumF}) and (\ref{antilin}) are trivial consequences of what proved, which is tantamount
to verify (a) and (b). $\Box$\\

\noindent {\bf Proof of Proposition \ref{Propembeddingscri}.}
The proofs of items (a),(b),(d),(e) as well as (f) are very similar to those of the corresponding items 
in Proposition \ref{Propembedding}, so they will be omitted. We instead focus our attention on (c),
whose proof is similar to the same point (c) of Proposition \ref{Propembedding}, though with some relevant 
differences. Let us ake $\psi,\psi' \in \sS(\scrim)$ and let us decompose them as 
 $\psi= \psi_0 + \psi_1$, $\psi'= \psi'_0+ \psi'_1$ where $\psi_0,\psi_1 \in C_0^\infty(\scrim; \bR)$ while
 $\psi'_0,\psi'_1$ are supported in $(0,+\infty) \times \bS^2$. Since $\sigma\doteq \sigma_\scrim$ and
  $\langle, \rangle = \langle,\rangle_\scri$, it holds
 $$\sigma(\psi,\psi') = \sigma(\psi_0,\psi'_0)+\sigma(\psi_0,\psi'_1)+\sigma(\psi_1,\psi'_0)+\sigma(\psi_1,\psi'_1)\:.$$
Exactly as in (c) of the Proposition \ref{Propembedding}, we conclude that
\beq\label{ONE} 
\sigma(\psi_0,\psi'_0)= -2Im \langle \sK_\scrim \psi_0,\sK_\scrim \psi'_0 \rangle\:.\eeq 
With reference to $\sigma(\psi_1,\psi'_1)$, we have instead: 
$$
-2Im \langle \sK_\scrim \psi_1,\sK_\scrim \psi'_1 \rangle =
-2Im \langle F_{(v)}(\psi_1), F_{(v)}(\psi'_1) \rangle = -2 Im \lambda_\scrim(\psi_1,\psi'_1)\:.
$$
If we make both use of (\ref{antilin}), and of the fact that the above identity can be used for 
$\psi_1,\psi_1'$ as established in (b) of Proposition \ref{PropMain5}, we have 
$$
-2Im \langle \sK_\scrim \psi_1,\sK_\scrim \psi'_1 \rangle = 
-2 Im \int_{\bR_+\times \bS^2} \overline{\dot{\psi_1}} \dot{\psi_1'} 2hdh \wedge d\bS^2
-2 Im \int_{\bR_+\times \bS^2} \overline{\dot{\psi_1} \dot{\psi_1'}} 2hdh \wedge d\bS^2\:,
$$
where $\dot{\psi}(h,\theta,\phi)$ is the $x$-Fourier-Plancherel transform of $\psi=\psi(v(x),\theta,\phi)$.
The last term in the right-hand side can be omitted for the following reason. If we look at (\ref{AGG}), we 
see that $-i0^+$ can be replaced by $+i0^+$ without altering the result, 
 since the functions in the numerator have disjoint supports. This 
 is equivalent to say that,  in the right-hand side of (\ref{ddd}), the last term can be replaced 
 with its complex conjugation without affecting the final result. 
 Finally, this means that the identity written above can be equivalently recast as
 $$-2Im \langle \sK_\scri \psi_1,\sK_\scrim \psi'_1 \rangle = 
-2 Im \int_{\bR_+\times \bS^2} \overline{\dot{\psi_1}} \dot{\psi_1} 2hdh \wedge d\bS^2
-2 Im \overline{\int_{\bR_+\times \bS^2} \overline{\dot{\psi_1} \dot{\psi_1}} 2hdh \wedge d\bS^2}\:.$$
 As a consequence the last term can be dropped, so that:
\beq\label{TWO}
-2Im \langle \sK_\scri \psi_1,\sK_\scrim \psi'_1 \rangle = 
-4 Im \int_{\bR_+\times \bS^2} \overline{\dot{\psi_1}} h\dot{\psi'_1} dh \wedge d\bS^2
= 2i \int_{\bR\times \bS^2} \overline{\dot{\psi_1}} h\dot{\psi'_1} dh \wedge d\bS^2
= \sigma(\psi_1,\psi'_1)\:.\eeq
In the last passage we have used that $-ih\dot{\psi'_1}$ is the $x$-Fourier transform of 
$\partial_x\psi'_1$, and that the integration in $\sigma(\psi_1,\psi'_1)$ can be performed in the variable 
$x$ since the singularity of the coordinates at $x=0$ is irrelevant, the supports of $\psi_1$ and $\psi'_1$ 
being away from there. One should also notice that
these functions are real so that $\overline{\dot{\psi_i}(h,\theta,\phi)} = \dot{\psi_i}(-h,\theta,\phi)$.
Let us consider the term $\sigma(\psi_0,\psi'_1)$ the other, $\sigma(\psi_1,\psi'_0)$ can 
be treated similarly. To this end, let us decompose $\psi'_1 = \phi'_0 + \phi'_1$ in order that 
$\phi'_0 \in C_0^\infty(\scrim; \bR)$
 and the support of $\phi'_1$ is disjoint from that of $\psi_0$. Therefore:
 $\sigma(\psi_0,\psi'_1) = \sigma(\psi_0,\phi'_0) + \sigma(\psi_0,\phi'_1)
= -2Im \langle \sK_\scri \psi_1,\sK_\scrim \phi'_1 \rangle + 
\sigma(\psi_0,\phi'_1)\:.$
Since we shall prove that:
\beq\label{ULTIMA}
\sigma(\psi_0,\phi'_1) = 0 = -2Im \langle \sK_\scrim \psi_0,\sK_\scri \phi'_1 \rangle\:,\eeq
 we also have that
 $$\sigma(\psi_0,\psi'_1) = -2Im \langle \sK_\scrim \psi_0,\sK_\scrim \psi'_1 \rangle\:,
\quad \mbox{and similarly,}\quad \sigma(\psi_1,\psi'_0) = -2Im \langle \sK_\scrim \psi_1,\sK_\scrim \psi'_0 \rangle\:, 
$$
 which, together (\ref{ONE}) and (\ref{TWO}) implies the validity of (c) by bi-linearity:
 $$\sigma(\psi,\psi') = -2Im \langle \sK_\scrim \psi,\sK_\scrim \psi' \rangle\:.$$
 To conclude, it is enough to prove (\ref{ULTIMA}). The left-hand side vanishes since the supports of the 
 functions  $\psi_0,\phi'_1$ are disjoint by construction. Hence it remains to verify that
  $Im \langle \sK_\scri \psi_0,\sK_\scrim \phi'_1 \rangle =0$.
 If it were $\supp(\psi_0) \subset (-\infty, 0)\times \bS^2$ and $\supp(\phi'_1) \subset (0, +\infty)\times 
\bS^2$, one would achieve $Im \langle \sK_\scrim \psi_0,\sK_\scrim \phi'_1 \rangle =0$
 through the same argument used in the corresponding case 
(that of $\sigma(\psi_1,\psi'_3)$) in the proof of (c) of the Proposition \ref{Propembedding}. To wit, one
should employ a sequence of real smooth functions which tends to $\phi'_1$ in the topology of 
$\lambda_\scrim$ and with compact supports all enclosed in $(0, +\infty)\times \bS^2$. Such a sequence  
exists in view of Proposition \ref{PropMain5}.
As a matter of fact, we can focus our attention to the lone case $\supp(\psi_0)\subset (-\infty, 0)\times
\bS^2$ and $\supp(\phi'_1)\subset (0, +\infty)\times\bS^2$, thanks to the following lemma which will also
play a pivotal role in the proof of (b) of  Theorem \ref{Main4scri}.

\lemma\label{ULTIMOlemma} 
{\em For $\psi \in \sS(\scrim)$ and $L\in \bR$, let $\psi_L \in \sS(\scrim)$ denote the function with
$\psi_L(v,\theta,\phi) \doteq \psi(v-L,\theta,\phi)$ for all $v\in \bR$ and $\theta,\phi\in \bS^2$. With the given definition for 
$\sK_\scrim : \sS(\scrim) \to \sH_\scrim$, it holds:
\beq\label{ULTIMOlemmaEq}
\left(\sK_\scrim\psi_L\right)(k,\theta,\phi) = e^{-iLk} \left(\sK_\scrim\psi\right)(k,\theta,\phi)\:, \quad 
\forall (k,\theta,\phi) \in \bR_+\times \bS^2\:.
\eeq}

\noindent {\em Proof of Lemma \ref{ULTIMOlemma}}.
Per definition, if $\psi\in \sS(\scrim)$ is fixed,  $\sK_\scrim \psi = F_{(v)} \psi_0 + F_{(v)} \psi_-$, 
where $\psi = \psi_0+ \psi_-$ with $\psi_0 \in C_0^\infty(\scrim; \bR)$
and $\psi_-\in \sS(\scrim)$ with $\supp(\psi_-)\subset (-\infty,0) \times \bS^2$.
Let us fix $L \in \bR$ and let us
notice that, the very definition of $F_{(v)}$ on $C_0^\infty(\scrim;\bR)$ yields
$$F_{(v)} (\psi_0)_L =  e^{-iLk} F_{(v)}\psi_0\:.$$
To conclude it is sufficient to establish that it also holds:
\beq\label{stanco}
F_{(v)} (\psi_-)_L =  e^{-iLk} F_{(v)}\psi_-\:.
\eeq
Since the definition of $\sK_\scrim \psi$ does not depend on the chosen decomposition 
$\psi = \psi_0+ \psi_-$, we can fix 
$\psi_0$ and $\psi_-$ such that the support of $(\psi_-)_L$ is still included in $(-\infty, 0)$.
This holds true for every $\psi_-$ if $L\leq 0$, but it is not straightforward for $L > 0$ and,
in this case, the support of $\psi_-$ has to be fixed sufficiently far from $0$).
To establish (\ref{stanco}), let us use the coordinate $x=-\sqrt{-v}$ for $v<0$. The singularity
at $v=0$ does not affect the procedure since the supports 
of all the involved functions do not include it. We know, thanks to the proof of Proposition \ref{PropMain5},
that there exists a sequence $C^\infty_0(\scrim; \bR) \ni \psi_n \to \psi_-$, all supported in $\supp(\psi_-)
\subset (-\infty,0)\times \bS^2$; 
the convergence is here meant both in in the topology of $H^1(\scrim)$ and in that of $\lambda_{\scrim}$.
Per direct inspection, one sees that, for the above-mentioned sequence it holds
$\supp (\psi_n)_L \subset \supp(\psi_L) \subset (0,+\infty)\times \bS^2$ and
$\psi_n \to \psi_-$ entails $(\psi_n)_L \to (\psi_-)_L$ for 
$n\to +\infty$ in the topology of $H^1(\scrim)_x$. According to (b) in Proposition \ref{PropMain5}, this also
implies that the convergence holds in the topology of $\lambda_\scrim$. Since $F_{(v)}$ is continuous 
with respect to the last mentioned topology, we get, as $n\to +\infty$:
$$
e^{-iLk} F_{(v)}\psi_n = F_{(v)} (\psi_n)_L \to F_{(v)} (\psi_-)_L\:.
$$
On the other hand, since $F_{(v)}\psi_n \to F_{(v)}\psi_-$ in $L^2(\bR_+\times \bS^2, kdk \wedge d\bS^2)$, 
it trivially holds as $n\to +\infty$:
$$
e^{-iLk} F_{(v)}\psi_n \to e^{-iLk} F_{(v)}\psi_-\:,
$$
and thus
$$
e^{-iLk} F_{(v)}\psi = F_{(v)} (\psi_-)_L\:,
$$
 which implies (\ref{stanco}), concluding the proof. $\Box$\\
 
\noindent To conclude the proof of (c), we note that, in view of (\ref{ULTIMOlemmaEq}),
it must hold $-2Im \langle \sK_\scrim \psi_0,\sK_\scri \phi'_1 \rangle
= -2Im \langle \sK_\scrim (\psi_0)_L,\sK_\scrim (\phi'_1)_L \rangle$ for every $L \in \bR$. Therefore, we 
can fix $L$ so that $\supp (\psi_0)_L \subset (-\infty, 0)\times \bS^2$ and $\supp((\phi'_1))_L) \subset (0, 
+\infty)\times\bS^2$, obtaining, as said before,
$$-2Im \langle \sK_\scrim \psi_0,\sK_\scrim \phi'_1 \rangle
= -2Im \langle \sK_\scrim (\psi_0)_L,\sK_\scrim(\phi'_1)_L \rangle=0\:.$$ 
This implies (\ref{ULTIMA}) and, hence, (c). $\Box$\\
 
\noindent{\bf Proof of Proposition \ref{distrib}.} The first assertion arises per direct inspection of 
definition \ref{omegaHdef} and, thus, we need only to prove that 
 the (complexified) functionals on $C^\infty_0(\mM; \bC)\times C^\infty_0(\mM; \bC)$,  $\Lambda_{\cH}$ and $\Lambda_{\Im^-}$ are 
separately distributions in $\mD'(\mM\times \mM)$. To this end, it suffices to show that the maps $f\mapsto\Lambda_i(f,\cdot)$
and $g\mapsto\Lambda_i(\cdot,g)$ are weakly continuous, {\em i.e.},  they tend to $0$ when tested with any 
sequence of functions $h_j\in C^
\infty_0(\mM; \bC)$ which converges to $0$ in the topology of test functions. Here and hereafter 
the subscript $i$ stands either for $\cH$ or for $\Im^-$. 
According to theorem 2.1.4 of \cite{Hormander}, such statement entails that both $\Lambda_i(f,\cdot)$ and
$\Lambda_i(\cdot,g)$ are distributions in $\mD'(\mM)$, hence they are sequentially continuous. Once
established, one can, therefore, invoke the Schwartz' integral kernel theorem to conclude that $\Lambda_i\in
\mD'(\mM\times \mM)$.  In view of the complexification procedure, it is sufficient 
to consider only the case of real valued test functions.\\
Let us start with $\Im^-$; in this case $\la_{\Im^-}$ has the explicit form \eqref{antilin} 
in proposition \ref{propidscri} and, thus, one can take into account a generic decomposition \eqref{dec0scri}
generated by a smooth function $\eta$ supported on $\bR^*_-\times\bS^2$ and equal to one for $v<v_0<0$. 
Due to the continuity property, discussed at point (f) of Proposition \ref{Propembedding}, 
$$
| \Lambda_{\Im^-}(f,h)|=|\la_{\scrim}(\varphi^f_\scrim,\varphi^h_\scrim)| \leq C\; \| \varphi^f_\scrim  \|_\scrim^\eta \; \| \varphi^h_\scrim  \|_\scrim^\eta\;.
$$
We recall that, for every $\varphi\in\sS(\scrim)$,  $\|\varphi\|_\scrim^\eta$ is defined in 
\eqref{normeScri} as the sum 
$$
\|\varphi\|_\scrim^\eta=\|\eta \varphi\|_{H^1(\scrim)_x}+\|(1-\eta) \varphi\|_{H^1(\scrim)_v}\;.
$$
Continuity is tantamount to show that, for $f,g\in C_0^\infty(\mM;\bR)$, if $f\to 0$ with a 
fixed $g$ (or $g\to 0$ with a fixed $f$) in the topology of $C^\infty_0(\mM; \bC)$, both Sobolev norms,
above written, tend to zero. Let us start from the second one. For a given compact $K\subset\mM$, as in the 
proof of Lemma \ref{lemma1}, let us fix a sufficiently large globally hyperbolic spacetime $\mN\subset
\widetilde{\mM}$ which is  equipped with the metric $\widetilde{g}$ and which extends $\mM$ partly around 
$\Im^-$. Furthermore $\mN$ must include $K$ and $\mN\cap\Im^-$ should encompass all the points with $v \geq v_0$, reached by the closure in $\mM \cup \scrim$ 
of  $J^-(K;\mM)$. 
If we notice that the causal propagator $E_{P_{
\widetilde{g}}}$ is a continuous map from $C^\infty_0(\mM; \bR)\to C^\infty (\mN;\bR)$, one has that 
$(1-\eta)\varphi^f_\scrim$ and all the $v$-derivatives uniformly vanishes as $f \to 0$ in the topology of 
$C^\infty_0(\mM; \bC)$. Since, per constriction, all the functions $(1-\eta)\varphi^f_\scrim$
have support in a common compact of $\bR \times \bS^2$ determined by $\eta$ and $J^-(K;\mM))$, also 
$\|(1-\eta) \varphi_\scrim^f\|_{H^1(\scrim)_v}$ tends to zero in view of the integral expression of the 
Sobolev norm.\\
To conclude, in order to deal with the contribution $\|\eta \varphi_\scrim^f\|_{H^1(\scrim)_x}$, let us 
notice that, according to proposition \ref{PropDR} (point (b) in particular), 
the restriction of a solution of the D'Alembert wave 
equation on $\Im^-$ decays on null infinity, for $|v|$ greater than a certain $|v_0|$, as $\frac{C_f}{\sqrt{1+
|v|}}$, while its $v$-derivative as $\frac{C_f}{1+
|v|}$, where $C_f$ tends to $0$ as $f\to 0$ in the topology of $C^\infty_0(\mM;\bC)$. 
Hence per direct inspection, if we work with the coordinate $x$, also$\|\eta \varphi_\scrim^f\|_{H^1(\scrim)_
x}$ vanishes as $f$ tends to zero in the topology of $C^\infty_0(\mM;\bC)$.\\
The case of $\cH$ can be dealt with in the same way using the continuity presented in point $(f)$ of 
proposition $\ref{Propembedding}$ 
and the appropriate decay estimates of the wave functions presented in proposition \ref{PropDR}. As before, 
one can reach the conclusion that both $\Lambda_{\cH}(f,\cdot)$ and $\Lambda_{\cH}(\cdot,g)$ lie in $\mD'(
\mM)$.
$\qed$\\

\noindent{\bf Proof of Proposition \ref{RapidDecay}.} 
Let us start considering 
 $\|\varphi^{f_{p}}_\scrim 
\|_\scrim$ as defined in \eqref{normeScri} for some generic decomposition based on the choice of the function
$\eta$). Here $f\in C^\infty_0(\mM)$ and $p$ lie in a conic neighbourhood $V_{k_x}$ of $k_x$, we are going to
specify. The procedure we shall employ can be similarly used also for $\|\chi'\varphi^{f_{p}}_{\cH}\|_
{\cH^-}$ to show that it is rapidly decreasing in $p$. Furthermore we recall that $f_{p}\doteq f e^{i
\langle p,\cdot\rangle}$, while $\varphi^{f_{p}}_\scrim$ is the smooth limit towards of  $\scrim$ of $E_{P_g}f_{p}$, 
where $E_{P_g}$ is  the causal propagator of $P_g$ as in \eqref{KG}. 
Furthermore $\varphi^{f_{p}}_\scrim $, together with its 
derivative along the global null coordinate $v$ is known to decay at $-\infty$ according to 
the estimates \eqref{stime2} in Proposition \ref{PropDR}, in turn based on the work of \cite{DR05}, {\it 
i.e.},
$$
|\varphi^{f_{p}}_\scrim  |\leq \frac{C_3}{\sqrt{1+|v|}}\;,
\qquad
|X(\varphi^{f_{p}}_\scrim ) |\leq \frac{C_4}{{1+|v|}}\;.
$$
Here $X$ still stands for the smooth Killing vector field on the conformally extended Kruskal spacetime,
coinciding with $\partial_v$ on $\scrim$. 
They yields that the norm $\|\varphi^{f_{p}}_\scrim \|_\scrim$,
defined as in \eqref{normeScri}, is controlled by the above coefficients $C_3$ and $C_4$ which depend on $\varphi^{f_{p}}$, since the norms 
of the remaining universal functions smoothed about $i^-$ are finite.
Hence, we shall analyse them explicitly and we notice that all the relevant results in \ref{PropDR} can be
straightforwardly extended to the complex case. Our goal is to establish that the coefficients 
 $C_3$ and $C_4$ are rapidly decreasing in  $p$ when computed for $\varphi^{f_p}$ in the given hypotheses about $x$ and $k_x$. \\
As a starting point, let us consider the case in which $x\in I^+(\cB; \mM)$, $\cB$ being the bifurcation.
In order to study this, as well as all other scenarios, we make use of the results and of the techniques
available in \cite{DR05} of which we shall adopt nomenclatures and conventions. In this last cited paper it 
is manifest, that, up to a term depending on the support of initial data, the dependence on the wave function
in $C_3$ and $C_4$ is factorised in the square root of the so-called coefficient $\tilde{E}_5$, namely 
formula (5.4)  in \cite{DR05}. After few formal manipulations, the relevant expression can be (re)written as 
an integral over the constant 
time surface $\Si_1\subset\mW$, unambiguously individuated, in the coordinates $(t,r,\theta,\phi)$, 
as the locus $t=-1$. Here we consider $t=-1$ because we are interested in the decay property in a 
neighbourhood of $i^-$. Hence
\begin{gather}
\tilde{E}_5(\varphi)=
\sum_{i=1..2} \int_{\Si_1} T_{\mu\nu}(\Om^i\varphi)\;  n^\mu n^\nu d\mu (\Si_1)+
\sum_{i=1..4} \int_{\Si_1} T_{\mu\nu}(\Om^i\varphi)\;  K^\mu n^\nu d\mu (\Si_1)+ 
\nonumber
\\
+
\sum_{i=1..5} \int_{\Si_1} T_{\mu\nu}(\Om^i\varphi)\;  X^\mu n^\nu d\mu (\Si_1),
\label{E2}
\end{gather}
where $n$ is the vector orthogonal to $\Si_1$, pointing towards the past, and normalised as $g^{\mu\nu}n_\mu n_\nu=-1$,  
$K\doteq  v^2\frac{\pa}{\pa v}+u^2\frac{\pa}{\pa u}$ is the so-called {\em Morawetz vector field}, $X$ is 
the timelike Killing vector field $\frac{\pa}{\pa t}$, whereas
$d\mu
(\Si_1)$ is the metric induced measure on $\Si_1$. Furthermore, 
$$
T_{\mu\nu}(\varphi) = \frac{1}{2}\left(\pa_\mu \overline{\varphi} \pa_\nu \varphi + \pa_\nu \overline{\varphi} \pa_\mu \varphi \right)
-\frac{1}{2} g_{\mu\nu} \at  \pa_\la \overline{\varphi} \pa^\la \varphi\ct,
$$ 
stands for the 
stress-energy tensor computed with respect of the solution $\varphi$, while $\Om^2\doteq r^2\displaystyle{
\not}{\nabla}\displaystyle{\not}{\nabla}$ is the squared angular momentum operator, $\displaystyle{\not}{\nabla}$, being the 
covariant derivative induced by the metric \eqref{g}, normalised with $r=1$, on the orbits of $SO(3)$ isomorphic to $\bS^2$. 
  We remark both that the 
above expression can be found in theorem 4.1 in \cite{DR08} and, more important to our purposes, that the 
integrand is a (hermitian) quadratic combination of a finite number of derivatives of $\varphi^{f_{p}}$ on $\Si_1$. Furthermore, 
since $J^-(supp(f_p); \mM)\cap \Si_1$ is compact,  the integrand in \eqref{E2} does not vanish at most on
a compact set and, thus, the overall integral can be bounded by  a linear combination of products of  
the sup of the absolute value of derivatives of $\varphi^{f_{p}}$ up to a certain order, all evaluated on $\Si_1$.
Let us notice that, all the remaining functions in the integrand which define $\tilde{E}_5$,
barring the said products of derivatives, are continuous and, thus, bounded on the compact set where 
$\varphi^{f_{p}}$ does not vanish on $\Sigma_1$.\\
Let us thus focus on $\varphi^{f_{p}}$ itself as well as on both the initially chosen $x\in I^+(\cB;\mM)$ and
$k_x$. If one uses global coordinates, we identify an open relatively compact set $\cO$ which contains both 
the support of $f$ and that of the function $\rho$ we shall introduce in
$\bR^4$ by means of a local coordinate patch. In this way every vector $p \in \bR^4$ can be viewed as an 
element of the cotangent space at any point in that set.
It is also always possible to select $f\in C_0^\infty(\mM;\bR)$ with $f(x)=1$  
 and with a sufficiently small support, such that 
every inextensible geodesic starting from $supp (f)$, with cotangent vector equal to  $k_x$, intersects $\cH$ 
in a point with coordinate $U>0$. 
Hence, we can always fix $\rho\in C^\infty_0(\mK;\bR)$ such that (i) $\rho =1$ on $J^-(supp (f); \mM)\cap
\Si_1$ and (ii) the null geodesics emanating from $supp(f)$ with $k_x$ as cotangent vector do not meet the 
support of $\rho$. Furthermore, on account of the form of the wave front set of $E_{P_g}(z,z')$, now thought of in 
the whole Kruskal spacetime $\mK$, whose elements $(z,z',k_z, k_{z'})$ have always to fulfil $(z,k_z)\sim 
(z',-k_{z'})$, we realize that, with $(x,k_x)$ fixed as above and with the given definitions of $f$ and 
$\rho$,
$$
\left\{(x_1,x_2,k_1,k_x)\in T^*(\mM\times \mM)\:|\; x_1\in\supp(\rho),\;
x_2\in \supp(f),\;
k_1\in\bR^4
\right\}
\cap WF(E)=\emptyset\;.
$$
If we employ this result and if we remember the definition of wavefront set we can use Lemma  8.1.1 in 
\cite{Hormander}, though working in the coordinate frame initially fixed on the compact $\overline{\cO}$, to 
further adjust  $\rho$, $f$ while preserving the constraints already stated. In this way there exists
an open conical neighbourhood $V_{k_x}$ of $k_x$ in $T^*_x\mM$ such that for all $n,n' = 1,2,\dots$, one can 
find two nonnegative constants $C_n$ and $C'_n$ which fulfil
\beq 
|\widehat{\rho E f }(k_1,p)| \leq \frac{C_n}{1+|k_1|^n} \frac{C'_{n'}}{1+|p|^{n'}}, \label{Li}
\eeq
uniformly for $(k_1,p)\in(\bR^4\setminus\{0\})\times V_{k_x}$.
The searched bounds on the behaviour at large $|p|$ for $C_3$ and $C_4$, computed for $\varphi^{f_p}$ with 
$p$ in a open conical neighbourhood of $k_x$, arise in term of corresponding bounds of the derivatives 
$|\partial_x^a \overline{\varphi^{f_{p}}(x)} \partial_x^b \varphi^{f_{p}}(x)|$. One must take into account 
the explicit expression of both $C_3$ and $C_4$ as integrals over the relevant portion of $\Sigma_1$, which 
has finite measure because it is compact. 
Each factor $\partial_x^a \varphi^{f_{p}}(x)$ coincides with the inverse $k_1$-Fourier transform of  
$\widehat{\rho E f }(k_1,p)$ multiplied with powers of the components of $k_1$ up to a finite order which
depends on the considered degree of the derivative.
As a last step, to get rid of the $k_1$ dependence,  one
needs to integrate the absolute value over $k_1$, but the right hand side of (\ref{Li}) grants us that the 
overall 
procedure yields that the supremum of the integrand in \eqref{E2} is of rapid decrease in $p$ for all 
$p\in V_{k_x}$. \\
Nonetheless, the result is not yet conclusive since we still need to analyse the case in which the point 
$x$ lies in $\partial J^+(\cB; \mK)\cap \mM$, that is $x \in \cH_{ev}$. In such case, for every open cone 
$\Ga\in T^*_x\mM$ containing $k_x$, there exists $p\in \Gamma$ such that the inextensible geodesic which
starts form $x$ and it is tangent to $p$, meets the closure of $\Si_1$, hence reaching $\cB$.
Therefore, in order to apply the same argument as before, 
we need to modify the form of $\Si_1$ in the computation of \eqref{E2} in a neighbourhood of 
$\cB$. Therefore we need a slightly more refined estimate of the decay-rate of the solutions of 
\eqref{KG} on $\scrim$. 
This can be achieved if we adapt the proof of Theorem 1.1 in \cite{DR05} under the assumption that we modify 
the form of $\Si_1$, used to compute \eqref{E2} into that of another spacelike hypersurface, say $\Si'_1$, 
contained in $\overline{\mW}$ and such that it intersects $\cH$ at some negative value of the Kruskal null 
coordinate $U$. Hence it differs from $\Si_1$ only in a neighbourhood of $\cB$.

In the forthcoming discussion, we shall briefly review the arguments given in \cite{DR05} in order to show 
that it is really
possible to deform the initial surface $\Si_1$ on which the value of $\tilde{E}_5$ is computed,
preserving at the same time the decay estimates presented above as well as in \eqref{stime2}.
To this end we shall follow the discussion and the notation introduced in \cite{DR05} in order to obtain the
decay estimates in the neighbourhood of $i^+$. The desired estimates towards $i^-$ could be
obtained out of the time reversal symmetry. Let us start noticing that a central role in the analysis 
performed in \cite{DR05} is played by the flux generated by the Morawetz vector field
$K= v^2\frac{\pa}{\pa v}+u^2\frac{\pa}{\pa u}$.
Moreover, as explained in Section 9 of \cite{DR05}, the crucial estimates, 
are obtained out of the divergence, {\it a.k.a}, Stokes-Poincar\'e theorem, applied to the current 
$J^K_\mu(\varphi)$:
$$
J^K_\mu(\varphi)= K^\nu T_{\mu\nu}(\varphi) +  |\varphi|^2 \nabla_\mu \psi - \psi \nabla_\mu |\varphi|^2 \;,
\qquad \psi = \frac{t r^*}{4 r}\at 1-\frac{2m}{r} \ct
$$
which is generated by $K$ though with a modification due to total derivatives. If we follow such way of
reasoning, we can compute the mentioned flux between two spacelike smooth surfaces $\Si_1$ and $\Si_2$ in
$\mW$, identified respectively as the loci with fixed time coordinate $\{t=t_1\}$ and $\{t=t_2\}$, though
with $t_2>t_1$. The end point is  $$
\hat{E}^{K}_\varphi{(t_2)}=\hat{E}^{K}_\varphi{(t_1)}+ \hat{I}^K_\varphi(\cP),
$$
where $\hat{E}^{K}_\varphi{(t_2)}$  is the boundary term computed on $\Si_2$ and $\hat{I}^K_\varphi(\cP)$ is
the the volume term computed in the region $\cP\doteq J^+(\Si_1) \cap J^-(\Si_2)$. Let us notice that the 
integrand of the boundary terms $\hat{E}^K_\varphi(t_1)$ are everywhere positive, while, as it can be seen 
from Proposition 10.7 of \cite{DR05}, the one of the volume element $\hat{I}^K_\varphi(\cP)$ is negative
everywhere, but in the region $\cP\cap\{ r_0<r< R \}$ where the constant $r_0$ and $R$ (with $2m < r_0 < 3m 
<R$) are defined in section 6 of \cite{DR05}. For our later purposes, since we would like to eventually
deform both $\Si_1$ and $\Si_2$ in a neighbourhood of $\cB$, one should notice that the integrand is negative
on such a neighbourhood if chosen in the region $r<r_0$.

Since the pointwise decay estimate towards $i^+$ on $\scri$ can be obtained from $\hat{E}^K_\varphi(t)$, the
problem boils down to control the bad positive volume term in $\hat{I}^K_\varphi(\cP)$. Luckily enough, the
positive part of $\hat{I}^K_{\varphi}(\cP)$ can be tamed by  $t_2$ times $\hat{I}^{\bf X}_\varphi(
\cP)+\hat{I}^{\bf X}_{\Om \varphi}(\cP)$ where $\hat{I}^{\bf X}_\varphi(\cP)$ is the sum of the volume terms.
These arise out of the divergence theorem applied to the modified current generated by vectors like $X_\ell =
f_\ell(r^*)\frac{\pa}{\pa r^*}$ acting separately on an angular mode decomposition\footnote{Here $\ell(\ell+1
)$ is the eigenvalue of the angular momentum operator.}. We refer to Section 7 of \cite{DR05} for further
details on the construction of $\hat{I}^{\bf X}_\varphi(\cP)$ and to \cite{DR07} for recent results that do
not require a decomposition in modes.

Notice that, as discussed in proposition 10.2 of \cite{DR05} the boundary terms $| E^{\bf X}_{\varphi}
(t) |$ are always smaller then a constant $C$ times the conserved flux of energy $E_\varphi(t)$, with respect
to the Killing time $\frac{\pa}{\pa t}$. Hence, if we collect all these results, it is possible to write
\beq\label{EK}
\hat{E}^{K}_\varphi{(t)} \leq \hat{E}^{K}_\varphi{(t_1)}+ (t-t_1) C \at E_{\varphi^\chi}(t_1)+ E_{\Om
\varphi^\chi}(t_1) \ct,
\eeq
where $\Om$ is the square root of the angular momentum while $\varphi^\chi$ is a solution of the equation of
motion coinciding with $\varphi$ on $(t_1,t)\times (r_0,R)\times \bS^2$. This vanishes in a neighbourhood of 
$\cB$, as the one constructed in the proof of Proposition 10.12 in \cite{DR05}.
More precisely, for $t$ sufficiently close to $t_1$,
$\varphi^\chi$ can be chosen as the solution generated by the following compactly supported Cauchy data on 
$\Si_{t_1}$: $\varphi^\chi(t_1,r^*)=\chi(2 r^*/t_1) \varphi(t_1,r^*)$ and $\pa_t\varphi^\chi(t_1r^*)=\chi(2 
r^*/t_1)\pa_t \varphi(t_1,r^*)$, where $\chi$ is a
compactly supported smooth function on $\bR$ equal to 1 on $[-1,1]$ and vanishing outside $[-1.5,1.5]$.

As explained in Section 12.1 of \cite{DR05},
it can be shown that, if $t_2=1.1 t_1$ and $t_1$ is sufficiently large, then $E_{\varphi^\chi}(t_2)
\leq C\; t_2^{-2}\; \hat{E}^{K}\varphi {(t_2)}$ and this allows to obtain a better estimate then \nref{EK}, 
namely it yields
\beq\label{sommaintervalli}
t_2 \hat{I}^{\bf X}_\varphi(\cP)  \leq \frac{C}{t_2}\hat{E}^{K}_\varphi{(t_1)}+ C \at E_{\varphi^\chi}(t_1)+ E_{\Om
\varphi^\chi}(t_1) \ct,
\eeq
which is valid for $t_2=1.1 t_1$ in particular.
The estimate for a generic interval $t-t_1$ can be obtained, along the lines of Section 12.1 of \cite{DR05},
dividing $t-t_1$ in sub interval $t_{i+1}= 1.1 t_{i}$ and eventually summing the estimates
\eqref{sommaintervalli} over $i$. In such a way it is possible to obtain
$$
t \hat{I}^{\bf X}_\varphi(\cP)  \leq C\hat{E}^{K}_\varphi{(t_1)}+ C \log(t) \at E_{\varphi^\chi}(t_1)+ E_{\Om
\varphi^\chi}(t_1) \ct,
$$
for a generic interval. As a final step, if we apply
the same reasoning for $t\hat{I}^{\bf X}_{\Omega \varphi}(\cP)$ and if we use both of them to control 
$\hat{I}^K_\varphi$, we obtain a better estimate for $\hat{E}^K$ then the one \nref{EK}, namely
\beq\label{EK2}
\hat{E}^{K}_\varphi{(t)}=C \hat{E}^{K}_{\varphi}{(t_1)}+C\hat{E}^{K}_{\Om\varphi}{(t_1)}+ C \log(t) \at E_
{\varphi^\chi}(t_1)+ E_{\Om\varphi^\chi}(t_1) + E_{\Om\Om\varphi^\chi}{(t_1)}\ct,
\eeq
where $t$ in \nref{EK} is substituted by $\log(t)$, the price to pay in order to consider higher angular derivatives.

The $\log(t)$ can eventually be removed once again out of the same line of reasoning, using \nref{EK2} in 
place of \nref{EK} to improve \nref{sommaintervalli}. The end point is
$$
\hat{E}^{K}_\varphi{(t_2)} \leq C \at \sum_{n=0..3}\hat{E}_{\Om^n\varphi}{(t_1)} + \sum_{n=0..2}\hat{E}^{K}_{
\Om^n\varphi}{(t_1)}\ct \leq \tilde{E}_5(\Si_1).
$$
We would like to stress that,
since the integrand $I^K_\varphi(\cO)$ is positive whenever $\cO$ is a small neighbourhood of $\cB$,
the very same results can be obtained out of a modification of the surfaces $\Si_1$ and $\Si_2$ in such a way
that they are still spacelike while they intersect the horizon $\cH_{ev}$ at positive $V$ equal to $V_0$; in
this new framework the form of $\tilde E_5(\Si_1')$ is left unaltered with respect to \eqref{E2}, though it
is computed on a modified surface $\Si_1'$. The decay estimate towards $i^+$ on $\scri$ can eventually be
obtained as in Section 13.2 of \cite{DR05}. At this point, out of time reversal, we can employ a similar
argument as before in order to get the rapid decrease in $p$ of $\|\varphi^{f_{p}}_\scrim \|^\eta_\scrim$.

The horizon case can be dealt in a similar way and, in such case, the pointwise decay on $\cH^-$ can be shown 
to be controlled by an integral similar to the one defining $\tilde E_5$, though here it is again computed on
the modified surface. In order to establish the mentioned peeling off rate, it is, however,
necessary to consider another flux, namely that generated by a vector field $Y$ which approaches $\frac{1}{1-
\frac{2m}{r}}\pa_u$ on the horizon $\cH$, as described in Section 8 of \cite{DR05}. In this framework, even
if the integrand of the volume term $\hat{I}^Y_\varphi$, associated with $Y$, is negative in a region formed
by the compact interval $[\hat r_0,R]$, it can be controlled in a similar way as previously discussed for $\hat{I}^{\bf X}_\varphi$.
$\qed$ \\

\vsp
 
\noindent {\bf Proof of Lemma \ref{Nozero}.}  
As a starting point, let us recall that $\Lambda_U$  is a weak-bisolution of \eqref{KG}, whose antisymmetric part 
is nothing but the causal propagator $ E_{P_g}$ in $\mM$. The wave-front set of $E_{P_g}$ is well-known 
\cite{Rada} and it contains only pair of non-vanishing light-like covectors, so that:
$$
(x,y,k_x,0)\notin WF(E_{P_g}) \;,\qquad   (x,y,0,k_y)\notin WF(E_{P_g})\;.
$$
Therefore, whenever $(x,y,k_x,0)\in WF(\Lambda_U)$, also $(y,x,0,k_x)$ must lie in $WF(\Lambda_U)$ and 
{\em vice versa}; otherwise
 the wavefront set of the antisymmetric part of $\Lambda_U$, which is nothing but $E_{P_g}$, would contain a forbidden  element $(x,y,k_x,0)$. This
allow us to focus only on an arbitrary, but fixed $(x,y,k_x,0)\in T^*(\mM\times \mM)\setminus\{0\}$ and we
need to show that it does not lie in $WF(\Lambda_U)$. 
Furthermore we know, thanks to {\em Part 1} of the proof of theorem \ref{maxt}, that $\Lambda_U$ is of Hadamard
form in $\mW$ and, thus, the statement of this lemma holds if $x,y \in \mW$. 
We shall hence focus on the case of $x\in\mM \setminus \mW$ and $y\in\mW$, the remaining ones will be treated
later. In this scenario, it suffices to consider only those $k_x$ such that there are no representatives of 
$B(x,k_x)$ lying in $\mW$, otherwise we would be falling in the already discussed case using a propagation of 
singularities argument. This restriction yields, however, that a representative 
$(q,k_q)\in B(x,k_x)$ exists such that $q\in \cH^+\cup \cB$. Summarising, we are going  to prove 
that $(x,y,k_x,0)$ is a direction of rapid decreasing for $\Lambda_U(f_{k_x},h)$, for some functions $f,g\in 
C_0^\infty(\mM; \bR)$ with $f(x)=h(y)=1$, provided that both
$x\in\mM \setminus \mW$, $y\in\mW$ and a representative $(q,k_q)\in B(x,k_x)$ exists such
that $q\in \cH^+\cup \cB$. As before, $f_{k_x}\doteq fe^{i\langle k_x,\cdot\rangle}$ and $\vphi^h \doteq Eh$.\\
In this scenario, let us pick a partition of unit $\chi+\chi'=1:\cH\to\bR$ where 
$\chi\in C^\infty_0(\cH; \bR)$ and  $\chi= 1$ in a neighbourhood of $q$. Hence
\beq\label{dec2}
\Lambda_{U}(f_{k_x},h)=\la_{\cH} (\chi \varphi^{f_{k_x}}_{\cH},\varphi^h_{\cH})+\la_{\cH}(\chi'\varphi^{
f_{k_x}}_{\cH},\varphi^h_{\cH})+\la_{\scrim}(\varphi^{f_{k_x}}_{\scrim},\varphi^h_{\scrim}).
\eeq
The second and third terms are rapidly decreasing in $k_x$ because they are respectively dominated by 
$C \|\chi' \varphi^{f_{k_x}}_{\cH} \|_{\cH^-} \cdot \| \varphi^h_{\cH}\|_{\cH^-}$ and 
$C' \|\varphi^{f_{k_x}}_{\scrim} \|_\scrim \cdot \| \varphi^h_{\scrim}\|_\scrim$, $C$ and $C'$ being
positive constants, which, in turn, are rapidly decreasing in $k_x$ due to Proposition \ref{RapidDecay}. The
norms $\|\cdot\|_{\cH}$ and $\|\cdot\|_\scrim$ are those respectively defined in \eqref{norme} and in
\eqref{normeScri}.
Therefore, we need only to establish that $k_x$ is of rapid decreasing for 
$\la_{\cH} (\chi \varphi^{f_{k_x}}_{\cH},\varphi^h_{\cH})$. This can be done 
by the same procedure as that used at the end of the {\bf case A} in the proof of Theorem \ref{maxt} leading to (\ref{aggI}),
 to prove the rapid decrease of $k_x \mapsto \la_{\cH} (\chi \varphi^{f_{k_x}}_{\cH},\varphi^{h_{k_y}}_{\cH})$
for a fixed $k_y$ and assuming $k_y=0$ there (that part of the proof is independent form the lemma we are proving here,
 while this lemma is used elsewhere therein). \\
 
Let us now treat the case $y\in\mM \setminus \mW$ and $x\in\mW$, and let us prove that $(x,y,k_x,0)\not\in WF
(\La_U)$ following procedures analogous to those exploited in \cite{SV00} . 
To this end we adopt an overall frame where a coordinate, indicated by $t$, is tangent to $X$ and the 
remaining three coordinates are denoted as 
$\overline{x}$. In this setting, the pull-back action of the one-parameter group generated by $X$ 
acts trivially as $(\beta_\tau f)(t,\overline{x}) = f(t-\tau,\overline{x})$.
To start with, let us notice that, due to the restriction \nref{PST1}, the cases of $k_x$ 
spacelike or timelike can be immediately ruled out, so we are left to consider $k_x\in T^*_x(\mM\setminus\mW)
\equiv \bR^4$ of null type. Hence we can exploit the splitting 
$k_x=({k_x}_t,\overline{k_x})$, where we have isolated the $t$-component from the three remaining ones 
$\overline{k_x}$.  \\
For $k_x$ as before, let us  consider the two non-null and non-vanishing covectors $q=(0,\overline{k_x})$ and
$q'=(-{k_x}_t,0)$. In view of \eqref{PST1} $(x,y,q,q')\not\in WF(\La_U)$, 
hence, out of (c) of Proposition 2.1 of \cite{Verch}, there exists an open neighbourhood $V'$ of $(q,q')$,
as well as a function $\psi'\in C^\infty_0(\bR^4\times\bR^4; \bC)$ with $\psi'(0,0)=1$, such that for all 
$n\geq 1$,
\beq\label{bastardissimo}
\sup_{k,k'\in V'} 
\left|\int d\tau d\tau' d \overline{x}' d\overline{y}' \; 
\psi'(x',y') \; 
e^{i \la^{-1}(k_t\tau + {\overline{k}} \overline{x}')}
e^{i \la^{-1}(k_t'\tau' + {\overline{k}'} \overline{y}')}
\Lambda_{U}(\beta_\tau \otimes\beta_{\tau'} (F^{(p)}_{(\overline{x}', \overline{y}),\la}))\right| \leq C_n 
\la^n,
\eeq
which holds for every $ 0<\la<\la_n$, where both $C_n\geq 0$ and $\lambda_n>0$ are suitable constants.
In the preceding expression we have employed the notation $x'=(\tau,\overline{x}')$, and 
$y'=(\tau',\overline{y}')$ where we have highlighted the $t-$component. 
Moreover  
$F^{(p)}_{(\overline{x}', \overline{y'})\lambda}(z,u)$ is defined as follows 
$$
F^{(p)}_{(\overline{x}',\overline{y}'),\lambda}(z,u)\doteq F(x+\lambda^{-p}(z-\overline{x}'-x),  y+\lambda^{-p}(u-\overline{y}'-y) )\;, \qquad F \in 
C^\infty_0(\mM\times \mM; \bC)\;,\qquad \widehat{F}(0,0)=1\:,
$$
$ \widehat{F}$ being the standard Fourier transform.
At this point we can make a clever use of the translation invariance of $\Lambda_U$ under the action of 
$\beta_{-\tau-\tau'}\otimes\beta_{-\tau-\tau'}$ in order to infer that
$\Lambda_{U}(\beta_\tau\otimes \beta_{\tau'}(F^{(p)}_{(\overline{x}', \overline{y}'),\la}))$ 
is equal to $\Lambda_{U}(\beta_{-\tau'}\otimes \beta_{-\tau}(F^{(p)}_{(\overline{x}', \overline{y}'),\la}))$,
Hence, from \eqref{bastardissimo}, it arises that, for all $p\geq 1$:
$$
\sup_{k,k'\in V} 
\left|\int d\tau d\tau' d \overline{x}' d\overline{y}' \; 
\psi'(x',y') \; 
e^{i \la^{-1}(k_t\tau + {\overline{k}} \overline{x}')}
e^{i \la^{-1}(k_t'\tau' + {\overline{k}'} \overline{y}')}
\Lambda_{U}(\beta_{-\tau'} \otimes\beta_{-\tau} (F^{(p)}_{(\overline{x}', \overline{y}),\la})))
\right| \leq C_n \la^n,
$$
if $0<\lambda<\lambda_n$. The found result implies that \nref{bastardissimo} also holds 
if one replaces (i)  $\psi'$ with $\psi(x',y')\doteq \psi((\tau',\overline{x}'),(\tau,\overline{y}'))$
and  (ii) $V'$ with $V=\{ (-k'_t, \overline{k}),(-k_t, \overline{k'}) ) \in \bR^4\times \bR^4 \:|\:  ((k_t,
\overline{k}),(k'_t, \overline{k'})) \in V' \}$. This is an open neighbourhood of $(k_x,0)$ as one can 
immediately verify since  $V' \ni (q,q')$,  so that both $V \ni (k_x,0)$,  and the map 
 $ \bR^4\times \bR^4  \ni ((k_t, \overline{k}),(k'_t, \overline{k'}))  \mapsto  (-k'_t, \overline{k}),(-k_t, 
 \overline{k'}) ) \in \bR^4 \times \bR^4$ is a homeomorphism, it being linear and bijective. 
If one exploit once more  proposition 2.1 of \cite{Verch}, it yields that $(x,y,k_x,0) \not \in WF(\La_U)$ as
desired.\\ In order to conclude the proof, we need to analyse the last possible case, namely both $x, y\in 
\mM \setminus \mW$.  If a representative of either $B(x,k_x)$ or $B(y,k_y)$ lies in $T^*\mW$, 
we fall back in the previous analysis. Hence, we need only to focus on the scenario
where no representatives of both $B(x,k_x)$ and  $B(y,k_y)$ lies in $T^*\mW$. In this case, we can make use 
of an argument  substantially identically to the one used in the analysis above, {\it i.e.}, if we introduce
a partition of unit on ${\cH}$ for both variables. In this way we have a decomposition like \eqref{dec2} with
two more terms which can be analysed exactly as the others, thus leading to the wanted statement. $\qed$

\end{document}